\documentclass[aps,pre,twocolumn,longbibliography]{revtex4-1}
\usepackage{graphicx}  
\usepackage{dcolumn}   
\usepackage{bm}        
\usepackage{amssymb}   
\usepackage{amsmath}
\usepackage{mathtools}

\begin{document}

\title{Low-dimensional model of the large-scale circulation of turbulent Rayleigh-B{\'e}nard convection in  a cubic container} 

\author{Dandan Ji}
\affiliation{Department of Physics, Yale University, New Haven, CT 06511, USA}

\author{Eric Brown}
\email{ericmichealbrown@gmail.com}
\affiliation{Department of Mechanical Engineering and Materials Science, Yale University, New Haven, CT 06511, USA}

\date{\today}

\begin{abstract}
We  test the ability of a low-dimensional turbulence model to predict how dynamics of large-scale coherent structures such as convection rolls change in different cell geometries. The model consists of stochastic ordinary differential equations, which were derived from approximate solutions of the Navier-Stokes equations.   We test the model using Rayleigh-B\'enard convection experiments in a cubic container, in which there is a single convection roll known as the large-scale circulation (LSC). The model describes the motion of the orientation $\theta_0$ of the LSC as diffusion in a potential determined by the shape of the cell.  The model predicts advected oscillation modes, driven by a restoring force created by the non-circular shape of the cell cross-section.  We observe the corresponding lowest-wavenumber predicted advected oscillation mode in a cubic cell, in which the LSC orientation $\theta_0$ oscillates around a corner, and a slosh angle $\alpha$ rocks back and forth, which is distinct from the higher-wavenumber advected twisting and sloshing oscillations found in circular cylindrical cells.  Using the Fokker-Planck equation to relate probability distributions of $\theta_0$ to the potential, we find that the potential has quadratic minima near each corner with the same curvature in both the LSC orientation $\theta_0$ and slosh angle $\alpha$, as predicted.  To quantitatively test the model,  we report values of diffusivities and damping time scales for both the LSC orientation $\theta_0$ and temperature amplitude for the Rayleigh number range $8\times10^7 \le Ra \le 3\times 10^9$.  The new oscillation mode around corners is found above a critical Ra $=4\times10^8$.  This critical Ra  appears in the model as a  crossing of an underdamped-overdamped transition. The natural frequency of the potential, oscillation period, power spectrum, and critical Ra for oscillations are consistent with the model if we adjust the model parameters by up to a factor of 2.9, and values are all within a factor of 3 of model predictions. However, these uncertainties in model parameters are too large to correctly predict whether the system is in the underdamped or overdamped state at a given Ra.  Since the model was developed for circular cross sections,  the  success of the model at predicting the potential and its  relation to other flow properties for a square cross section -- which has different flow modes than the circular cross section --  suggests that such a modeling approach could be applied more generally to different cell geometries that support a single convection roll.
\end{abstract}

\maketitle    


\section{Introduction}
While turbulent flows are often thought of as irregular and erratic, large-scale coherent flow structures are commonplace in turbulence.  Examples of such structures include convection rolls in the atmosphere, oceans, and many other geophysical flows.  Such structures and their dynamics can play a significant role in heat and mass transport. 

 A particular challenge is to predict the dynamical states of these large-scale flow structures, and how they change  with different boundary  geometries,  for example in the way that local weather patterns depend on the topography of the Earth's surface.  The Navier-Stokes equations that describe flows are impractical to solve for such turbulent flows, so low-dimensional models are desired. It  has long been  recognized that the dynamical states of large-scale coherent structures are similar to those of low-dimensional dynamical systems  models \cite{Lo63} and stochastic ordinary differential equations   \cite{BA07a, TB07, TMS14, RMBM15}.  However, such models tend to be descriptive rather than predictive, as parameters are typically fit to each observation, rather than derived from fundamentals such as the Navier-Stokes equations  \cite{HL96}. In particular, such dynamical systems models tend to fail at  quantitative predictions of new dynamical states in regimes outside where they were parameterized.  Our goal is to develop and test a general low dimensional model that can quantitatively predict the different dynamical states of large-scale coherent structures in different geometries.   

We  test the application of a low-dimensional model to different geometries in the model system of turbulent Rayleigh-B\'enard convection.  In Rayleigh-B\'enard convection, a  fluid is heated from  below and cooled from above to generate buoyancy-driven flow  \cite{AGL09,LX10}. This system exhibits robust large-scale coherent structures that retain a similar organized flow structure over a long time. For example, in containers  of aspect ratio near 1, a large-scale circulation (LSC) forms. This LSC consists of localized blobs of coherent fluid known as plumes.  The plumes collectively form a single convection roll in a vertical plane  that can be identified with appropriate averaging over the flow field or timescales on the order of a turnover period of the circulation \cite{KH81}.  This LSC spontaneously breaks the symmetry of symmetric containers, but turbulent fluctuations cause the LSC orientation $\theta_0$ in the horizontal plane to meander spontaneously and erratically and allow it to sample all orientations \cite{BA06a}.  While the LSC exists nearly all of the time, on rare occasions these fluctuations lead to spontaneous cessation  followed by reformation of the LSC \cite{BA06a, XX07}.  In circular-cross-section containers, the LSC exhibits an oscillation mode \cite{HCL87, SWL89, CGHKLTWZZ89, CCL96, TSGS96, CCS97, QYT00, QT01b, NSSD01, QT02, QSTX04, SXT05,TMMS05} consisting of twisting and sloshing \cite{FA04, XZZCX09, ZXZSX09, BA09}, and in some cases a jump-rope-like oscillation \cite{VHGA18}.  The Coriolis force causes a  rotation of the LSC orientation \cite{BA06b,ZLW17, SLZ16}. 

There are several low-dimensional models for various aspects of LSC dynamics \cite{SBN02, Be05, FGL05, RPTDGFL06, BA08a, CV11, PS15, GKKS18, VFKSSV19}.  Of these, only a few attempt to address geometry dependence  of dynamics.  In one, an approximate analytical model was applied to ellipsoidal containers, which predicted an oscillation with a geometry-dependent restoring force \cite{RPTDGFL06}.  However the methods used could not be applied to other geometries.  Brown \& Ahlers proposed a model of diffusive motion in a potential well, which is  formulated in a way that it can make predictions for a wide variety of container geometries based on predictions of the potential as a function of the container cross-section geometry  \cite{BA07a, BA08a, BA08b}.

The model of Brown \& Ahlers and its extensions have successfully described most of the known dynamical modes of the LSC in circular horizontal cross-section containers including the meandering, cessations, and oscillation modes described above \cite{BA08a, BA08b, BA09,ZLW17, SLZ16}, with the exception of the recently observed jump rope mode which has not yet been modeled. The combination of twisting and sloshing oscillation modes  \cite{FA04, XZZCX09, ZXZSX09, BA09} can alternatively be described in this model as a single advected oscillation mode with two oscillation periods per LSC turnover period \cite{BA09}.  Predictions are typically quantitatively accurate within a factor of 2, but can be more accurate when more fit parameters are used \cite{AAG11}.  

While a circular cross section is the trivial case of that model when it comes to geometry dependence (i.e.~the geometry-dependent term was equal to zero), the model of Brown \& Ahlers has also successfully described some states and dynamics that are unique to non-circular cross section containers.  In a container with rectangular horizontal cross-section, the model describes the preferred alignment of the LSC orientation $\theta_0$ with the longest diagonals, and an oscillation of $\theta_0$ between nearest-neighbor diagonals \cite{SBHT14}.  The model also successfully predicted the existence of a stochastic switching of the LSC orientation between adjacent corners of a cubic container \cite{LE09, BJB16, FNAS17, GKKS18, VSFBFBR16, VFKSSV19}, including a quantitative prediction of the frequency of events, and how it varies with the flow strength \cite{BJB16}.

In this manuscript, we test a couple of geometry-dependent predictions of the model that have not yet been tested quantitatively.  One of these predictions is how the  probability distribution of the LSC orientation is affected by the container geometry, which directly connects to the potential that provides a forcing to the LSC orientation via the Fokker-Planck equation. In a cubic cell, for example, the geometry is predicted to lead to four preferred orientations aligned with the diagonals of the cell \cite{BA08b}.  While it has been known since before the development of the model  the LSC tends to align with the diagonals of a cubic container \cite{Zimin1978,ZML90,QX98, VPCG07}, the probability distribution has not yet been quantitatively compared to the prediction, and thus the potential has not been characterized.  The second quantitatively untested prediction of the model relates to the advected oscillation mode of the LSC around corners of a cell \cite{BA08b, SBHT14}.  While oscillations around the longest diagonals of a cross-section have now been observed in horizontal cylinders \cite{SBHT14} and a cube \cite{GKKS18}, quantitative tests of the model have not yet been reported regarding the frequency and power spectrum.   This test includes a prediction of a different structure of the advected oscillation mode in non-circular cross sections which has not yet been observed: an oscillation in $\theta_0$ in-phase at all heights and sloshing oscillation out of phase in the top and bottom halves of the cell  \cite{BA09}.

The remainder of this manuscript is organized as follows.  The low-dimensional model of Brown \& Ahlers is summarized in Sec.~\ref{sec:model}.  Details of the experimental apparatus design, calibrations, and methods used to characterize the LSC orientation are described in Sec.~\ref{sec:methods}.   Probability distributions to characterize the model potential and compare to predictions are presented in Sec.~\ref{sec:potential}.  The detailed prediction for the advected oscillation and measured power spectra are presented in Sec.~\ref{sec:powerspec}.  Predictions of the oscillation structure and the correlation functions used to characterize it are presented  in Sec.~\ref{sec:oscillation_structure}.  The frequency of measured oscillations for different Rayleigh numbers to test the conditions for the existence of the oscillations is presented  in Sec.~\ref{sec:period}.  Measurements of the different model parameters to apply the model to cubic containers and make comparisons to predictions and  measurements in other geometries are presented in Appendix 1.

  \section{Review of the low-dimensional model}
\label{sec:model}

In this section we summarize the model of Brown \& Ahlers, including its geometry-dependent potential \cite{BA08a}.  The model consists of a pair of stochastic ordinary differential equations,  using the empirically known,  robust LSC structure as an approximate solution to the Navier-Stokes equations to obtain equations of motion for parameters that  describe the LSC dynamics.  The effects of fast, small-scale  turbulent fluctuations are separated from the slower, large-scale motion when  obtaining this approximate solution,  then added back in as a stochastic term in the low-dimensional model.   The flow strength in the direction of the LSC is characterized by the temperature amplitude $\delta$, corresponding to half the temperature difference from the  the upward-flowing side of the LSC to the  downward-flowing side.   The equation of motion for $\delta$ is
\begin{equation}
\dot\delta = \frac{\delta}{\tau_{\delta}} - \frac{\delta^{3/2}}{ \tau_{\delta}\sqrt{\delta_0}} + f_{\delta}(t)\ .
\label{eqn:delta_model}
\end{equation}  
The first  forcing term on the right side of the equation corresponds  to buoyancy, which strengthens the LSC, followed by damping, which weakens the LSC.    $\delta_0$ is the stable fixed point value of $\delta$ where buoyancy and damping balance each other, and $\tau_{\delta}$  is a damping timescale for changes in the strength of the LSC.    $f_{\delta}(t)$ is a stochastic forcing term representing the effect of small-scale turbulent fluctuations and is modeled as Gaussian white noise with diffusivity $D_{\delta}$.  

The equation of motion for the LSC orientation $\theta_0$ is
\begin{equation}
\label{eqn:theta_model}
\ddot{\theta}_0 = - \frac{\dot{\theta}_0\delta}{\tau_{\dot{\theta}}\delta_0} - \nabla V_g(\theta_0,\alpha) + f_{\dot{\theta}}(t) \ . 
\end{equation}   
The first term on the right side of the equation is a damping term where  $\tau_{\dot{\theta}}$ is a damping time scale for changes of orientation of the LSC. $f_{\dot{\theta}}$ is another stochastic forcing term with  diffusivity $D_{\dot\theta}$.  $V_g$ is a potential which physically represents the pressure of the sidewalls acting on the LSC.  Equation \ref{eqn:theta_model} is mathematically equivalent to diffusion in a potential landscape $V_g(\theta_0, \alpha)$. 

\begin{figure}
\includegraphics[width=0.475\textwidth]{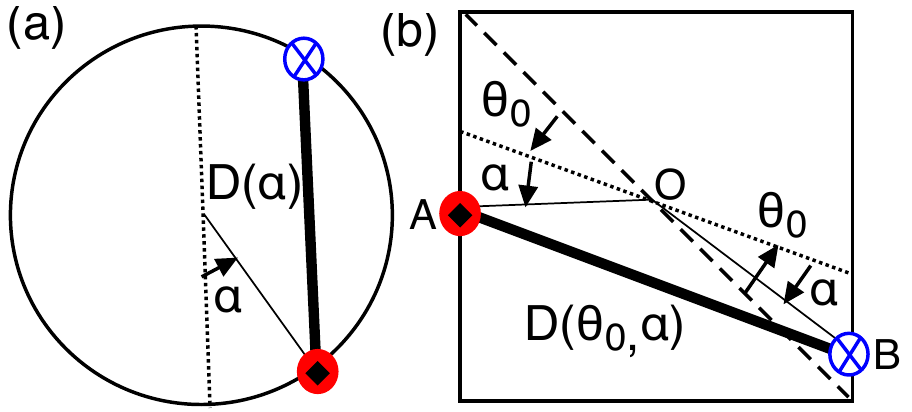}
\caption{(a) Top view of a circular cross-section container. (b) Top view of a square cross-section container.  In each case, the circulation plane of the LSC is indicated by the thick solid line, and the hot  side is indicated by the circle with a diamond in the center, while the cold side is indicated by the circle with a cross in the center.  The slosh angle $\alpha$  is defined by the  lateral displacement of the plane of circulation away from a line going through the cell center (dotted line).  The orientation of the LSC is defined as the angle $\theta_0$ of the hot side of the center line (dotted line) relative to a corner in a cube (dashed line).    The length of the circulation plane $D(\theta_0,\alpha)$ across a horizontal cross-section determines the model potential in Eq.~\ref{eqn:potential}.
}
\label{fig:angle_definitions}
\end{figure} 

\begin{figure}
\includegraphics[width=0.475\textwidth]{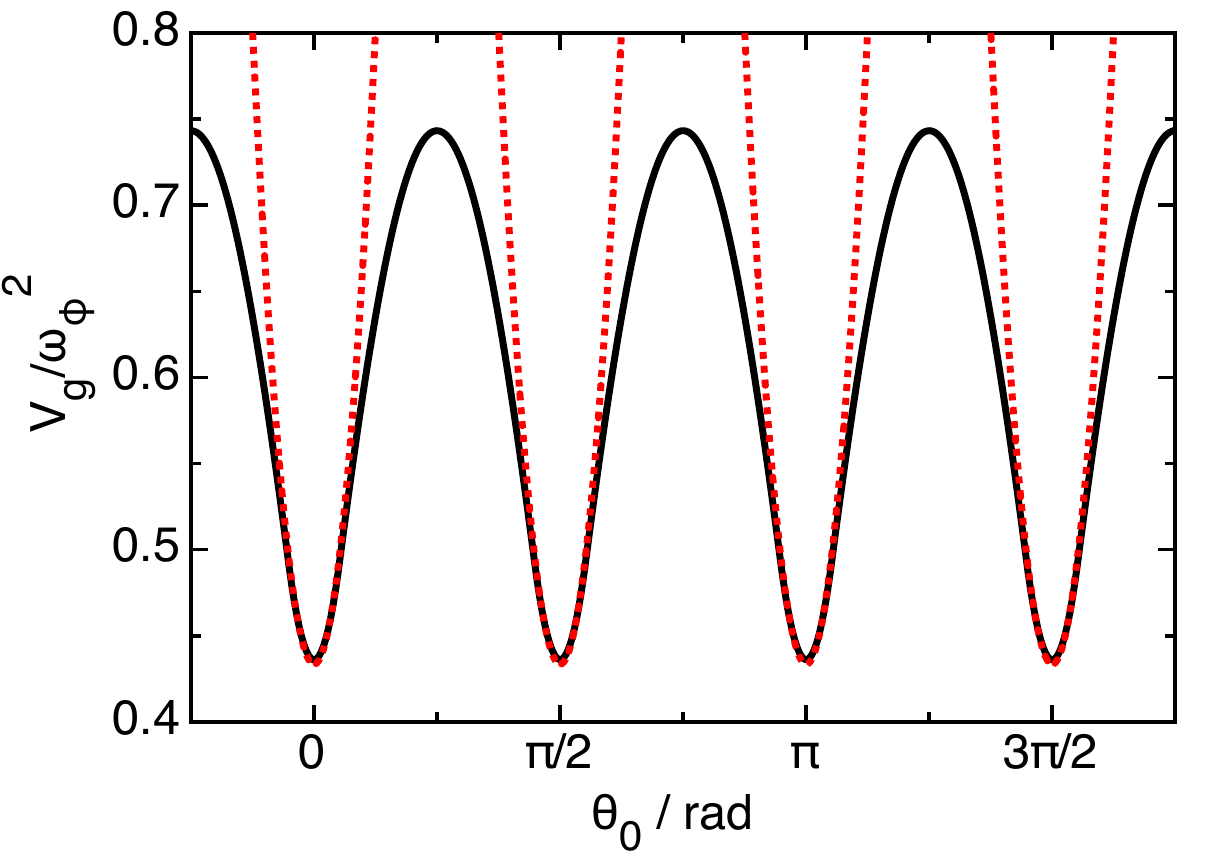}
\caption{The model potential $V_g(\theta_0)$ for a cubic cell when $\alpha=0$, normalized by $\omega_{\phi}^2$ to make it purely a function of container geometry.  Solid line: exact expression from Eq.~\ref{eqn:potential}. Dotted lines: quadratic approximations around each potential minimum.  The  potential minima correspond to the four corners of the cell. Eq.~\ref{eqn:theta_model} describes diffusive fluctuations of $\theta_0$ in this potential, in which $\theta_0$ can occasionally cross the barriers to switch between corners, or oscillate around a potential minimum. 
}
\label{fig:potential}
\end{figure}

The potential is given by
\begin{equation}
V_g(\theta_0,\alpha) = \left<\frac{ 3\omega_{\phi}^2 H^2 }{4 D(\theta_0,\alpha)^2}\right>_{\gamma}
\label{eqn:potential}
\end{equation} 
where $\omega_{\phi}$ is the angular turnover frequency of the LSC, and $H$ is the height of the container \cite{BA08b}.  $D(\theta_0,\alpha)$ is the distance across a horizontal cross-section of the cell, as a function of $\theta_0$ and $\alpha$ as shown in Fig.~\ref{fig:angle_definitions}.  The  notation $\langle ...\rangle_{\gamma}$  represents a smoothing of the potential over a range of $\gamma=\pi/10$ in $\theta_0$ due to the  non-zero width of the LSC \cite{SBHT14}.   Equation \ref{eqn:potential} includes an  update to the numerical coefficient, given here for aspect ratio 1  containers \cite{SBHT14}.  This expression assumes a mean solution for $\delta=\delta_0$ as an approximation to ignore any $\delta$-dependence in the potential.  The diameter function $D(\theta_0,\alpha)$ can be evaluated for any cross-sectional geometry, with the caveat that in this form of the model the geometry must support a single-roll LSC.  In practice, this requires aspect ratios of order 1, and assumes the horizontal cross-section variation affects the flow structure more than any variation with height.  Since $D(\theta_0,\alpha)$ can be calculated for a given geometry as illustrated in Fig.~\ref{fig:angle_definitions}, then  $V_g(\theta_0,\alpha)$ can be predicted explicitly (for example, shown in Fig.~\ref{fig:potential} for a square cross section \cite{BA08b,BJB16}).  Equation \ref{eqn:theta_model} can then be solved statistically for any cross-section that has a single-roll LSC.   Predictions exist for $\delta_0$, $\tau_{\dot\theta}$, and $\tau_{\delta}$ using approximation solutions of the Navier-Stokes equations \cite{BA08a}. The only parameters that have not been predicted are the constants $D_{\dot\theta}$ and $D_{\delta}$,  which in principle could be predicted from models of turbulent kinetic energy in future work.   In practice, $D_{\dot\theta}$ and $D_{\delta}$, $\tau_{\dot\theta}$, and $\tau_{\delta}$ can all be obtained from independent measurements of the mean-square displacement of $\dot\theta_0$ and $\delta$ over time, and $\delta_0$ can be obtained from the peak of the probability distribution of $\delta$ \cite{BA08a}.  Short term measurements from sparse data can then be used as input to the model to make predictions of more complex dynamics such as oscillation structures and stochastic barrier crossing.

Equations  \ref{eqn:delta_model} and \ref{eqn:theta_model} have mainly been applied to describe LSC dynamics in circular-cross-secction containers.   Most of the time, $\delta$ fluctuates around its stable fixed point at $\delta_0$ in Eq.~\ref{eqn:delta_model}. Occasionally, cessations occur when $\delta$ drops to near zero.  When these cessations occur, the damping term n Eq.~\ref{eqn:theta_model} weakens, allowing larger diffusive meandering in $\theta_0$ \cite{BA08a}. Additional forcing terms in Eqs.~\ref{eqn:delta_model} and \ref{eqn:theta_model} due to tilting the cell relative to gravity produced a potential minimum in the plane of the tilt \cite{BA08b}.  If a cell is tilted far enough,  this can provide enough of a restoring force to overcome damping, and an oscillation can then be driven by turbulent fluctuations  which provide a broad spectrum noise to drive oscillations at the  resonant frequency  of the potential \cite{BA08b}.  Oscillations observed in leveled circular-cross-section containers are a combination of twisting \cite{FA04, FBA08} and  sloshing \cite{XZZCX09, ZXZSX09} oscillation modes.  This combination can alternatively be described as a single advected oscillation mode based on an extension of Eq.~\ref{eqn:theta_model} to include advection in the direction of the LSC motion \cite{BA09}.  The  restoring force comes from the potential of Eq.~\ref{eqn:potential} increasing as the diameter $D(\alpha)$ changes as the slosh angle $\alpha$ oscillates, as illustrated in Fig.~\ref{fig:angle_definitions}a.   In this case,  while there is still a  turbulent background driving  a broad spectrum of frequencies, advection in a  closed loop limits the solutions with constructive interference to those that have integer multiples of  the frequency of the LSC turnover, and only even-order modes have a restoring force in $\alpha$ for a circular cross section (corresponding to the twisting and sloshing modes).

The geoemetry-dependent potential $V_g(\theta_0,\alpha)$ can produce different LSC dynamics.  For example, oscillations can occur around a potential minimum at a corner of a cell \cite{SBHT14,GKKS18}.  Large fluctuations can lead to the LSC crossing potential barriers  between wells if the fluctuation energy level $D_{\dot\theta}\tau_{\dot\theta}$ is not too much smaller than the potential barrier height $\Delta V_g$ \cite{BJB16}.  In rectangular cross-section cells, the potential barrier heights shrink along the shorter sides of the rectangle, leading to more frequent barrier crossings \cite{SBHT14}, and periodic oscillations between neighboring potential minima also occur if $D_{\dot\theta}\tau_{\dot\theta}$ is larger than $\Delta V_g$ \cite{SBHT14}.     A cubic cell was chosen to study oscillations around corners because it has the potential minima around corners to provide a restoring force for oscillations, while there are no lowered potential barriers between neighboring corners, which suppresses periodic oscillations between neighboring corners. Detailed predictions for the potential and oscillations in a cubic geometry are given in Secs.~\ref{sec:potential}, \ref{sec:powerspec_model}, and Appendix 3.

  

\section{The experimental apparatus and methods}
\label{sec:methods}

\subsection{Apparatus}
\begin{figure}
\includegraphics[width=0.475\textwidth]{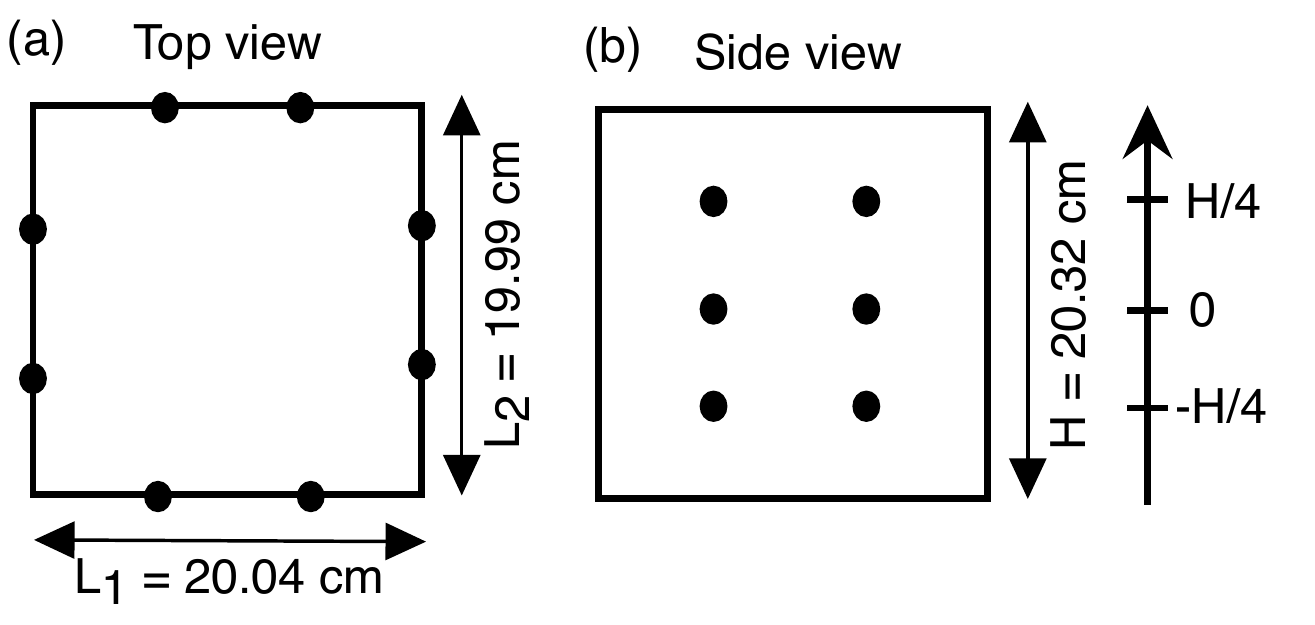} 
\caption{A schematic of the  experimental setup. (a) Top view.  (b) Side view. Thermistor locations on the sidewall are indicated by circles.
}
\label{fig:setup}
\end{figure}   

 The design of the apparatus was based on an earlier circular cylindrical cell  \cite{BNA05}, but with a cubic flow chamber instead. It is the same apparatus used in Ref.~\cite{BJB16}, but described here for the first time in detail. The  flow chamber is nearly cubic, with height  $H = 20.32$ cm, and horizontal dimensions measured at the top and bottom plates of $20.04$ cm and $19.99$ cm, as is shown in Fig.~\ref{fig:setup}. 

To control the temperature difference $\Delta T$ between the top and bottom of the cell, water was circulated through top and bottom plates, each by a temperature-controlled water bath.  The plates were aluminum, with double-spiral water-cooling channels as in \cite{BFNA05}, except that the inlet and outlet of each plate were adjacent to minimize the  spatial temperature variation within the plates.    Each plate  had 5 thermistors  to record control temperatures, with one at the center and one on the diagonal between the center and each corner of the plate.    The top and bottom plate were parallel within 0.06$^{\circ}$.

The sidewalls were plexiglas to thermally insulate the cell from the surroundings.  Three out of four of the sidewalls had a thickness of 0.55 cm.  The fourth sidewall (referred to as the middle wall) was shared with another identical  flow chamber to be used in future experiments.   The middle wall had a thickness of $0.90$ cm to thermally insulate the chambers from each other. 

  The flow chamber was further insulated from the room  as in Ref.~\cite{BFNA05} with 5 cm thick closed-cell foam around the cell, surrounded on the sides by a copper shield with water at temperature $23.00\pm0.02^{\circ}\mathrm{C}$ circulating through a pipe welded to the shield.  The shield was surrounded by an outer layer of 2.5 cm thick open-cell foam.

 To measure the LSC, thermistors were mounted in the sidewalls as in \cite{BA06b}.   There were 3 rows thermistors at heights $-H/4$, $0$ and $H/4$  relative to the middle height, as shown in Fig.~\ref{fig:setup}.  In each row there were 8 thermistors lined up vertically and equally spaced in the angle $\theta$  measured in  a horizontal plane, as shown in Fig.~\ref{fig:setup}a.    
 
 On the 3 outer walls, the thermistors were mounted in blind holes drilled into the sidewall from outside, 0.05 cm away from the fluid surface.   To mount the thermistors in the middle wall, two grooves were cut on each side of the middle wall  in which to place the thermistors and run wiring through the grooves  and out the holes in the top plate. The thickness of the width of the middle wall  between the grooves was reduced to $0.27$ cm.  The remaining space in the grooves was  filled with silicon sealant so the flow chambers remained  thermally insulated from each other.  The silicon sealant  protruded out of the grooves by as much as 0.17 cm  over a surface area 1.78 cm x 0.40 cm.  Silicon sealant was also used to seal the four edges where the sidewalls meet along the height of the cell which stuck out less than 0.1 cm in a region within 0.5 cm of the edge along the wall.  The sidewalls bowed out by up to 0.07 cm at the mid-height near the middle wall.  The top and bottom plate each had a small hole of diameter 0.17 cm, and the middle wall had a hole of diameter 0.2 cm in a corner, which was mostly above the level of the top plate, for filling, degassing, and pressure release of the flow chamber.   All of the aforementioned variations away from a perfectly cubic cell can introduce asymmetries that in principle can affect the flow dynamics \cite{BA06b}.  However, we will confirm in Sec.~\ref{sec:potential} that the cubic shape is the dominant geometric factor, and that non-uniformities in the plate temperature are the largest source of asymmetry in measurements (Secs.~\ref{sec:platetemp}, \ref{sec:shape_imperfections}).
 
 The alignment of the LSC can also be controlled with a small tilt of the cell relative to gravity \cite{KH81}.  Unless otherwise specified, we report measurements with the cell tilted at an angle of $\beta = (1.0\pm0.2)^{\circ}$ relative to gravity, in the plane along a diagonal at orientation $\theta_\beta= 0\pm0.03$ rad, to lock the LSC orientation near one diagonal and suppress diagonal-switching.   
 
The working fluid was degassed and deionized water with mean temperature of 23.0$^{\circ}\mathrm{C}$, for a Prandtl number $Pr = \nu/\kappa = 6.41$, where the  $\nu=9.36\times10^{-7}$ m$^2$/s is the kinematic viscosity and $\kappa=1.46\times10^{-7}$ m$^2$/s is the thermal diffusivity. The Rayleigh number is given by $Ra= g\alpha_T \Delta T H^3/\kappa\nu$ where $g$ is the acceleration  of gravity, and $\alpha_T =  0.000238$/K is the thermal expansion coefficient.  Unless otherwise specified, we report measurements at $\Delta T=18.35$ $^{\circ}$C  for $Ra = 2.62\times10^9$.

\subsection{Thermistor calibration}
\label{sec:calibration}

The thermistors in the top and bottom plates were calibrated together inside a water bath.   The thermistors embedded in the sidewalls were calibrated inside the cell every few months relative to the plate temperatures.  The calibrations identified systematic errors in temperature measurements of $T_{err}= 4$ mK, mostly due to drift in thermistor behavior between calibration checkpoints.  Fluctuations of recorded temperature in equilibrium had a standard deviation of 0.7 mK, corresponding to the random error on temperature measurements.   Deviations from the calibration fit function were typically 6 mK for the plate thermistors, which are much smaller than systematic effects of the top and bottom plate temperature non-uniformity (see Sec.~\ref{sec:platetemp}).

\subsection{Obtaining the LSC orientation $\theta_0$} 
\label{sec:lsc}

\begin{figure}
\includegraphics[width=.475\textwidth]{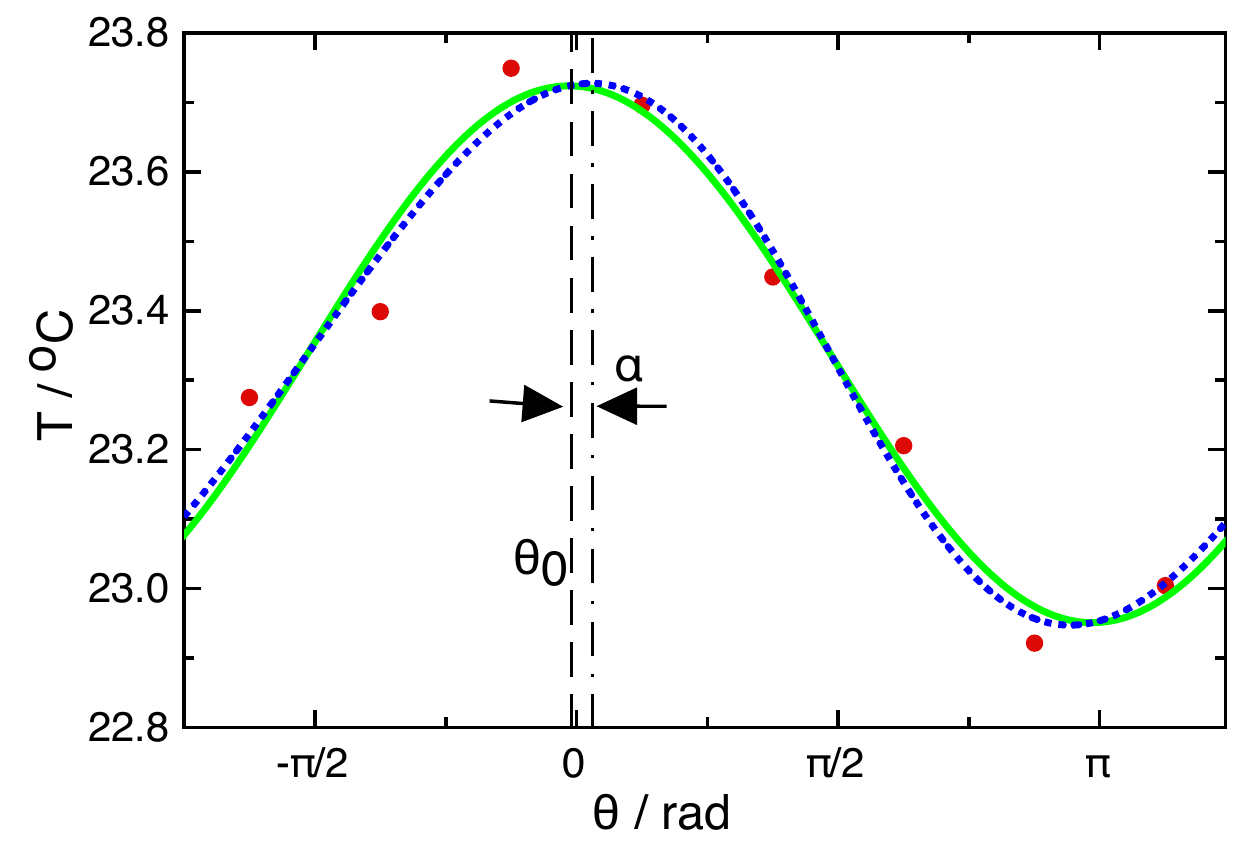} 
\caption{Example of the temperature profile measured by the sidewall thermistors. Solid curve:  cosine fitting  which gives the LSC orientation $\theta_0$ and amplitude $\delta$.  Dotted curve:  fit of the cosine term plus the 2nd order sine term $A_2$.  The shift  in peak locations (vertical lines) between the two curves defines the slosh angle $\alpha$. 
}
\label{fitting_example}
\end{figure}

To characterize the LSC in a noisy background of turbulent fluctuations, we  the function 
\begin{equation}
T = T_0 + \delta cos(\theta - \theta_0)
\label{eqn:delta}
\end{equation}
to obtain the orientation $\theta_0$ and strength $\delta$ of the LSC \cite{BA06a}.  This is fit to measurements of the 8 sidewall thermistor temperatures in the same row every 9.7 s.  An example of the temperature profile measured by the sidewall thermistors  for one time step at the middle row for $Ra = 2.62\times10^9$ is shown in Fig.~\ref{fitting_example} along with the cosine fit.  Deviations from the cosine fit at one instant are mostly due to local temperature fluctuations from turbulence.  Propagating the error on the thermistor measurements from Sec.~\ref{sec:calibration} leads to  systematic uncertainties of  $\Delta\theta_0 = T_{err}/\delta\sqrt{8} =1.5/\delta$ mK/rad and $\Delta\delta = T_{err}/\sqrt{8} = 1.5$ mK, corresponding to 0.003 rad and 0.3\% errors, respectively, at our typical Ra $=2.62\times10^9$.  Throughout this manuscript, we present data for the middle row thermistors only, unless otherwise stated.

\subsection{Obtaining the slosh angle $\alpha$}
\label{sec:alpha}

Some oscillations of the LSC shape can be characterized in terms of the slosh angle $\alpha$ \cite{XZZCX09, ZXZSX09, BA09}.  We use the same definition of the slosh angle $\alpha$ as in Ref.~\cite{BA09} based on the shift in the extrema of the profile $T(\theta)$ away from $\theta_0$. The  temperature profile is quantified by the Fourier series 
\begin{equation} 
T = T_0 + \delta \cos(\theta - \theta_0) + \sum_{n = 2}^{4} A_n \sin[n(\theta - \theta_0)]  
\label{eqn:fourier_series_temp}
\end{equation} 
 at each point in time, where the Fourier moments are
\begin{equation}
A_n = \frac{1}{4}\sum_{i = 1}^{8}{[T_i - T_0 - \delta \cos(\theta - \theta_0)]\sin[n(\theta - \theta_0)]} \ .
\label{eqn:a_n}
\end{equation}
Here the sum over $i$ corresponds to the sum over different thermistors.  Moments are calculated only up to 4th order due to the Nyquist limit with 8 temperature measurement locations.  We do not include cosine terms since they are relatively small \cite{JB20b}, and do not shift the extrema of the temperature profile much. $A_1$ is not included because it is trivially zero since we first fit to Eq.~\ref{eqn:delta}.   

Following the procedure of Ref.~\cite{BA09}, the slosh angle $\alpha$ is calculated as the shift of the positions of the extrema of the temperature profile from the sum of the $\delta$ term and $A_2$ term only in Eq.~\ref{eqn:fourier_series_temp}, which yields the implicit equation $\delta \sin\alpha = 2A_2 \cos(2\alpha)$ to solve for $\alpha$.  An example of this fitting is  shown as the dotted line in Fig.~\ref{fitting_example}.  Note that  $A_2$ is small compared to the variation around the fit due to turbulent fluctuations, so measurements of $A_2$ include a lot of noise, but we will see in Sec.~\ref{sec:oscillation_structure} that an oscillation structure can still be determined from a power spectrum of $A_2$, as was found for cylindrical cells \cite{BA09}.  The absolute error on $A_2$ is the same as the error on $\delta$, since both are the result of fitting the same temperature profile. Propagating the temperature measurement error of $\Delta T_{err}=4.1$ mK results in the systematic error $\Delta \alpha \approx 2 \Delta A_2 / \delta = 0.007$ rad in the small $\alpha$ limit (since $A_2$ is much smaller than $\delta$) at Ra $=2.62\times10^9$.  The ratio of errors on $\Delta\alpha/\Delta\theta_0 = 4\sqrt{2}\Delta A_2/\Delta T_{err} = 2.3$ indicates that $\alpha$ is more sensitive to temperature forcing than $\theta_0$.

\subsection{Effects of the plate temperature non-uniformity}
\label{sec:platetemp}

\begin{figure}
\includegraphics[width=.475\textwidth]{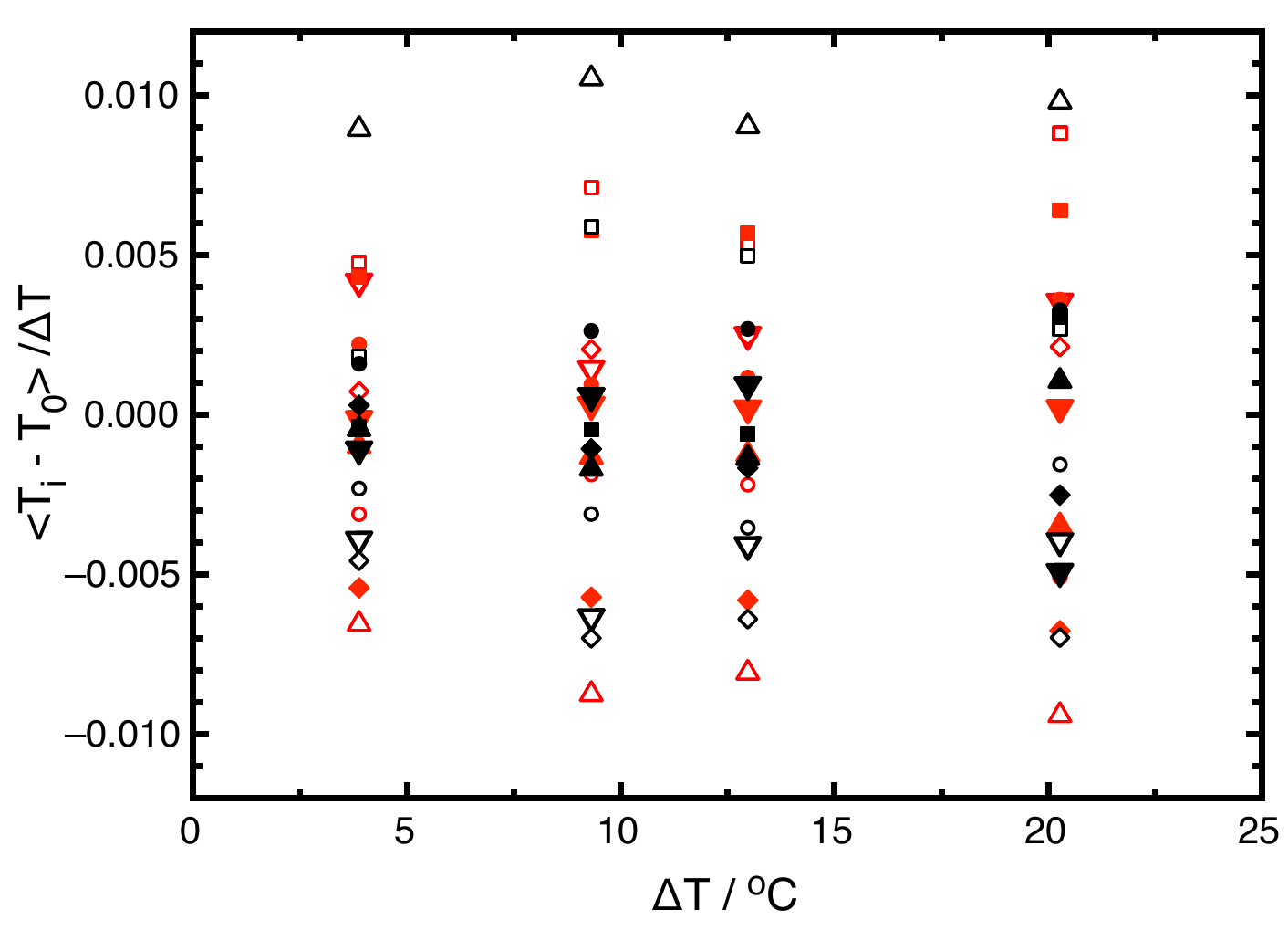} 
\caption{(color online) The  spatial variation of the temperature  in the top and bottom plates $\langle T_i - T_0 \rangle/\Delta T$.  Different symbols represent different plate thermistors. Red symbols: thermistors in the bottom plates.  Black symbols: thermistors in the top plates. Solid symbols represent thermistors in the cell we report data from in this paper, while open symbols correspond to data in the adjacent cell. The down-pointing triangle, up-pointing triangle, square and circle symbols represent the plate thermistors at orientations $0$, $\pi/2$, $\pi$ and $3\pi/2$, respectively, while the diamonds correspond to thermistors at the center of their plates. The standard deviation of temperatures in each plate is 0.005$\Delta T$.  }
\label{plate_stat}
\end{figure}

We quantify the variation  of the top and bottom plate temperatures because it has been shown that  even a slight variation in temperature profile can have a large affect on the alignment of the LSC \cite{BA06b}. The  spatial variation of the temperature within each plate is shown in Fig.~\ref{plate_stat}. These measurements were done in a level cell ($\beta = 0 \pm 0.003$ deg). We show the mean temperature of each thermistor $T_i$ normalized as $\langle T_i - T_0\rangle/\Delta T$, where $T_0$ is the average temperature of the five thermistors in the same plate. The standard deviation of $T_i-T_0$ is 0.005$\Delta T$, which is more uniform than previous experiments \cite{BFNA05}. It can be seen in Fig.~\ref{plate_stat} that the ratio for each thermistor is consistent for different $\Delta T$, indicating this is a systematic effect coming from imperfections in the plates and cooling channel design.  


\begin{figure}[]
\centering 
\includegraphics[width=.475\textwidth]{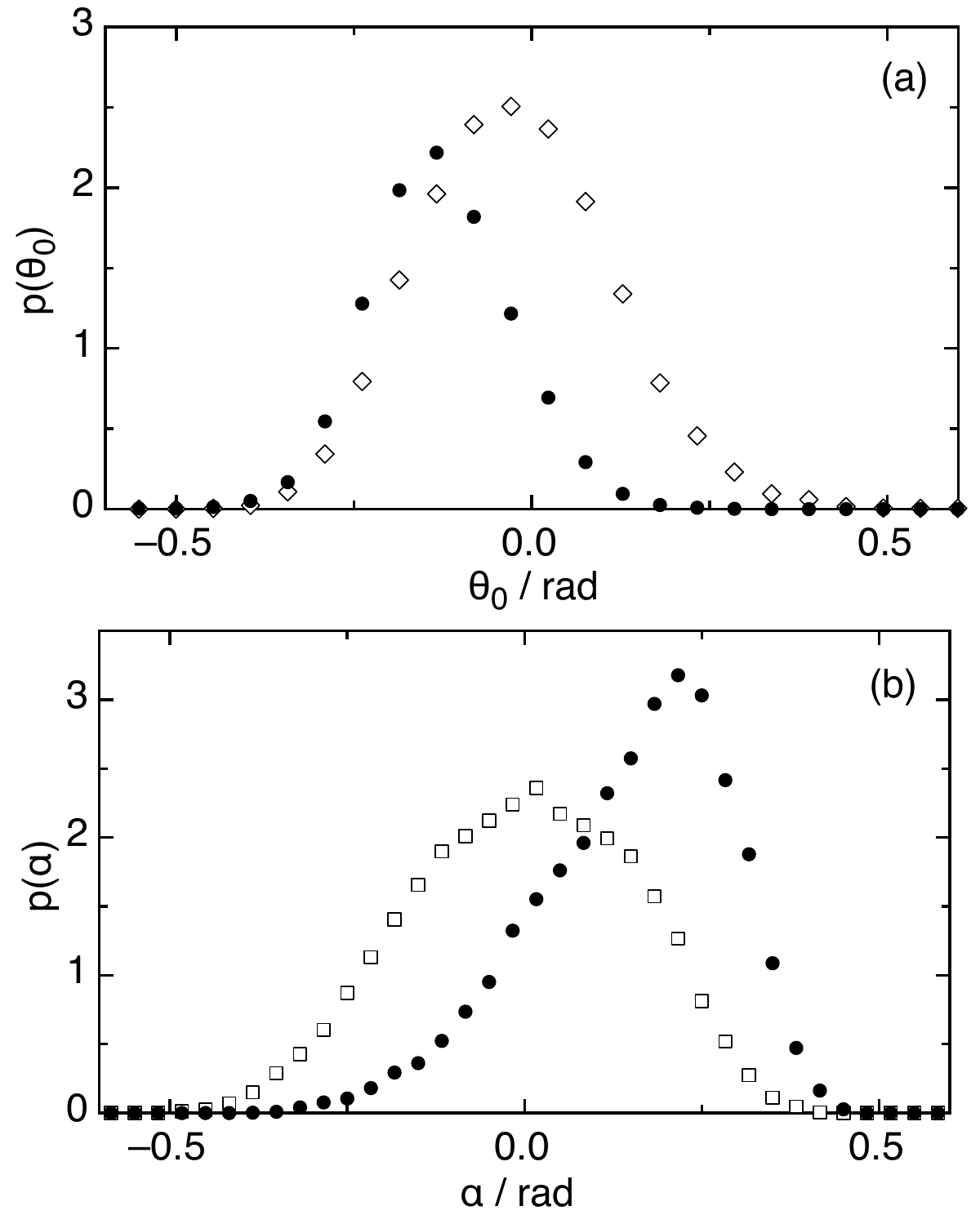} 
\caption{Probability distributions of (a) $\theta_0$ and (b) $\alpha$ for forward flow (open symbols) and reversed flow (solid symbols) in the cooling channels of the top and bottom plates. The plate temperature distributions significantly affect both the peak location and shape of the probability distributions of $\theta_0$ and $\alpha$.
}
\label{theta0_switching_inoutlet}
\end{figure}

To determine how much the temperature profile in the top and bottom plates affects the LSC, we switched the flow direction of the temperature-controlled water in the top and bottom plates.   Figure \ref{theta0_switching_inoutlet} shows probability distributions $p(\theta_0)$ in panel (a) and $p(\alpha)$ in panel (b) for forward and reversed flow in the plates, when $\Delta T=18.35^{\circ}$ C.  The peak location of $p(\theta_0)$ shifted by 0.1 rad and the peak of $p(\alpha)$ shifted by 0.2 rad, and even which of the corners the LSC sampled changed (only data from the corner which was sampled by both datasets are shown in Fig.~\ref{theta0_switching_inoutlet}). The larger shift in $p(\alpha)$ is a consequence of the larger sensitivity of $\alpha$ to changes in temperature by a factor of 2.3 (Sec.~\ref{sec:alpha}).   Additionally, $p(\alpha)$ became skewed to one side after the change in flow direction, indicating that non-uniformity in the plate temperature can also change the shape of the distributions.  Assuming the standard deviation of plate temperatures of $0.005\Delta T$ extends into the LSC, we estimate  $\Delta\theta_0 = 0.005\Delta T /\delta\sqrt{8} = 0.08$ rad for $\Delta T=18.35^{\circ}$ C, close to the observed shift in the  peak of $p(\theta_0)$ in Fig.~\ref{theta0_switching_inoutlet}. This confirms that significant effects on $p(\theta_0)$ and $p(\alpha)$ observed in Fig.~\ref{theta0_switching_inoutlet} are due to the non-uniformities in the plate temperatures, and the temperature variation in the plate extends into the LSC.

\section{The potential $V_g$}
\label{sec:potential}

\subsection{Relating $V_g(\theta_0)$ to the probability distribution of $\theta_0$}
\label{sec:4well_ptheta}
 
 We start our analysis of the potential in one dimension to focus on the 4-well potential in $\theta_0$. For now we ignore $\alpha$ (i.e.~set $\alpha=0$), although we will include $\alpha$ in our analysis starting Sec.~\ref{sec:potential_generalization}.  We test the prediction of the potential $V_g(\theta_0)$ given in Eq.~\ref{eqn:potential} by measuring the probability distribution $p(\theta_0)$.    The two are related by the steady-state solution of the Fokker-Planck equation for Eq.~\ref{eqn:theta_model}, assuming overdamped stochastic motion in a potential, and approximating $\delta=\delta_0$, resulting in \cite{SBHT14} 

\begin{equation}
-\ln p(\theta_0)=\frac{V_g(\theta_0)}{D_{\dot\theta}\tau_{\dot\theta}} \ .
\label{eqn:lnptheta}
\end{equation}

\begin{figure}
\includegraphics[width=.475\textwidth]{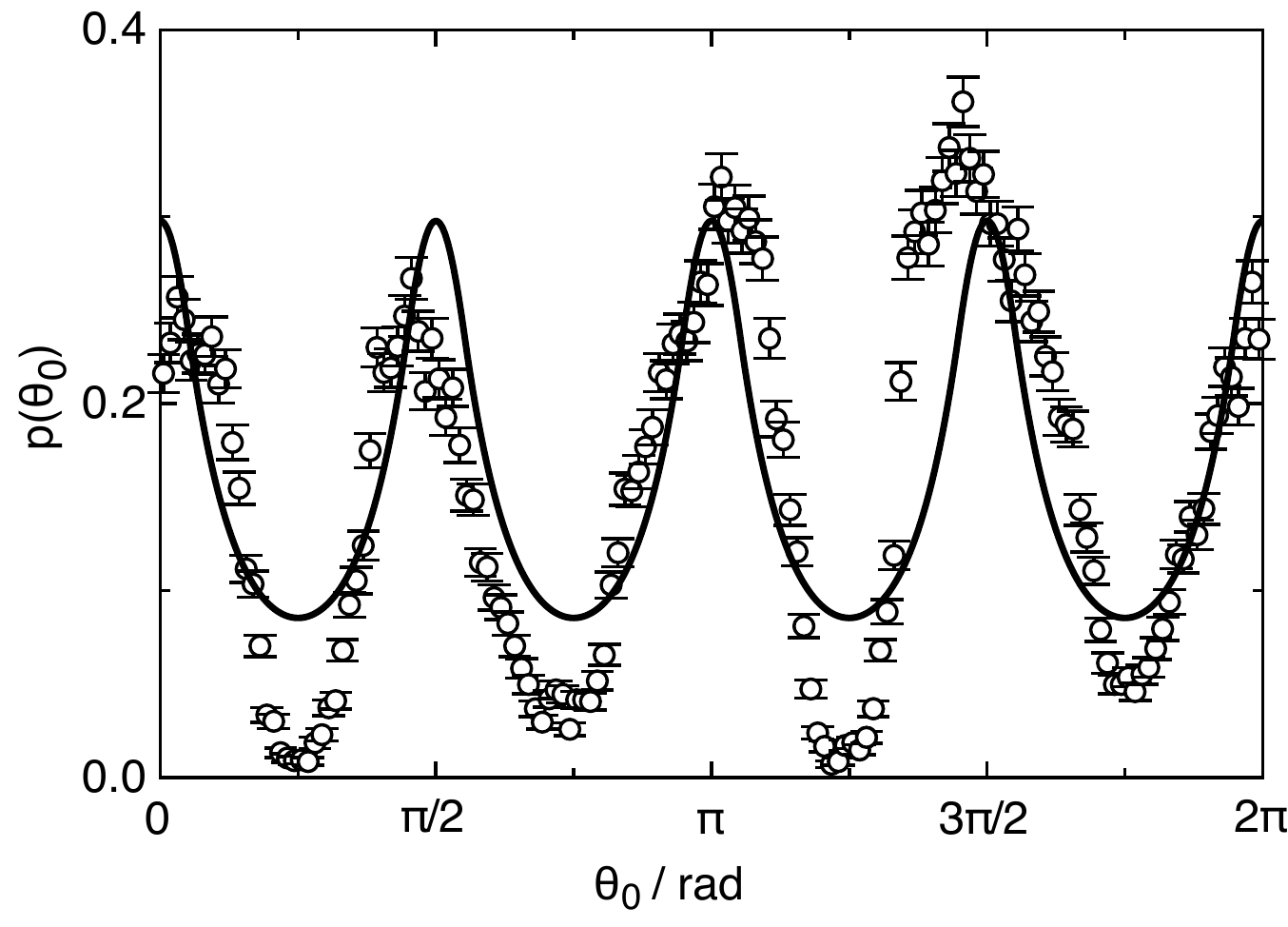} 
\caption{Circles: A measured probability distribution $p(\theta_0)$ with 4 wells.   Solid line:  model prediction from Eq.~\ref{eqn:lnptheta} with $\tau_{\dot\theta}$ as an adjustable fit parameter. This shows that the predicted 4-peaked $p(\theta_0)$ can be obtained in experiments,  although it required extreme effort required to carefully tilt the cell to cancel out other sources of asymmetry.
}
\label{fig:ptheta_4well}
\end{figure} 

To test this prediction, we searched for a 4-peaked $p(\theta_0)$ by carefully leveling the cell.  Due to the extreme  sensitivity of $p(\theta_0)$ to weak asymmetric forcings mainly from the non-uniformity in the plate temperature, we usually find $p(\theta_0)$ is locked mainly in 1 corner, even when we nominally level the cell (see Fig.~\ref{theta0_switching_inoutlet}).   To get such a uniform, 4-peaked $p(\theta_0)$ required a months-long effort to tune the tilt angle $\beta$  to just the right value to mostly cancel out other sources of asymmetry.   We show  a  measurement of $p(\theta_0)$ with 4 peaks in Fig.~\ref{fig:ptheta_4well} at Ra$=4.8\times 10^8$ and $\beta=0.0005 \pm 0.0009$ rad (This is the same dataset used in Ref.~\cite{BJB16}).  Even with the extreme effort to obtain a 4-peaked $p(\theta_0)$, this data is not entirely ergodic, as crossings of the LSC orientation from corner to corner only occurred 48 times during the time series, for an average of 12 crossings of each potential barrier and sampling each potential well 12 times,  with an approximately $1/\sqrt{12}= 30\%$ error on the relative peaks and minima of $p(\theta_0)$.  The corresponding prediction of $p(\theta_0)$ from Eq.~\ref{eqn:lnptheta} is shown for comparison where we used the measured values $D_{\dot\theta}= 2.37\times 10^{-6}$ rad$^2$/s$^3$ and $\mathcal T = 286$ s  for a different dataset at the same nominal parameter values \cite{BJB16},  and fit $\tau_{\dot\theta}= 45.5$ s (a factor of 2.6 larger than the measured value at the same nominal parameter values).  The comparison indicates that the general shape of the 4-peaked $p(\theta_0)$ can be achieved in experiment, although the extreme sensitivity to  asymmetries makes this more of a special limiting case than a typical case.  

\subsubsection{Natural frequency $\omega_r$}
\label{sec:omega_r}

In anticipation of analytical solutions for oscillation modes, we write an approximation of $V_g(\theta_0)$ in terms of a first order expansion around each corner for small $\theta_0$ as in Ref.~\cite{SBHT14}

 \begin{equation}
 V_g(\theta_0) \approx \frac{1}{2}\omega_r^2\theta_0^2 + V_{0} \ ,
 \label{eqn:quadratic}
 \end{equation}
 
 \noindent where $\omega_r$  corresponds to the natural frequency  of oscillations around the potential minimum.  $\omega_r$ is predicted from Eq.~\ref{eqn:potential} to be \cite{BJB16}
 
 \begin{equation}
\omega_r = \sqrt{\frac{15}{\pi}}\omega_{\phi} \ ,
\label{eqn:omega_r}
\end{equation}

\noindent  where $\omega_{\phi}= 2\pi/\mathcal{T}$ and $\mathcal{T}$ is the turnover time of the LSC.  Figure \ref{fig:potential} shows the quadratic approximation of the potential in comparison to the full solution.  

\begin{figure}
\includegraphics[width=.475\textwidth]{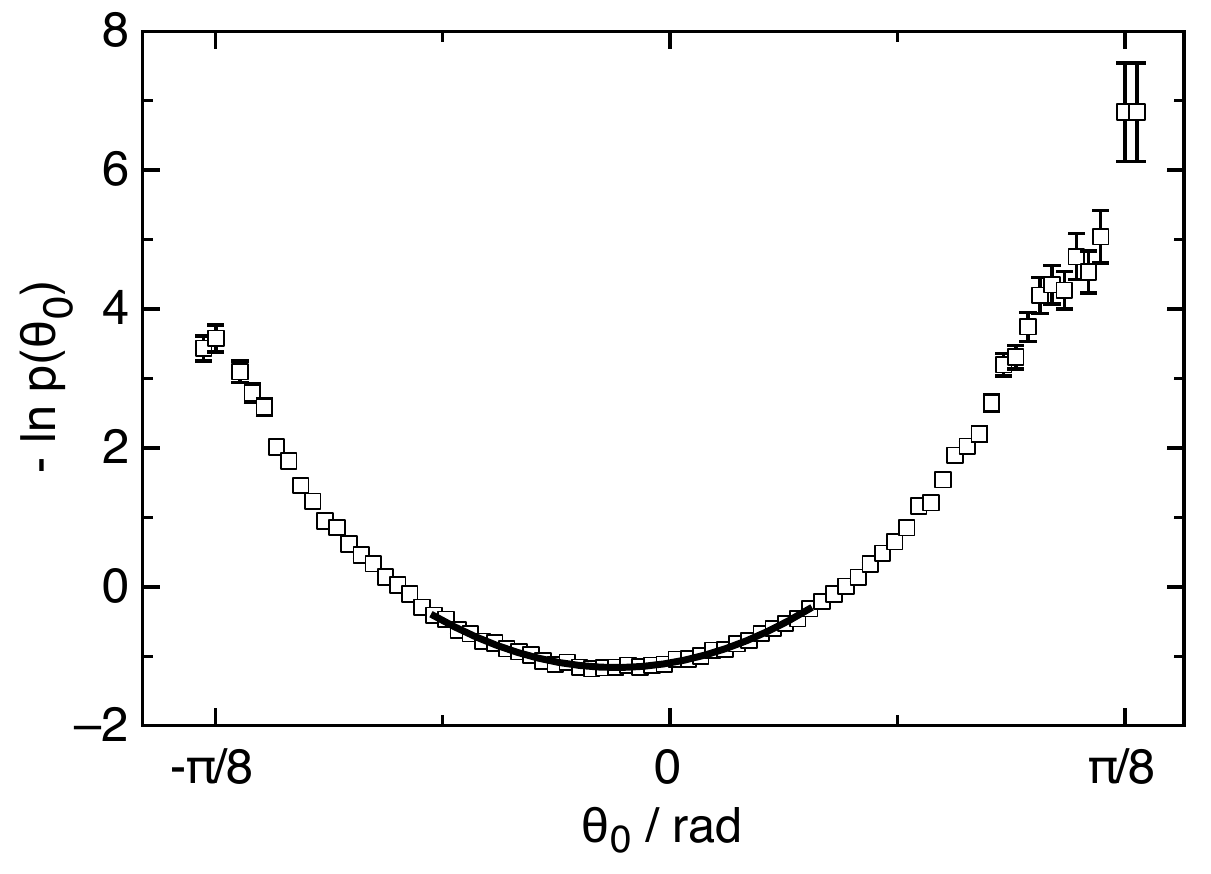} 
\caption{The probability distribution $p(\theta_0)$ around a single corner, rescaled as $-\ln p(\theta_0)$ to be equal to the predicted dimensionless potential $V_g(\theta_0)/D_{\dot\theta}\tau_{\dot\theta}$.  Solid line:  quadratic fit  to the data, indicating agreement with the prediction of Eq.~\ref{eqn:quadratic}.
}
\label{fig:pdf_potential}
\end{figure}

For a quantitative test of the predicted quadratic potential minimum, we compare to measurements where $p(\theta_0)$ is locked around a single corner.  In most of our experiments, we find long intervals where the LSC  orientation is locked around a single corner due to the large potential barriers, and further enforce this orientation-locking with a standard tilt of the cell by $\beta=1^{\circ}$ at Ra $=2.62\times10^9$.      The measured $-\ln p(\theta_0)$ for this case is shown in Fig.~\ref{fig:pdf_potential}.  We find that the peak of $p(\theta_0)$ is offset slightly from the corner, by  0.05 rad, consistent with the 0.1 rad uncertainty on the peak position due to the  non-uniform plate temperature.
 To compare the shape of $p(\theta_0)$ to the prediction of Eq.~\ref{eqn:lnptheta}, we combine Eq.~\ref{eqn:lnptheta} and Eq.~\ref{eqn:quadratic} into a fit function, and include an offset $\theta_p$ to account  for the shift in the peak of $p(\theta_0)$ from the corner:

\begin{equation}
-\ln p(\theta_0)=\frac{1}{2} \frac{\omega_r^2}{D_{\dot\theta}\tau_{\dot\theta}}(\theta_0-\theta_p)^2 + \frac{V_{0}}{D_{\dot\theta}\tau_{\dot\theta}}\ .
\label{eqn:lnptheta_omega_r}
\end{equation}

\noindent  To obtain the curvature $\omega_r^2/D_{\dot\theta}\tau_{\dot\theta}$, we fit Eq.~\ref{eqn:lnptheta_omega_r} to data in Fig.~\ref{fig:pdf_potential}.  The input errors for fitting $-\ln p(\theta_0)$ were propagated errors on $p(\theta_0)$ assuming a Poisson distribution, and the fit is reported over the largest range of data which had a reduced $\chi^2 \approx 1$.  The fit range of $0.3$ rad is shown as the curve in Fig.~\ref{fig:pdf_potential}.   This range includes 81\% of the measured data, and 95\% of the data is in a range where the quadratic approximation of the potential from Eq.~\ref{eqn:quadratic}  is within 5\% of the exact calculation of $V_g(\theta_0)$ from Eq.~\ref{eqn:potential} (illustrated in Fig.~\ref{fig:potential}).  This indicates that the quadratic shape of the potential is correctly predicted within the measurement resolution over the range where most of the data is found.  The fit of Eq.~\ref{eqn:lnptheta_omega_r} to the data in Fig.~\ref{fig:pdf_potential} yields the curvature $\omega_r^2/D_{\dot\theta}\tau_{\dot\theta}$, which is  multiplied by the measured values of $D_{\dot\theta}$ and $\tau_{\dot\theta}$ obtained from data at the same Ra in Figs.~\ref{fig:Dthetadot} and \ref{tauthetadot} of Appendix 1 to obtain $\omega_r$.

\subsubsection{Ra-dependence of $\omega_r$}

Values of $\omega_r$ are obtained by fitting $-\ln p(\theta_0)$ at different Ra. Fits were done with a fixed fit range equal to 2.3 times the standard deviation of the distribution (consistently 80\% of the data).  For Ra  $<8\times10^8$, the reduced $\chi^2$ increases from 1 up to 16 at the lowest Ra, coinciding with an increased asymmetry of the probability distribution around the  potential minimum.  This indicates an inconsistency with the predicted quadratic probability distribution at lower Ra.  The larger $\chi^2$ may result from the increased correlation time of $\theta_0$ at higher Ra \cite{BA08a}, 
which makes the data in $p(\theta_0)$ less independent of each other, and the Poisson distribution a less good approximation for the error.
 
\begin{figure}
\includegraphics[width=.475\textwidth]{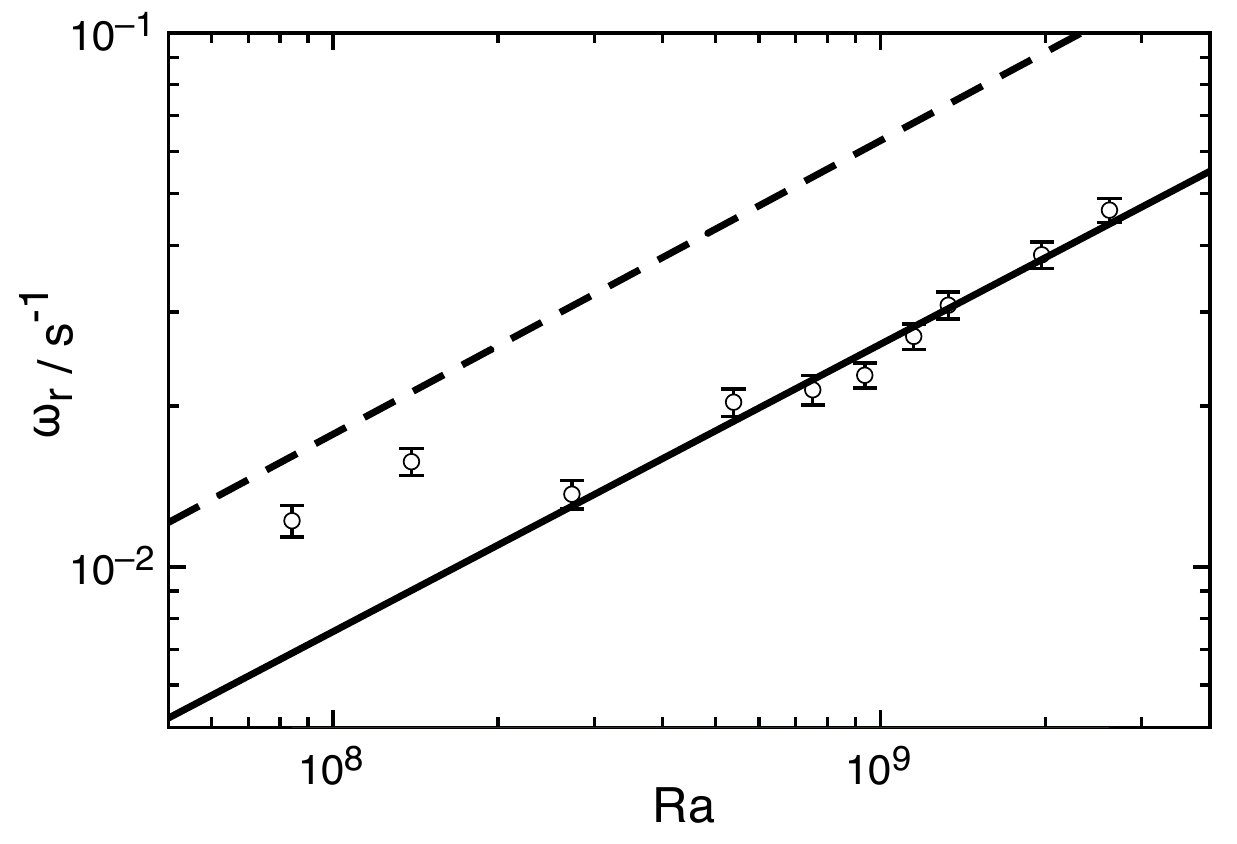} 
\caption{The natural frequency $\omega_r$ of the potential as a function of Ra, obtained from fitting $-\ln p(\theta_0)$.  Solid line: power law fit to the data for Ra $>2\times10^8$.  Dashed line:  The predicted potential from Eq.~\ref{eqn:quadratic} using the measured  turnover time $\mathcal{T}$.  $\omega_r$ scales with the inverse of $\mathcal{T}$ as predicted for Ra $>2\times10^8$, although the magnitude of $\omega_r$ is 2.6 times smaller than the prediction in this range.
}
\label{curvature_Ra_mid}
\end{figure} 

Values of $\omega_r$ are shown as a function of Ra in Fig.~\ref{curvature_Ra_mid}.  The errors on $\omega_r$ shown in Fig.~\ref{curvature_Ra_mid} include the error on the fits and errors propagated from the scatter of the data of 9.6\% on $D_{\dot\theta}$ from Fig.~\ref{fig:Dthetadot} and 3.2\% on $\tau_{\dot\theta}$ from Fig.~\ref{tauthetadot}.  For comparison, we also plot the model prediction for the natural frequency $\omega_r$ in Fig.~\ref{curvature_Ra_mid}.  The prediction is obtained by plugging in the fit of the  measured turnover time $\mathcal{T}$ vs.~Ra from Fig.~\ref{turnoverTimeVSRa} into Eq.~\ref{eqn:omega_r} to obtain $\omega_r =  7.71\times10^{-7}Ra^{0.55\pm0.05}$ s$^{-1}$.  A power law fit to data for Ra $>2\times10^8$ yields $\omega_r = 3.7\times10^{-7}Ra^{0.54\pm0.03}$ s$^{-1}$, shown in Fig.~\ref{curvature_Ra_mid}.    The errors on the power law exponent overlap for the prediction and data, indicating consistency with the predicted scaling relation for Ra $>2\times10^8$.  The prediction is on average 2.6 times the measured data in this range of Ra.  Such an error is typical of this modeling approach \cite{BA08a, BA08b, BA09}, as it makes significant approximations about the shape of the LSC, scale separation between the LSC and small-scale turbulent fluctuations, and the distribution of turbulent fluctuations.

 To confirm that the small tilt of  $\beta=1^{\circ}$ did not bias the data, we applied the same procedure to obtain $\omega_r^2/D_{\dot\theta}\tau_{\dot\theta}$ by fitting data at $\beta=0$ for Ra $=2.62\times10^9$, normalizing the fit curvature of $-\ln p(\theta_0)$ by values of $D_{\dot\theta}$ and $\tau_{\dot\theta}$ measured at $\beta=0$.  We found consistent values of $\omega_r$ at both tilt angles
  within the 5\% error.  On the other hand, the non-uniformity in plate temperature has a significant effect on $p(\theta_0)$; the change in flow direction in the plates shown in Fig.~\ref{theta0_switching_inoutlet}a caused a 20\% change in the fit value of $\omega_r$.

\subsubsection{Barrier crossing}

\begin{figure}
\includegraphics[width=.475\textwidth]{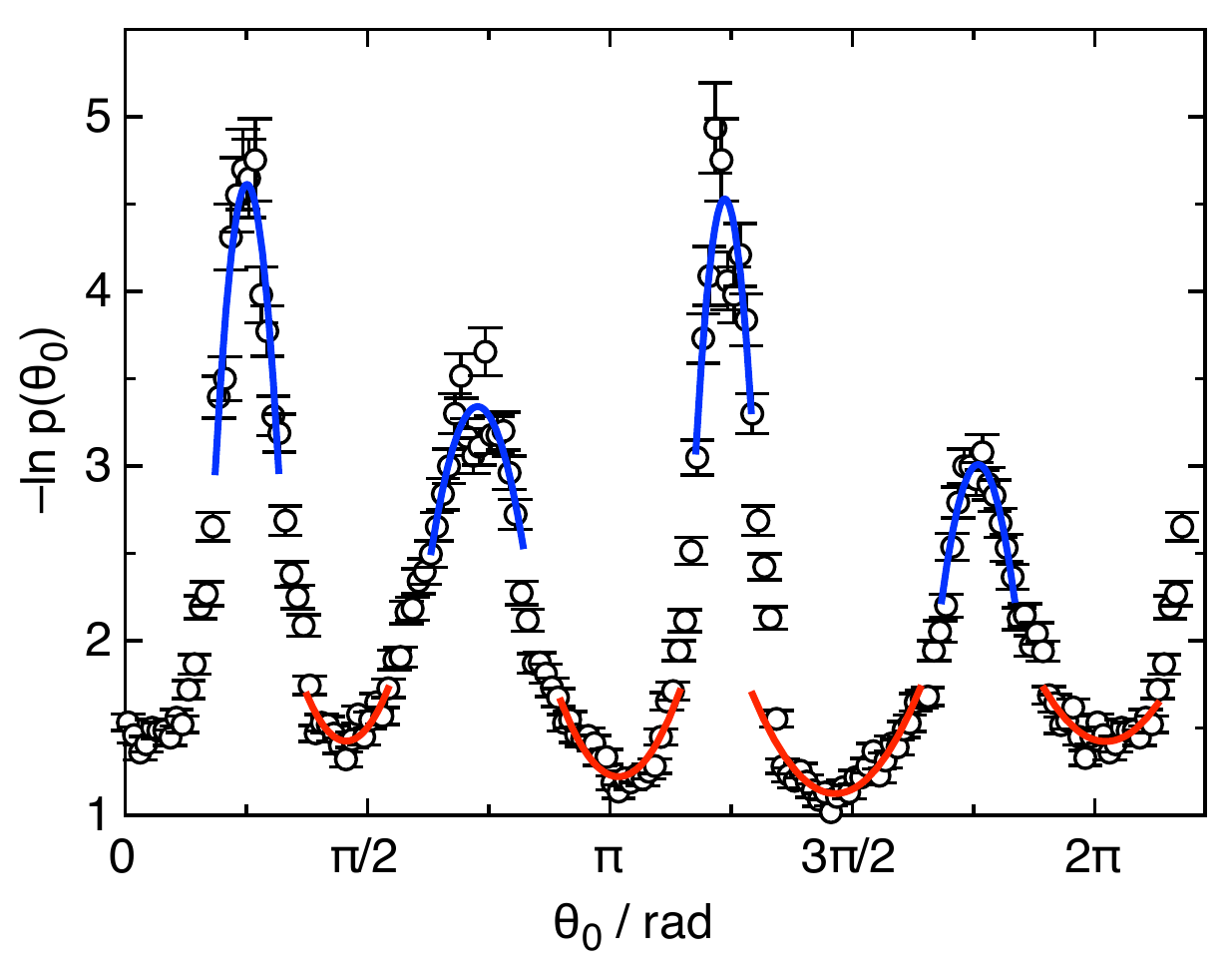} 
\caption{The measured dimensionless potential $-\ln p(\theta_0)$ when the cell is carefully tilted to cancel most sources of asymmetry.  Curves: quadratic fits of the potential minima and maxima to obtain $\omega_r$, $\omega_{max}$ and $\Delta V_g$.
}
\label{fig:ptheta_4well_fit}
\end{figure} 

Other aspects of the shape of the potential are relevant to calculating the rate of  the LSC orientation $\theta_0$ crossing a potential barrier $\Delta V_g$ from one corner to another.  We previously reported the overall barrier crossing rate in \cite{BJB16}, and here we show measurements of the geometric parameters used in Kramer's formulation \cite{Kr40}.   This requires not only the natural frequency $\omega_r$ at the potential minimum, but also a quadratic fit around the maximum of the potential with a corresponding frequency $\omega_{max}$, and the potential barrier height $\Delta V_g$ \cite{BJB16}.  These barrier crossing events are non-existent in most of our datasets, such as those represented in Fig.~\ref{fig:pdf_potential} where the LSC is locked into a narrow range of $\theta_0$ around a single corner due to the tilt of the cell.  Instead, $\omega_{max}$ and $\Delta V_g$ can be obtained from more ergodic data such as in Fig.~\ref{fig:ptheta_4well} where the cell was tilted carefully to make each potential well nearly equally likely.  In a previous article, we showed that accurate predictions of the rate of barrier crossing $\omega$ could be made using a Kramers' formulation
\begin{equation}
\omega =  \frac{ \omega_r\omega_{max}\tau_{\dot{\theta}} }{2\pi} \exp \left( - \frac{ \Delta V_g }{D_{\dot{\theta}} \tau_{\dot{\theta}}} \right) \,
\label{eqn:switching_rate}
\end{equation}  
where we predicted $\omega_{max}=\sqrt{3/2}\omega_{\phi}$, $\Delta V_g = (3/8)(1-\gamma/2)\omega_{\phi}^2$ from Eq.~\ref{eqn:potential}, and with values of $\tau_{\dot\theta}$, $D_{\dot\theta}$, and $\omega_{\phi}$  obtained from independent measurements \cite{BJB16}.  Here, we show the intermediate step that relates the parameters $\omega_{max}$ and $\Delta V_g$ directly to the potential.   To do so, we convert the probability distribution in Fig.~\ref{fig:ptheta_4well} to $-\ln p(\theta_0)$ and plot it in Fig.~\ref{fig:ptheta_4well_fit}.  Figure \ref{fig:ptheta_4well_fit} also shows fits of Eq.~\ref{eqn:lnptheta_omega_r} to data around each potential minimum to obtain $\omega_r^2/D_{\dot\theta}\tau_{\dot\theta}$, and analogous fits around each potential maximum of 
\begin{equation}
-\ln p(\theta_0)=-\frac{1}{2} \frac{\omega_{max}^2}{D_{\dot\theta}\tau_{\dot\theta}}(\theta_0-\theta_{max})^2 + \frac{V_0+\Delta V_g}{D_{\dot\theta}\tau_{\dot\theta}} \ .
\label{eqn:lnptheta_omega_max}
\end{equation}
To obtain the natural frequencies $\omega_r$ and $\omega_{max}$, the fit curvatures of $\omega_r^2/D_{\dot\theta}\tau_{\dot\theta}$ and  $\omega_{max}^2/D_{\dot\theta}\tau_{\dot\theta}$ are multiplied by the measured parameters for  $\tau_{\dot\theta}=17.5$ s and $D_{\dot\theta} = 2.37\times10^{-6}$ rad$^2$/s$^3$ for data at the same nominal parameter values \cite{BJB16}, and averaged together for the four fits of each.  Averaging over the 4 minima or maxima helps reduce the bias introduced from the plate temperature non-uniformity, which may affect the curvature of each minimum and maximum differently.  This yields $\omega_r=0.015$ s$^{-1}$ (3 times smaller than the prediction of Eq.~\ref{eqn:omega_r}, similar to Fig.~\ref{curvature_Ra_mid}), and $\omega_{max}=0.046$ s$^{-1}$ (70\% larger than the prediction \cite{BJB16}). The potential barrier $\Delta V_g$ was obtained by averaging the vales of the fits of Eq.~\ref{eqn:lnptheta_omega_max}, while fixing $V_0$ to have the same value for the fits of Eq.~\ref{eqn:lnptheta_omega_r} and \ref{eqn:lnptheta_omega_max}, resulting in $\Delta V_g=1.1\times10^{-4}$ s$^{-2}$ (30\% smaller than the prediction \cite{BJB16}).  These parameters overestimate the measured barrier crossing rate $\omega$ by a factor of 2.
This confirms that the relevant quantifiable features of the potential $V_g$ can all be predicted within a factor of 3.

\subsubsection{Quantitative comparisons of sources of asymmetry}
\label{sec:shape_imperfections}

Asymmetries in the dynamics of the LSC are often attributed in part to imperfections in the shape of the container, slight tilt of the cell, temperature profile in the plates, and Earth's Coriolis force \cite{BA08b}.  Here we quantitatively compare how large a contribution some of these make to the potential.  The largest imperfection in the shape of our flow cell is the epoxy sticking out of the middle wall to protect thermistors by $\Delta D= 0.17$ cm.  The predicted potential difference due to this epoxy at its most extreme position relative to the potential difference from the cubic shape is $\Delta V/\Delta V_g \approx 4\Delta D/L = 8\%$.
In terms of the dimensionless potential $V/D_{\dot\theta}\tau_{\dot\theta}$, the change from potential minimum to maximum due to the cubic geometry is $-\Delta\ln p(\theta_0)=2.6$ based on the average of fits in Fig.~\ref{fig:ptheta_4well_fit}, so the effect of the wall shape imperfection is expected to be $-\Delta\ln p(\theta_0)=0.2$.  For comparison, the difference due to the change in flow direction in Fig.~\ref{theta0_switching_inoutlet} is $-\Delta\ln p(\theta_0)=3.3$ at the orientation where the difference is largest, much larger than the wall shape imperfection, and even larger than the effect of the cubic geometry.  This explains why in a nominally leveled cell we observed a single-peaked probability distribution of $\theta_0$  instead of 4 peaks -- the plate temperature profile is dominating the potential.  Only when we tilted the cell to cancel out most of the effect of the plate temperature nonuniformity did we observe the 4-peaked probability distribution.  

When we calculate values of $\omega_r$,  part of the curvature of $-\ln p(\theta_0)$ we measure might come from the plate temperature non-uniformity, other sources of asymmetry, and non-ergodic statistics.  Thus, the values we report may be affected by this.  For example, the values of $\omega_r$ measured at different potential minima in Fig.~\ref{fig:ptheta_4well_fit} have a standard deviation of 36\%, while the errors on individual fits average 15\%.  This suggests the other 21\% of the variation in measured $\omega_r$ comes from asymmetries in the cell, which corresponds to a systematic error in reported values of $\omega_r$ from single-corner measurements such as Fig.~\ref{fig:pdf_potential}.  A similar difference of 20\% in $\omega_r$ is obtained from data before and after switching the flow direction in the top and bottom plates in Fig.~\ref{theta0_switching_inoutlet}, confirming this asymmetry in the measured potential could come from the plate temperature non-uniformity.

\subsection{Generalization of the potential to $V_g(\theta_0,\alpha)$}
\label{sec:potential_generalization}

 While the geometry-dependence of the potential $V_g$ (Eq.~\ref{eqn:potential}) was first introduced as a function of $\theta_0$ \cite{BA08a}, and the slosh displacement represented by the angle $\alpha$  is responsible for the sloshing oscillation in circular cylindrical cells \cite{BA09}, the potential has not been calculated as a function of both parameters before.   To obtain a  potential in terms of both $\theta_0$ and $\alpha$ using the method of Brown \& Ahlers \cite{BA08b}, we first calculate the horizontal cross-section length $D(\theta_0, \alpha)$ in a square horizontal cross-section with displacements in both $\theta_0$ and $\alpha$ as illustrated in Fig.~\ref{fig:angle_definitions}b.  This purely geometric calculation is an extension to a previous calculation for a rectangular cell for $\theta_0$ \cite{SBHT14}, with the additional variable $\alpha$.  A full derivation of the pathlength $D(\theta_0,\alpha)$ is given in  Appendix 2.  For analytical calculations near corners, where most of the data lies,  the expression simplifies in the small angle limit for both $\theta_0$ and $\alpha$ to
\begin{equation}
D(\theta_0,\alpha)^2 \approx 2H^2(1 - |\theta_0+\alpha| - |\theta_0-\alpha|) \ .
\label{eqn:diameter_linear}
\end{equation}
 
\noindent  We next convert $D(\theta_0,\alpha)$ to a potential via Eq.~\ref{eqn:potential}, which includes smoothing over the width $\gamma=\pi/10$ to account for the finite width of the LSC as in  Ref.~\cite{SBHT14}.  To smooth the potential with two parameters, we integrate over the orientations of the hot side of the LSC $\alpha_h = \theta_0 + \alpha$, and the cold side of the LSC $\alpha_c =  \theta_0 - \alpha$, as ilustrated in Fig.~\ref{fig:angle_definitions}. The bounds of the integrals are from $\alpha_h - \gamma/2$ to $\alpha_h + \gamma/2$ on the hot side, and from $\alpha_c - \gamma/2$ to $\alpha_c + \gamma/2$ on the cold side. We change the variables in Eq.~\ref{eqn:diameter_linear} to $\alpha_h$ and $\alpha_c$ when substituting into Eq.~\ref{eqn:potential} for the integral, and then change the variables back to $\theta_0$ and $\alpha$ to obtain
 
  \begin{equation}
 V_g(\theta_0, \alpha) \approx \frac{15\omega_\phi^2 (\alpha^2 + \theta_0^2)}{2\pi}
 \label{eqn:potential_quadratic}
\end{equation}

\noindent in the limit of small $\alpha$ and $\theta_0$, and ignoring an addictive constant in a potential.   This quadratic potential leads to harmonic oscillator solutions to Eq.~\ref{eqn:theta_model} as a result of the smoothing over the range $\gamma$ in Eq.~\ref{eqn:potential}, with equal curvatures for the potentials in $\theta_0$ and $\alpha$, and thus equal natural frequencies for oscillations in each variable.  This contrasts with the potential for a circular cross section, in which $\theta_0$ does not appear in the potential.

\subsection{$p(\alpha)$}

 \begin{figure}
\includegraphics[width=.475\textwidth]{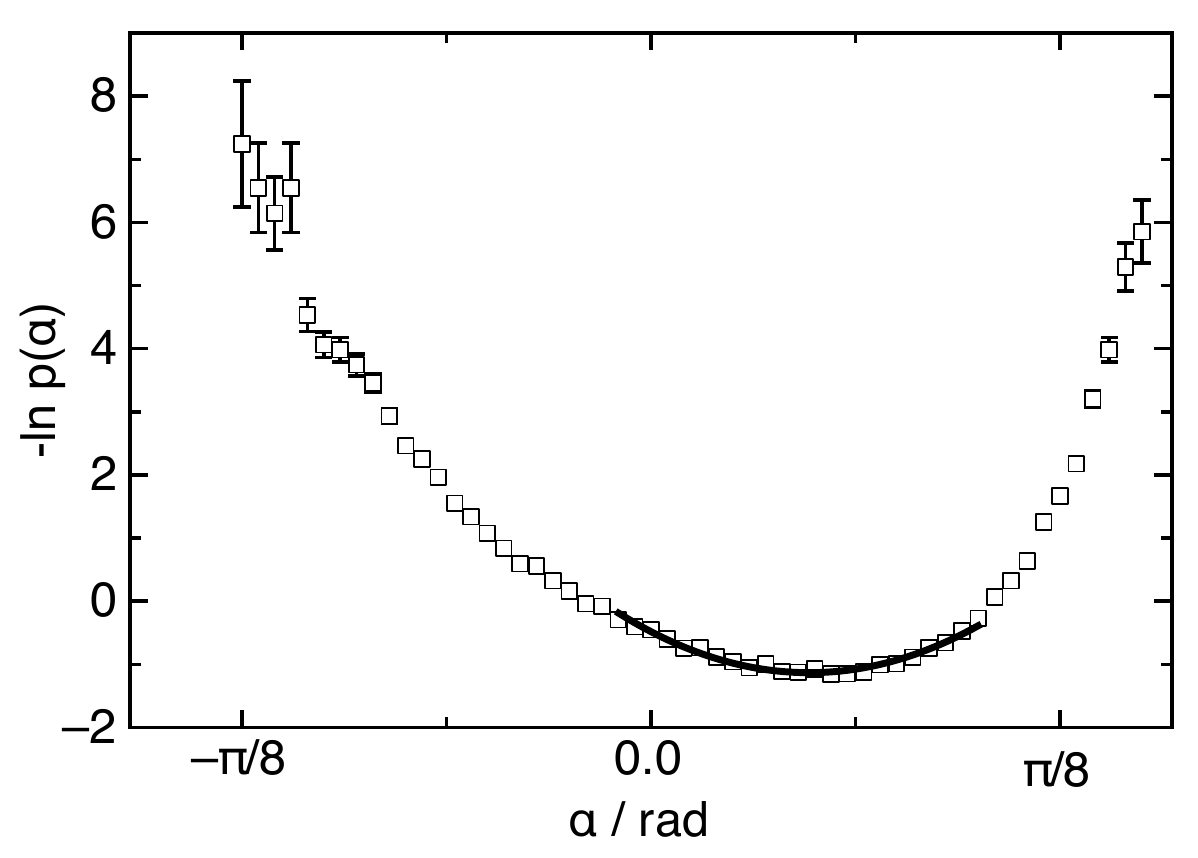} 
\caption{The probability distribution $p(\alpha)$, rescaled as $-\ln p(\alpha)$ to be equal to the predicted dimensionless potential $V_g(\alpha)/D_{\dot\theta}\tau_{\dot\theta}$.  
Line:  quadratic fit using the predicted shape from Eq.~\ref{eqn:potential_quadratic}. 
}
\label{fig:potential_alpha}
\end{figure} 

To test the prediction that the potential has the same form in $\theta_0$ and $\alpha$ in the small angle limit, we show $-\ln p(\alpha)$ in Fig.~\ref{fig:potential_alpha} as we did for $-\ln p(\theta_0)$ in Fig.~\ref{fig:pdf_potential}, using the same data set as in Fig.~\ref{fig:pdf_potential}.  Figure \ref{fig:potential_alpha} shows the potential minimum offset from the corner at 0 by almost 0.2 rad, within the deviation of 0.2 rad found when switching the direction of water flowing through the top and bottom plates in Fig.~\ref{theta0_switching_inoutlet}b.  The skewness in $p(\alpha)$ in Fig.~\ref{fig:potential_alpha} is also within the range observed in Fig.~\ref{theta0_switching_inoutlet}b.  Thus, these deviations of the peak locations from the corner and asymmetry in $p(\alpha)$ are likely due to the non-uniformity of the plate temperature, and are not interpreted to be significant results relevant to the idealized geometry of a cube.

A fit of a quadratic function to $-\ln p(\alpha)$ is shown in Fig.~\ref{fig:potential_alpha} using the same fit range of 0.3 rad as in Fig.~\ref{fig:pdf_potential}. This fit range includes 77\% of the measured data.  The reduced $\chi^2 = 8$  in this fit range using Poisson statistics. This larger $\chi^2$ than for the quadratic fit of $p(\theta_0)$ may be the result of a skewing of $p(\alpha)$ due to the plate temperature non-uniformity, which was found to have a larger effect on $\alpha$ than on $\theta_0$ by a factor of 2.3 (Sec.~\ref{sec:alpha}).

\begin{table}
\begin{tabular}{|r|r|}  
 \hline
 projection  & $\omega_r$ \\ \hline 
$p(\theta_0)$ &$47\pm9$ \\ \hline 
$p(\alpha)$ & $46\pm18$ \\ \hline 
$p(\theta_0, \alpha = \alpha_p)$ &$47\pm9$ \\ \hline 
$p(\alpha, \theta_0 = \theta_p)$ & $48\pm19$ \\ \hline 
$p(\theta_0-\theta_p=\alpha-\alpha_p)$ & $50\pm10$ \\ \hline 
$p[\theta_0-\theta_p=-(\alpha-\alpha_p)]$  & $41\pm8$ \\ \hline 
\end{tabular}
\caption{Comparison of the  natural frequency $\omega_r$  in units of mrad/s obtained from  quadratic fits of various projections of the two-dimensional  probability distribution $p(\theta_0,\alpha)$. Values are obtained from fits of $-\ln p(\theta_0)$ in Fig.~\ref{fig:pdf_potential} and of $-\ln p(\alpha)$ in Fig.~\ref{fig:potential_alpha}, as well as from fits of the negative logarithm of the  joint probability distribution $p(\theta_0,\alpha)$ along the 4 slices of the $(\theta_0,\alpha)$ plane shown in Fig.~\ref{fig:pdf_joint}.  The  natural frequency $\omega_r$ and thus the  curvature of the potential are  consistent along all different slices, in agreement  with the predicted  quadratic shape  of the potential in Eq.~\ref{eqn:potential_quadratic}, and indicate that distributions of $\theta_0$ and $\alpha$ are  independent of each other.
}
\label{tab:potential_curvatures}
\end{table}
 
The values of $\omega_r$ obtained from $p(\theta_0)$ and $p(\alpha)$ are shown in  Table \ref{tab:potential_curvatures}.  The curvatures fit in both Figs.~\ref{fig:pdf_potential} and \ref{fig:potential_alpha} were divided by measured values of  $D_{\dot\theta}\tau_{\dot\theta}$ from Fig.~\ref{fig:Dthetadot} to obtain $\omega_r$.  The dominant error is due the plate temperature non-uniformity, which was measured from the difference in $\omega_r$ obtained from data before and after switching the flow direction in the top and bottom plates in Fig.~\ref{theta0_switching_inoutlet}, to be  20\% for $p(\theta_0)$ and 40\% for $p(\alpha)$.  This error may be different for each distribution, so is considered an error for the purposes of comparisons within Table \ref{tab:potential_curvatures}.  In multiplying the curvature of $-\ln p(\theta_0)$ by $D_{\dot\theta}\tau_{\dot\theta}$, there is an additional systematic error of 10\% on $\omega_r$ that is the same for each distribution, so is not reported in Table \ref{tab:potential_curvatures}.  The values of $\omega_r$ are consistent for $p(\theta_0)$ and $p(\alpha)$ -- well within the error from the plate temperature non-uniformity --  as predicted from Eq.~\ref{eqn:potential_quadratic}.  

\subsection{$p(\theta_0,\alpha)$}
\label{sec:pdf_joint}

\begin{figure}[]
\includegraphics[width=.475\textwidth]{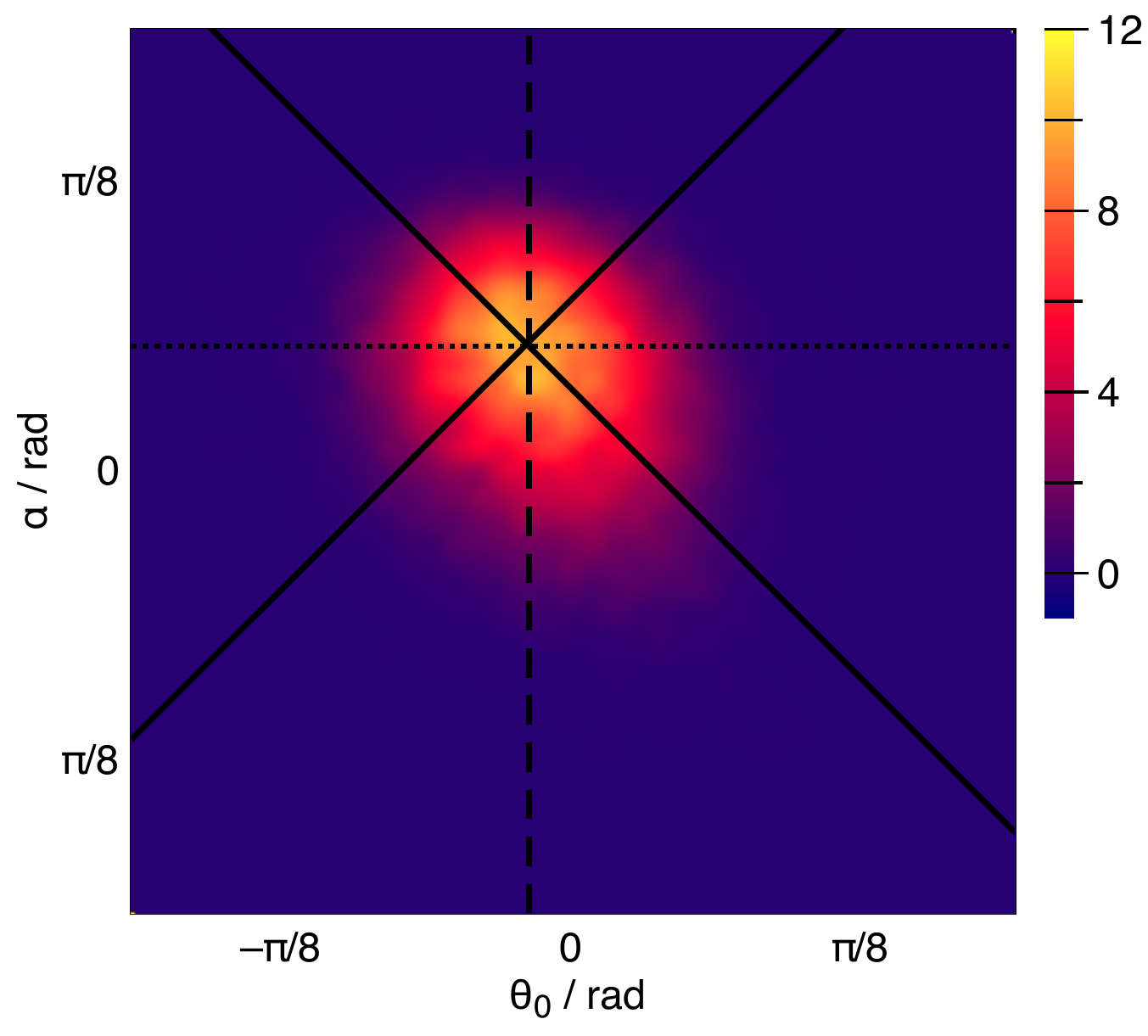} 
\caption{The joint probability distribution $p(\theta_0,\alpha)$. The distribution is close  to the circular shape  predicted from  the  quadratic approximation of Eq.~\ref{eqn:potential_quadratic}, but with a slightly oval shape.  The lines indicate slices along which one-dimensional  probability distributions are shown in Fig.~\ref{fig:potential_slices}, centered on the peak of  the probability distribution at coordinates $(\theta_p, \alpha_p)$.  
}
\label{fig:pdf_joint}
\end{figure} 

To test the quadratic potential approximation in Eq.~\ref{eqn:potential_quadratic} in both $\theta_0$ and $\alpha$ simultaneously, we calculate the joint probability distribution $p(\theta_0,\alpha)$.  We show $p(\theta_0,\alpha)$ in Fig.~\ref{fig:pdf_joint} as a 3-dimensional color plot.    The joint probability distribution is centered at $\theta_p = -0.06$ rad and $\alpha_p = 0.17$ rad, not at the origin, due mostly to the non-uniform plate temperature profile.  As a first approximation, Fig.~\ref{fig:pdf_joint} is close to the circular shape  predicted from  the  quadratic approximation of Eq.~\ref{eqn:potential_quadratic}. However,   there appears to be a slight oval shape to the distribution

 \begin{figure}[]
\includegraphics[width=.475\textwidth]{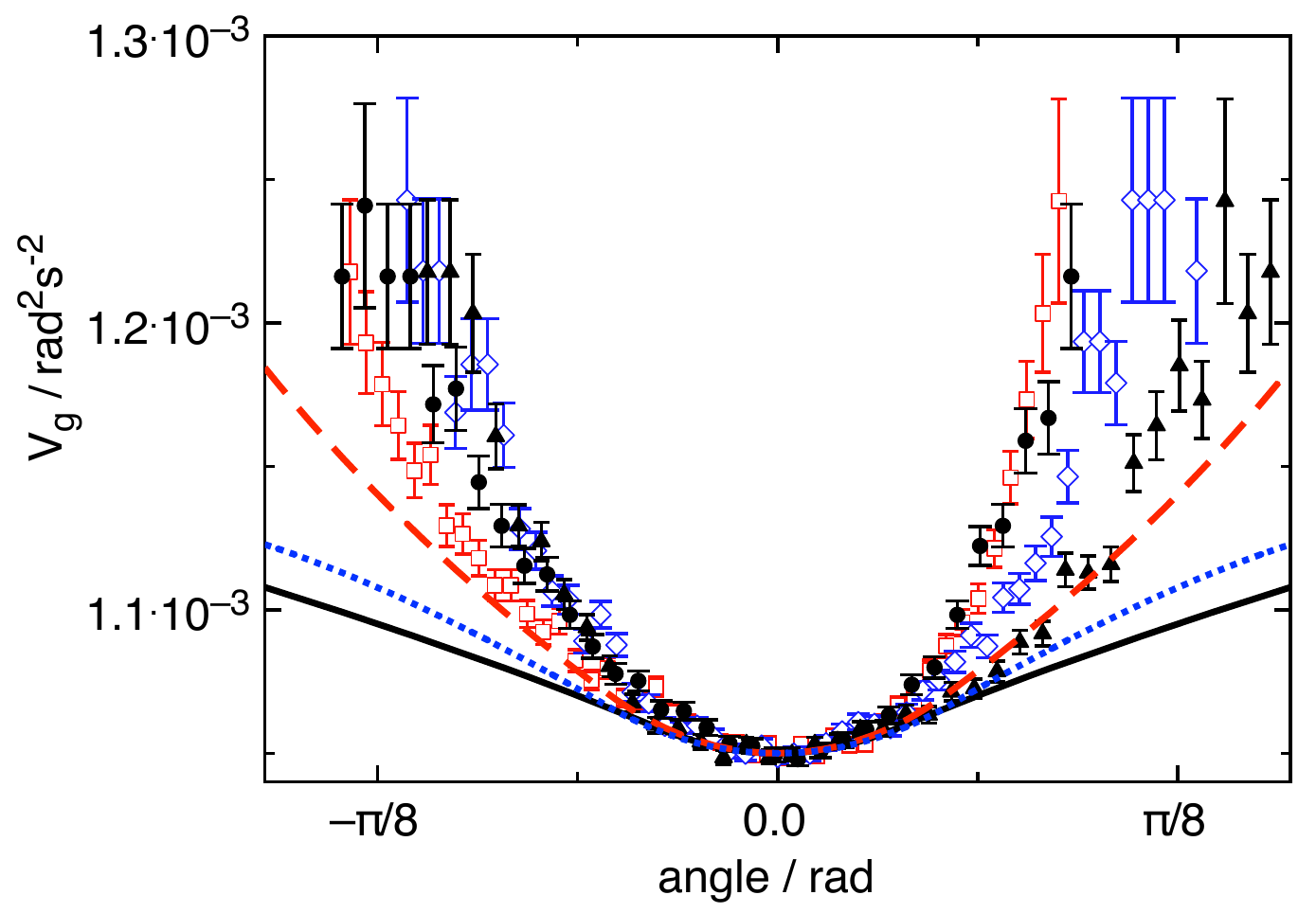} 
\caption{Comparison of the shape of the  predicted potential $V_g(\theta_0, \alpha)$ with the  negative logarithm of the measured probability distribution  along different slices of the $\theta_0$-$\alpha$ plane drawn in Fig.~\ref{fig:pdf_joint}.  Dotted line: predicted potential vs.~$\theta_0$. Dashed line: predicted potential vs.~$\alpha$. Solid line: predicted potential for $\theta_0 = \alpha$ and $\theta_0 = -\alpha$. All potential predictions are scaled have the natural frequency $\omega_r$ reduced by a factor of 2.9 to fit the measured data near the minimum.  Open diamonds:  measured $-\ln p(\theta_0)$. Open squares: measured $-\ln p(\alpha)$.  Solid triangles and circles: the measured probability distributions along $\theta_0 - \theta_p= -(\alpha - \alpha_p)$ and $\theta_0 - \theta_p= \alpha - \alpha_p$, respectively, corresponding to the solid lines in Fig.~\ref{fig:pdf_joint}. The collapse of the data near the potential minimum  confirms the form of the potential $V_g \propto \theta_0^2 + \alpha^2$ predicted in Eq.~\ref{eqn:potential_quadratic}.
}
\label{fig:potential_slices}
\end{figure} 

Since the three-dimensional plot in Fig.~\ref{fig:pdf_joint}  only allows for coarse comparisons, we plot slices of the two-dimensional  prediction of the potential $V_g(\theta_0,\alpha)$, and compare with the  measured probability distributions in Fig.~\ref{fig:potential_slices}.   The predictions shown are calculated from the full potential given in Appendix 2, not the linear approximation in Eqs.~\ref{eqn:diameter_linear} and \ref{eqn:potential_quadratic}, with the natural frequency $\omega_r$ a factor of 2.9 smaller than the prediction to fit the potential near the minimum and better compare the shape of the potentials.  The prediction of the full potential  retains the four-fold symmetry of a cube outside of the linear approximation,  although it does not retain azimuthal symmetry around the potential minimum in the $\theta_0$-$\alpha$ plane at large angles.  The  local maxima of the potential on any given circle in the $\theta_0$-$\alpha$ plane centered around the potential minimum are predicted to occur along the $\theta_0$-axis and $\alpha$-axis.   The local potential minima on these circles are predicted to be at angles of $\pm 45^{\circ}$ relative to the $\theta_0$ and $\alpha$-axes, illlustrated as the diagonal lines in Fig.~\ref{fig:pdf_joint}.  The slices of the predicted potential along these lines are shown in Fig.~\ref{fig:potential_slices}.  

Measured values of $-\ln p(\theta_0-\theta_p)$ at $\alpha=\alpha_p$, and $-\ln p(\alpha-\alpha_p)$ at $\theta=\theta_p$ are shown in Fig.~\ref{fig:potential_slices}.  These distributions are shifted to be centered around $\theta_p$ and $\alpha_p$  to better compare the shapes of the probability distributions to the predicted potential.  These distributions are taken as thin slices of the $\theta_0$-$\alpha$ plane with a width of 0.016 rad.   We also plot the measured  negative  logarithm of the probability distribution along the slices where $\theta_0 - \theta_p = \alpha - \alpha_p$ and $\theta_0 - \theta_p = - (\alpha - \alpha_p)$, corresponding to the diagonal lines shown in Fig.~\ref{fig:pdf_joint}.   The probability distributions all have similar curvature near the potential minimum,  but the tails consistently drop off faster than the predictions.  Most of the tails of the measured  probability distributions are asymmetric around their centers, likely due to the non-uniform plate temperature or other asymmetries of the setup, so it is difficult to compare the shapes of the measured potentials along different projections.

 We fit the quadratic function of Eq.~\ref{eqn:quadratic} to the four  probability distributions  shown in Fig.~\ref{fig:potential_slices}, using the same fit range of 0.3 rad as in Figs.~\ref{fig:pdf_potential} and \ref{fig:potential_alpha}, and centered on $\theta_p$ and $\alpha_p$.  The fit values are summarized in Table \ref{tab:potential_curvatures}.  All fits have a reduced $\chi^2\approx1$,  indicating that they are well-described by a quadratic function within $\pm0.15$ rad of the peak, where 80\% of the data lies.  Furthermore, values of $\omega_r$ are  consistent with each other along all of the slices of the probability distribution.  This  indicates the  predicted  quadratic potential of Eq.~\ref{eqn:potential_quadratic} is a good qualitative model for $p(\theta_0,\alpha)$ near the potential minimum.
Since the fit values of $\omega_r$ for the two slices at fixed $\alpha$ and $\theta_0$ are consistent with the fit results using all of the data, this indicates an independence of the parameters such that distributions of $\theta_0$ are  not conditional on $\alpha$,  and distributions of $\alpha$  not conditional on $\theta_0$.  This confirms that the quadratic potential lacks any coupling  terms between $\theta_0$ and $\alpha$ in the small angle limit, which implies the forcing on $\theta_0$ is independent of $\alpha$ (i.e.~$-\partial V_g(\theta_0,\alpha)/\partial\theta_0$ is not a function of $\alpha$), and similarly the forcing on $\alpha$ is independent of $\theta_0$.   This lack of coupling confirms that a correct analysis of the system can be obtained by analyzing it as a function of one parameter at a time.

\section{Power spectrum}
\label{sec:powerspec}

\subsection{Model for advected modes}
\label{sec:powerspec_model}
One of the possible dynamical consequences of a potential $V_g(\theta_0, \alpha)$ with a local minimum is oscillations in $\theta_0$ and $\alpha$ around that potential minimum.  In circular cylindrical containers, a combination of sloshing and twisting of the LSC structure around the plane of the LSC was found \cite{FA04, XZZCX09, ZXZSX09, BA09}, where the restoring force for the oscillation came only from the slosh displacement $\alpha$ for a circular cross section.  Since a cubic geometry leads to a restoring force in both $\theta_0$ and $\alpha$, it can potentially excite different modes of oscillation.  This section explains the process of calculating the  power spectrum for a square cross section from traveling wave solutions of  the advected oscillation model of Brown \& Ahlers  \cite{BA09}, and highlights the differences from a circular cross section.  

To obtain equations of motion that account for advection, we start with Eq.~\ref{eqn:theta_model} for $\theta_0$ with the quadratic potential approximation (Eq.~\ref{eqn:potential_quadratic}), here we also assume the independence of the forcings on $\theta_0$ and $\alpha$, which is now justified in Sec.~\ref{sec:pdf_joint}. We assume a mathematically similar equation of motion for $\alpha$, since Eq.~\ref{eqn:potential_quadratic} has the same  quadratic term for both variables. 

These equations of motion can be converted into Lagrangian coordinates to view them as traveling waves.  Traveling waves  can be described in terms of the angles of the hottest spot $\alpha_h = \theta_0+\alpha$ and coldest spot $\alpha_c = \theta_0-\alpha$ of the temperature profile in a horizontal cross-section  and  as they travel up and down the walls (illustrated in Fig.~\ref{fig:angle_definitions}.  In the stationary frame, these traveling wave superpose to produce our more traditional LSC angles
\begin{equation}
 \theta_0 = \frac{\alpha_h+\alpha_c}{2} \ ; \  \alpha= \frac{\alpha_h-\alpha_c}{2}
 \label{eqn:angle_conversions}
\end{equation}

To account for advection, we add advective terms to the equations of motion for $\alpha_h$ and $\alpha_c$, corresponding to upward and downward motion in $z$, respectively: $\mp (\omega_{\phi}/k_0)\partial\alpha_{h/c}(z,t)/\partial z$.   $\omega_{\phi}/k_0$ is  the circulation velocity, where $k_0$ is the wavenumber corresponding to the circulation. The  resulting equation of motion for the upward traveling wave is 
\begin{equation}
\ddot\alpha_h = -\frac{\dot\alpha_h}{\tau_{\dot\theta}} - \omega_r^2\alpha_h - \frac{\omega_{\phi}}{k_0}\frac{\partial\dot\alpha_h}{\partial z} + f_h(t) \ .  
 \label{eqn:model_alphah}
\end{equation}
Here we have used the approximation $\delta\approx \delta_0$ for simplicity.   A similar equation results for  the downward traveling wave in terms of $\alpha_c$, with subscripts $_h$ replaced by $_c$, and $z$ replaced by $-z$.  

Using the identities of Eq.~\ref{eqn:angle_conversions}, we convert the equations of motion for the traveling waves (i.e.~Eq.~\ref{eqn:model_alphah} and its analogy for $\alpha_c$) back to the stationary frame in terms of $\theta_0$ and $\alpha$ to obtain
\begin{equation}
\ddot\theta_0 = -\frac{\dot\theta_0}{\tau_{\dot\theta}} - \omega_r^2\theta_0 - \frac{\omega_{\phi}}{k_0}\frac{\partial\dot\alpha}{\partial z} + f_{\dot\theta}(t) \ .
\label{eqn:theta0_model_advected}
\end{equation}
A similar equation results for $\alpha$ with  $\theta_0$ and $\alpha$ transposed. These equations are similar to Eq.~\ref{eqn:theta_model}, but with an additional equation for $\alpha$, and each equation now has an advective term which couples the equations of motion for $\theta_0$ and $\alpha$.  In contrast, for a circular cross section, the coupling of these equations allowed the restoring force in $\alpha$ to drive oscillations in both $\alpha$ and $\theta_0$ to get the combined sloshing and twisting mode, as no separate restoring force was found for $\theta_0$ in that geometry \cite{BA09}.  In the cube, this  coupling is unnecessary to get oscillation of both parameters, as both parameters have their own restoring force due to the minimum in the potential $V_g(\theta_0,\alpha)$ in terms of each parameter.  

To find solutions to Eq.~\ref{eqn:theta0_model_advected}, we first solve the uncoupled Eq.~\ref{eqn:model_alphah} which has partial solutions in the form of traveling waves  given by 
\begin{equation}
\alpha_{h,n}(\omega,t)=a_n(\omega)\cos(nk_0 z - \omega t - \Phi_n(\omega)]
\label{eqn:alphah}
\end{equation}
and
\begin{equation}
\alpha_{c,n}(\omega,t)=a_n(\omega)\cos[-nk_0 z - \omega t - \Phi_n(\omega)+ \psi_n]
\label{eqn:alphac}
\end{equation}
where  $z$ is the height of the  thermistor rows relative to the midplane, $\Phi_n(\omega)$ accounts for any phase shifts between different frequencies or modes, and $\psi_n$ is a phase shift between the upward- and downward-traveling waves.   If plumes making up the LSC remain coherent over multiple turnover times, they can destructively interfere with each other when they loop around.  To satisfy the condition for constructive interference for a closed loop circulation, where $\alpha_h$ and $\alpha_c$ are two different segments of the same traveling wave, requires $\psi_n=(n+1)\pi$ with integer $n$, and the coefficients $a_n(\omega)$ and phases $\Phi_n(\omega)$ be the same for both $\alpha_h$ and $\alpha_c$.   Since there is a restoring force for both $\theta_0$ and $\alpha$,  nontrivial solutions are expected for all positive integers $n$.  In contrast, for a circular cross section there is only a restoring force for $\alpha$, producing only even-$n$  order modes \cite{BA09}.   Summing these  traveling wave solutions for $\alpha_h$ and $\alpha_c$ using the identities in Eq.~\ref{eqn:angle_conversions} results in standing waves in the lab frame given by  
\begin{equation}
\theta_{0,n}(\omega,t)=a_n(\omega)\cos(nk_0 z)\cos[\omega t - \Phi_n(\omega)] 
\label{eqn:theta_travelingwave}
\end{equation}
and
\begin{equation}
\alpha_n(\omega,t)=a_n(\omega)\sin(nk_0 z)\sin[\omega t - \Phi_n(\omega)] 
\label{eqn:alpha_travelingwave}
\end{equation}
for odd $n$.  The sines and cosines are switched for even order $n$ solutions. Plugging in these standing wave solutions with the time-dependence represented as a complex exponential into  Eq.~\ref{eqn:theta0_model_advected} as a function of frequency $\omega$ and assuming $f(t)$ is described by white noise with diffusivity $D_{\dot\theta}$ yields the power spectrum of $\theta_0$ at $z=0$  for a given mode of integer order $n$:
\begin{equation}
P_n(\omega) =   |a_n(\omega)|^2 = \frac{D_{\dot{\theta}}}{(\omega^2 - \omega_r^{2} - n\omega\omega_\phi)^2 + \left(\frac{\omega}{\tau_{\dot{\theta}}}\right)^2} \ .
\label{eqn:power_spec_theory}
\end{equation}

While there is an infinite series of modes for integer $n$, the $n=1$ mode is expected to be the dominant mode, since higher-frequency modes tend to have less peak power due to the larger magnitude of damping at higher frequency.

\subsection{Testing the power spectrum}

\begin{figure}
\includegraphics[width=.475\textwidth]{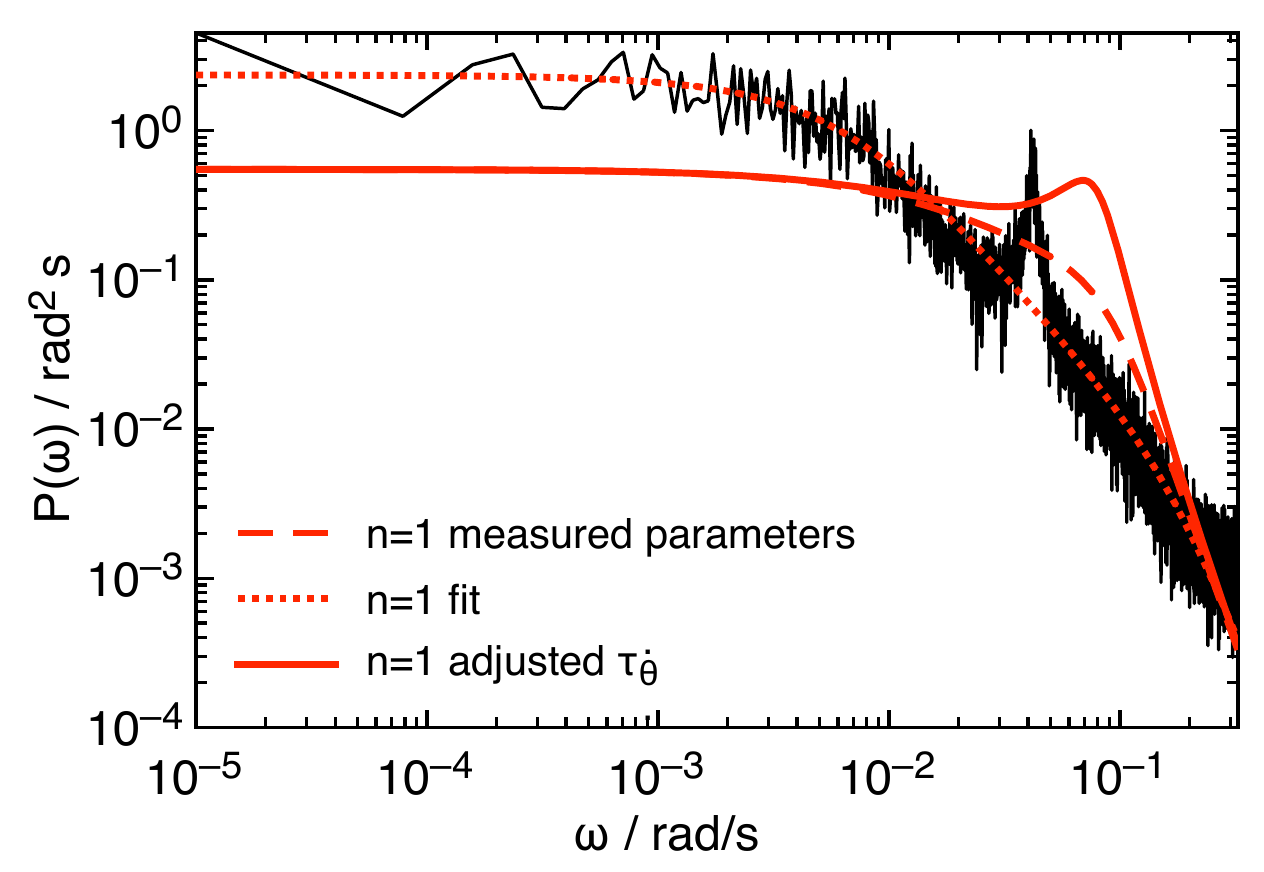} 
\caption{(color online) Power spectrum of $\theta_0$. Black curve: measured data.  Red dashed curve: prediction using  measured values of $D_{\dot\theta}$, $\tau_{\dot\theta}$, $\mathcal{T}$, and $\omega_r$ and $n=1$ as input.  This prediction captures the plateau of the power spectrum, but is in the underdamped  regime of the model  where there is no resonance peak.  Red dotted curve:  fit to the data. The fit parameters are all within a factor of about 2 of independently measured values, but do not include a resonance peak.    Red solid curve:  using the same measured values for input data as the red dashed curve but scaling $\tau_{\dot\theta}$ up by a factor of 2.3 to fit the resonant frequency as a function of Ra.  This small  change in parameter values is enough to move the  model from an overdamped to underdamped state, with resonance near the observed peak frequency, indicating that the model is consistent with the {\em observed} oscillation.  However, the uncertainties on parameter values are too large to make correct {\em predictions} that the system is in the underdamped state rather than the overdamped state.   
}
\label{power_spec_theta_mid}
\end{figure} 

To test the prediction of the power spectrum $P_1(\omega)$, we  show an example in Fig.~\ref{power_spec_theta_mid}  of the measured power spectrum for $Ra = 2.62\times10^9$ and $\beta=1^{\circ}$.  The power spectrum has a roll-off indicative of damping, and a peak  which  corresponds to a resonant frequency for oscillations.
The corresponding probability distribution in Fig.~\ref{fig:pdf_potential} confirms that these oscillations are nearly centered around a corner, as predicted.  

To test the  self-consistency of a stochastic ODE model to describe the power spectrum,  we calculate a prediction for $P_1(\omega)$ for the expected dominant $n=1$ mode from Eq.~\ref{eqn:power_spec_theory}, using the measured  value of $\omega_r$ obtained from fitting $p(\theta_0)$ (Sec.~\ref{sec:potential}),  along with the independently measured $\mathcal{T}$, $D_{\dot{\theta}}$, $\tau_{\dot{\theta}}$ (Appendix 1) and the definition $\omega_{\phi}=2\pi/\mathcal{T}$.  This prediction of $P_1(\omega)$ is shown  as the red dashed curve in Fig.~\ref{power_spec_theta_mid}.  The low-frequency plateau is about a factor of 3 away from the measured plateau.  The  resonant peak is missing for these parameter values,  although there is some power above the background near the natural frequency,  which is within 9\% of the measured oscillation frequency.  While the prediction is in the ballpark of the measured power spectrum, the parameter values are in a range such that the model does not predict the observed resonance peak.

To determine what model parameter range could be consistent with the data, we fit the predicted functional form of the power spectrum $P_1(\omega)$ from Eq.~\ref{eqn:power_spec_theory} to the measured $P(\omega)$, assuming a constant error (to balance out the higher logarithmic density of data at lower probability),
 and fixing $ n = 1$. The fit yields $D_{\dot{\theta}} = (3.9\pm5.2)\times10^{-6}$ rad$^2$/s$^3$, $\tau_{\dot{\theta}} = 5.7\pm3.6$ s, $\omega_{\phi} = 0.066\pm0.045$ rad/s, and $\omega_r = 0.036\pm0.012$ rad/s. This fit  is shown as the red dotted curve in Fig.~\ref{power_spec_theta_mid}.  The large errors are a consequence of having more free parameters than distinct features in the background, and the parameters being strongly coupled to each other.  For comparison, the values from measurements of the mean-square displacement and turnover time are $D_{\dot{\theta}} = (2.68\pm0.26)\times10^{-6}$ rad$^2$/s$^3$, $\tau_{\dot{\theta}} = 13.3\pm0.4$ s,  $\omega_{\phi} = 0.049\pm0.003$ rad/s from Appendix 1, and $\omega_r = 0.047\pm0.009$ rad/s from measurements of $p(\theta_0)$ in Sec.~\ref{sec:potential}. The fitted  parameter values are in general within a factor of about 2 of the measured values, which confirms some amount of self-consistency of the stochastic ODE model  within these generous errors that the model has required in some cases \cite{BA08a}. While these parameters fit the background well,  this fit still does not capture the peak of the power spectrum in Fig.~\ref{power_spec_theta_mid}.  This is because the model overestimates the width of the peak, so a least-squares fit results in a better fit by fitting only the background and ignoring the peak.

The linearized model of Eq.~\ref{eqn:power_spec_theory} also failed to capture the peak of the power spectrum of $\theta_0$ in a circular cylinder \cite{BA09}.  In that case, the discrepancy could be attributed to the variable damping in the original model due to the variation of $\delta$ in  the damping term of Eq.~\ref{eqn:theta_model} \cite{BA08a}.  Modeling the damping as random effectively reduces the damping by a factor to $1-S$ where $S=(\sigma_{\delta}/\delta_0)^2 \tau_{\delta}/\tau_{\dot\theta}$, and $\sigma_{\delta}=\sqrt{D_{\delta}\tau_{\delta}}$ is the standard deviation of $\delta$ in the linearized limit of Eq.~\ref{eqn:delta_model} \cite{BA09}.  This correction could move the state of the system from the overdamped to the underdamped regime if the damping adjustment $S$ is large enough.  For this transition to occur for our measured parameter values from Appendix 1 at $Ra = 2.62\times10^{9}$ requires $S>0.47$. However, we calculate  $S$ ranging from 0.01 to 0.16 as Ra decreases from $2.62\times10^9$ to $8.41\times10^7$  using the measured parameter values from Appendix 1.  This is not enough of a correction to move the system to the underdamped regime. This is a much smaller correction factor in a cube compared to the $S=0.5$ reduction found at higher Ra in a circular cylinder \cite{BA09}. This smaller correction is mostly a consequence of the smaller relative fluctuation strength $\sigma_{\delta}/\delta_0$ for this dataset relative to that of \cite{BA09}, also the reason cessations are much less likely for this data \cite{BJB16}. 

To  show how close the parameter values are to the underdamped regime, starting with the measured parameters values, we increase $\tau_{\dot\theta}$ to a factor of 2.3 times the measured value to move the system into the underdamped regime, shown as the red solid curve in Fig.~\ref{power_spec_theta_mid}.   The system is near enough  to the overdamped-underdamped transition  that a change in parameter values  by a factor of about 2 can cross this transition and  qualitatively change the dynamics.  While this variation in parameter values is consistent with typical errors of the model, this means that these uncertainties on parameter values are too large for the model to correctly predict the observation that the system is in the underdamped state rather than the overdamped state.


While $n=1$ is predicted to be the advected mode with the most peak power, the predicted dropoff in peak power averages about 30\% for each integer increase in $n$ (not shown) when using the measured parameter values $D_{\dot\theta}$, and $\omega_r$, and increasing $\tau_{\dot\theta}$ by a factor of 2.3 to obtain resonance.  Higher order modes are not resolvable in the steep and noisy rolloff of the measured data in Fig.~\ref{power_spec_theta_mid}. 


\section{Oscillation structure}
\label{sec:oscillation_structure}

In this section, we characterize the oscillation structure and compare it to model predictions of how it differs from oscillations in other cell geometries such as  twisting and sloshing in cells with circular cross section \cite{FA04, XZZCX09, ZXZSX09, BA09}, or rocking in horizontal cylinders \cite{SBHT14}.  We characterize the oscillation structure by phase shifts in correlation functions between  orientations or slosh angles at different rows of thermistors \cite{BA09}.  

Detailed predictions for the phase shifts of the correlation functions from the  standing-wave solutions of Eqs.~\ref{eqn:alpha_travelingwave} and \ref{eqn:theta_travelingwave} are shown in Appendix 3.  The predicted structure for the expected dominant $n=1$ mode is an oscillation where $\theta_0$ is in-phase at all  rows of thermistors, similar to that found in a tilted circular cylinder \cite{BA08b}, and $\alpha$ is out-of-phase at the top and bottom rows, corresponding to an LSC rocking back and forth around the horizontal axis in the LSC plane, similar to the rocking mode found in a horizontal cylinder \cite{SBHT14}.  

\subsection{Power spectra of $A_n$}

\begin{figure}
\includegraphics[width=.475\textwidth]{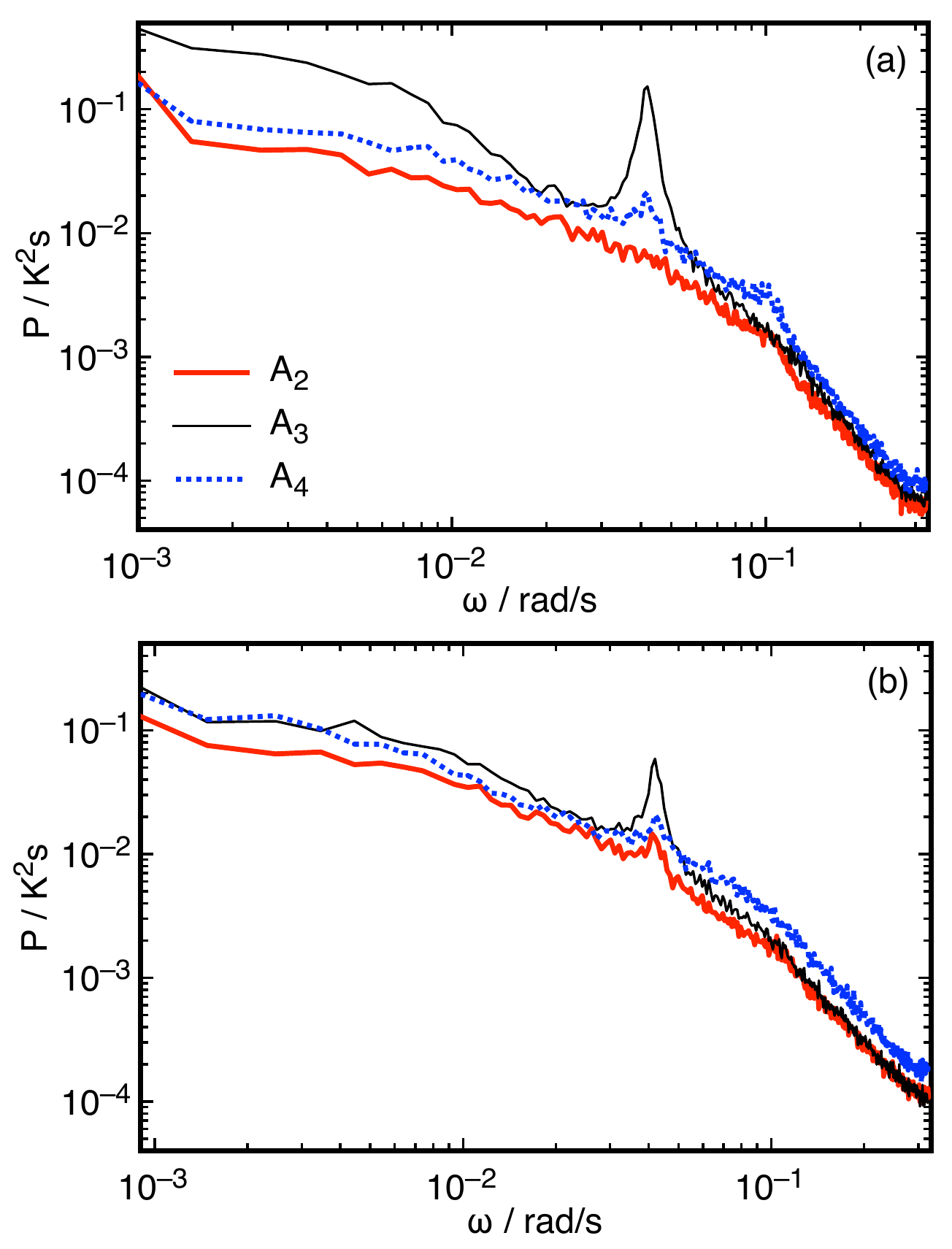}
\caption{(color online) The power spectra of the Fourier modes $A_n$ of the temperature profile for (a) the middle row of thermistors, and  (b) bottom row of thermistors. Red thick solid line: $A_2$. Thin black solid line: $A_3$. Blue dotted line: $A_4$.  
Each signal has a peak at the same frequency as the oscillation in $\theta_0$ except for $A_2$ at the middle row.  This  differs qualitatively from circular cylinders, where only $A_2$ was found to oscillate.  Oscillations in $A_2$ at the top and bottom row only in the primary peak correspond to the predicted $n=1$ mode.  Weaker higher-frequency peaks in $A_2$ at all rows correspond to the predicted $n=2$ mode.
}
\label{fig:power_spec_momentsA}
\end{figure}

Before showing correlation functions to identify the oscillation structure, we show power spectra of the Fourier moments $A_n$ of the temperature profile to identify which of these Fourier moments contribute to oscillation structure, and thus which $A_n$ should be considered in calculating correlation functions and phase shifts.  Power spectra of the Fourier moments $A_n$ of the temperature profile from Eq.~\ref{eqn:a_n} are shown in Fig.~\ref{fig:power_spec_momentsA}a for the middle row thermistors and panel b for the bottom row.  The top row power spectrum is not shown since we find it to be qualitatively similar to the bottom row, following the symmetry of the Boussinesq approximation.  Each signal except for $A_2$ at the middle row has a primary peak at $\omega=0.44$ rad/s, the same frequency as the oscillation in $\theta_0$ (Fig.~\ref{power_spec_theta_mid}).  

Since $\alpha$ is calculated from $A_2$ (Sec.~\ref{sec:alpha}), the oscillation of $A_2$ in Fig.~\ref{fig:power_spec_momentsA} at the top and bottom rows, but not at the middle row ($z=0$) at the primary frequency, is consistent with the predicted $n=1$ sloshing mode (Eq.~\ref{eqn:alpha_travelingwave}).  This oscillation is distinct from the $n=2$ mode in a circular cylindrical cell where $A_2$ oscillates at all 3 rows \cite{BA09}.   

The oscillations in $A_3$ and $A_4$  in Fig.~\ref{fig:power_spec_momentsA} were not found in circular cylinders \cite{BA09}.   $A_4$ contributes to a shift in the extrema of the temperature profile closer together, thus could also be interpreted as contributing to a sloshing oscillation.  The  peak in $A_3$ causes a shift in both extrema of the temperature profile in the same direction, so does not contribute to sloshing, but could be affect the interpretation of the LSC orientation.  The moment $A_3$ is the dominant oscillating mode in Fig.~\ref{fig:power_spec_momentsA}, and even has more power than the oscillation $\theta_0$ (comparing the integral of the peaks in Fig.~\ref{fig:power_spec_momentsA} with Fig.~\ref{power_spec_theta_mid}).  $A_3$ corresponds to an oscillation of the shape of the temperature profile that appears to be induced by the oscillation of $\theta_0$ around the corners of the cubic cell, which will be discussed in detail in a follow-up paper \cite{JB20b}.

Smaller, secondary peaks are observed in the power spectra in Fig.~\ref{fig:power_spec_momentsA} at about twice the frequency of the $n=1$ mode, where the $n=2$ mode is expected.  Observed oscillations in $A_2$ and $A_4$ at all 3 rows correspond to the sloshing component of the $n=2$ mode, and the oscillations in $A_3$ at the top and bottom rows, but not the middle row, could correspond to the twisting component of the $n=2$ mode.  These observations are in agreement with the expectations that this $n=2$ mode that has been found in circular cylinders \cite{BA09} is still expected to occur in a cube at approximately twice the frequency of the $n=1$ mode \cite{BA08b}, but with less power than the lower-frequency $n=1$ mode due to increased damping at higher frequency.  

\subsection{Definition of modified oscillation angles $\hat\theta_0$ and $\hat\alpha$ for measuring phase shifts}

Since $A_3$ and $A_4$ were not found to oscillate in a circular cylindrical cell \cite{BA09}, they were not used in the original definitions of $\theta_0$ or $\alpha$.  Since $A_3$ and $A_4$ could be interpreted as contributing to a the LSC orientation or slosh angle, respectively, we consider alternate definitions of the LSC orientation and slosh angle when calculating correlation functions that include these higher-order Fourier modes.  We assume that the orientations of the maximum $\hat\alpha_h$ and minimum $\hat\alpha_c$ of the temperature profile (Eq.~\ref{eqn:fourier_series_temp}) are the relevant orientations as far as oscillations are concerned.  The modified oscillation angles are defined as $\hat\theta_0=(\hat\alpha_h+\hat\alpha_c)/2$ and $\hat\alpha= (\hat\alpha_h-\hat\alpha_c)/2$ in analogy to Eq.~\ref{eqn:angle_conversions}, where $A_3$ contributes to $\hat\theta$, and both $A_2$ and $A_4$ contribute to $\hat\alpha$.  For consistency with previous work, we have used the previous definitions of $\theta_0$ and $\alpha$ in Secs.~\ref{sec:potential}, \ref{sec:powerspec}, and \ref{sec:parameters}.   We confirmed that the values of $\omega_r$ in Sec.~\ref{sec:potential}, for example, vary by only 20\% (within systematic errors) if $p(\hat\theta_0)$ is used to calculate $\omega_r$ instead of $p(\theta_0)$.

\subsection{Definition of correlation functions}

The correlation function between two signals $x(t)$ and $y(t)$ is defined as
\begin{equation}
C_{x,y}(\tau) = \frac{\langle (x(t) - \overline{x})(y(t-\tau) -  \overline{y})\rangle}{\sqrt{\langle (x(t)- \overline{x})^2\rangle \langle (y(t)- \overline{y})^2\rangle}} \ ,
\label{eqn:correlation}
\end{equation}
where both $\langle ...\rangle$ and  $\overline{x}$ denote time averages.  $x$ and $y$ can stand for the angles $\theta_m$, $\theta_t$, and $\theta_b$, which correspond to $\theta_0$ at the middle, top, and bottom rows of thermistors, respectively, and the angles $\alpha_m$, $\alpha_t$, and $\alpha_b$, which correspond to $\alpha$ at the middle, top, and bottom rows, respectively, or the modified versions of those angles.

\subsection{Measured correlation functions and comparison with prediction}
\label{sec:correlations}

In this subsection, we show the five independent cross-correlation functions where we found the most clear oscillations to measure all the phase shifts between the six independent time series of angles (orientation and slosh angles at 3 rows each).

\begin{figure}
\includegraphics[width=.475\textwidth]{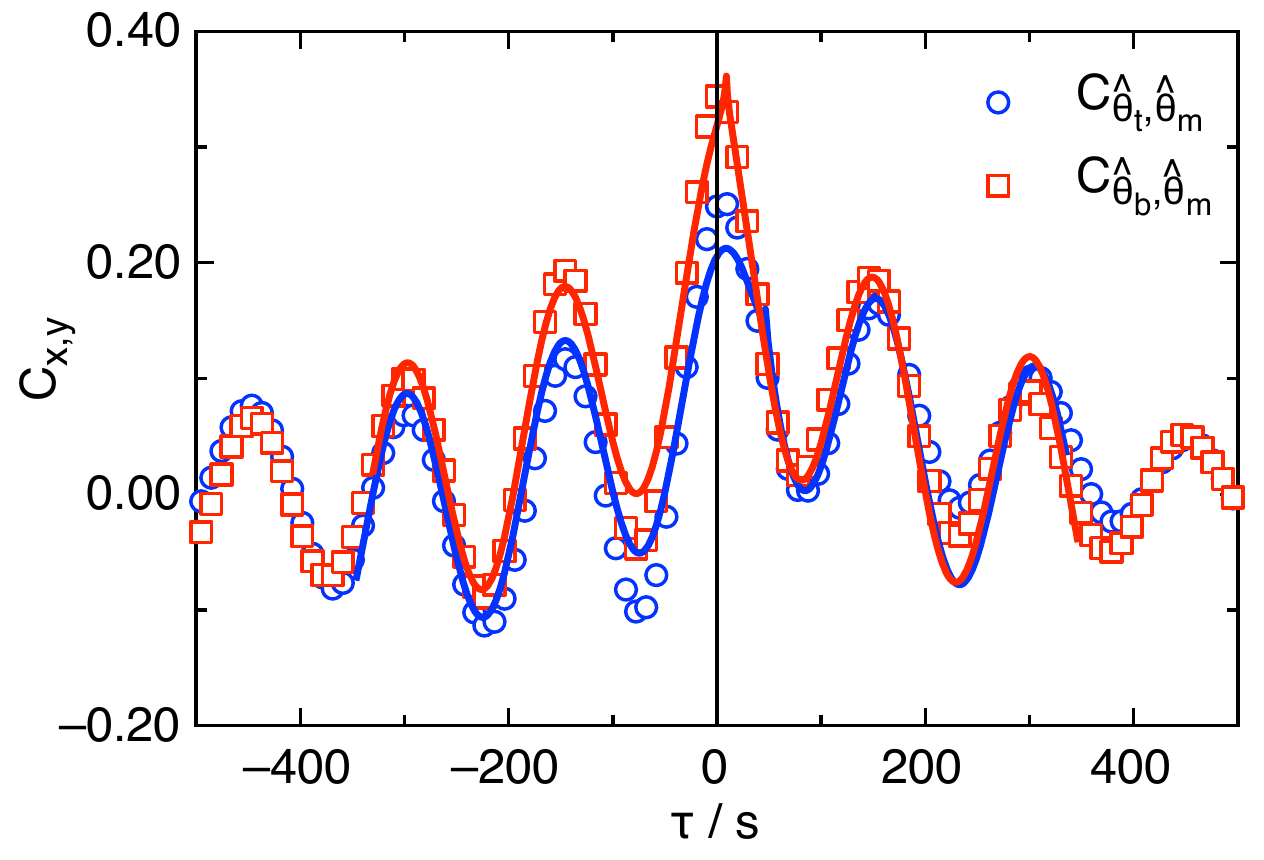} 
\caption{(color online) Cross-correlations between the modified LSC orientation $\hat\theta_0$ of different rows of thermistors, as indicated in the legend.  Lines: fits of Eq.~\ref{fitting_corr_period} to data of the same color to  determine  the phase shifts between signals and frequency of oscillation.  The alignment of the peaks at $\tau=0$ indicates that $\hat\theta_t$, $\hat\theta_b$ and $\hat\theta_m$ are oscillating in phase with each other, in agreement with the predicted $n=1$ advected oscillation mode.
}
\label{fig:corrtheta}
\end{figure}

Figure \ref{fig:corrtheta} shows the cross-correlations $C_{\hat\theta_b, \hat\theta_m}$ and $C_{\hat\theta_t, \hat\theta_m}$.  Since both signals in Fig.~\ref{fig:corrtheta} have evenly spaced peaks with the same spacing, then $\hat\theta_m$, $\hat\theta_b$, and $\hat\theta_t$ are all oscillating at the same frequency.   Since $C_{\hat\theta_b, \hat\theta_m}$ and $C_{\hat\theta_t, \hat\theta_m}$ have peaks near $\tau=0$, then $\hat\theta_t$, $\hat\theta_b$ and $\hat\theta_m$ are all in phase with each other. The positive correlations indicate the 3 rows tend to line up in a vertical plane.  This in-phase oscillation in $\hat\theta_0$ is in agreement with the predicted $n=1$ advected oscillation mode described in Appendix 3, and distinct from the $n=2$ twisting oscillation found in circular cylinders \cite{FA04, BA09}.  

The phase shift of $\hat \theta_0$ is dominated by the $A_3$ Fourier mode, which has the same phase at all 3 rows of thermistors.  
$\theta_0$ at the top and bottom rows is also in phase with $\hat\theta_0$, however $\theta_0$ at the middle row is found to 
be out $\pi$ rad of phase with the top and bottom rows and the $A_3$ mode.  These different phase shifts for different definitions of the LSC orientation indicate a more complex oscillation structure than predicted, which we will follow up on in a later paper \cite{JB20b}.


\begin{figure}
\includegraphics[width=.475\textwidth]{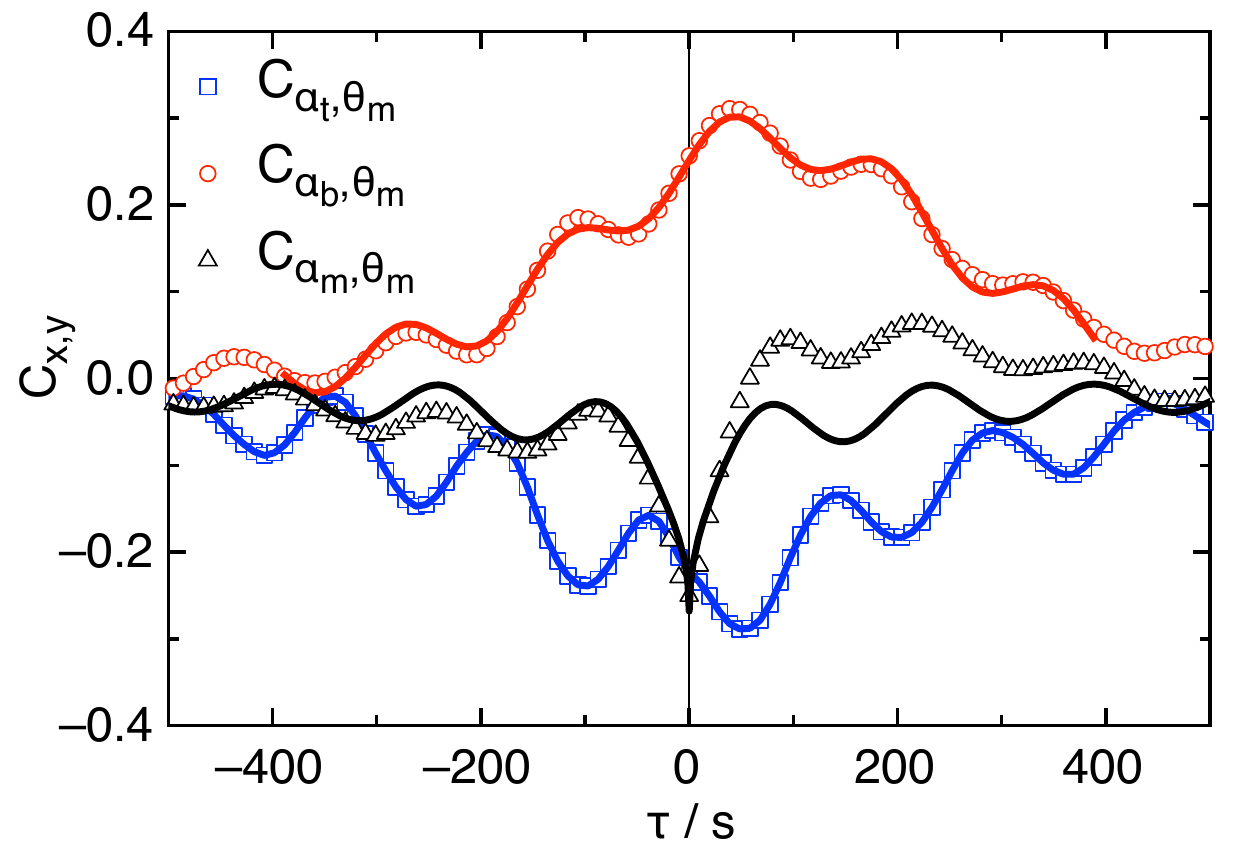} 
\caption{(color online) Cross-correlations between $\alpha$ at different rows of thermistors with $\theta_m$, as indicated in the legend.  Lines: fits of Eq.~\ref{fitting_corr_period} to data of the same color to  determine  the phase shifts between signals and frequency of oscillation.  The approximately $\pi/2$ rad phase shifts and opposite signs of $C_{\alpha_b,\theta_m}$ and $C_{\alpha_t,\theta_m}$ indicate that $\alpha_b$ and $\alpha_t$ are oscillating $\pi$ rad out of phase with each other, consistent with the $n = 1$ mode. The oscillation in $\alpha_m$ is not part of the $n=1$ mode, but indicates an asymmetry between the hot and cold sides of the LSC. 
}
\label{fig:corralphatheta}
\end{figure}

Figure \ref{fig:corralphatheta} shows phases shifts between $\alpha$ at different rows with $\theta_m$  to determine the phase shifts between different rows of $\alpha$.  The extrema of $C_{\alpha_b,\theta_m}$ and $C_{\alpha_t,\theta_m}$ have phases of approximately $\pi/2$ rad with opposite signs, corresponding to $\alpha_b$ and $\alpha_t$ oscillating $\pi$ rad out of phase with each other.   Regardless of which angle definitions we use, we find $\alpha_b$ and $\alpha_t$ are $\pi$ rad out of phase with each other.  The out-of-phase behavior of $\alpha_b$ and $\alpha_t$ is consistent with the $n = 1$ oscillation mode prediction in Appendix 3, again distinct from the $n=2$ mode found in circular cylindrical containers in which the different rows of $\alpha$ oscillate in phase with each other \cite{XZZCX09, ZXZSX09, BA09}.  
The equally spaced peaks in $C_{\alpha_m,\theta_m}$ in Fig.~\ref{fig:corralphatheta} indicate that $\alpha_m$ is also oscillating at the same frequency as other modes, however, this is not expected as part of the $n =  1$ mode.  The oscillation in $\alpha_m$ is much weaker than $\alpha_t$ and $\alpha_b$, as there was no resolvable peak in the power spectrum of $A_2$ in  Fig.~\ref{fig:power_spec_momentsA}(a).  The weak oscillation in $\alpha_m$ that is $\pi$ rad out-of-phase with $\theta_m$ corresponds to a weak sloshing on top of the $\theta_0$ oscillation such that $\alpha_c$ oscillates with a slightly larger amplitude than $\alpha_h$.  This asymmetry between $\alpha_h$ and $\alpha_c$ appears to be a new non-Boussinesq effect and not an asymmetry of the setup (see Appendix 4 for justification). 

\section{Oscillation period}
\label{sec:period}

\begin{figure}
\includegraphics[width=.475\textwidth]{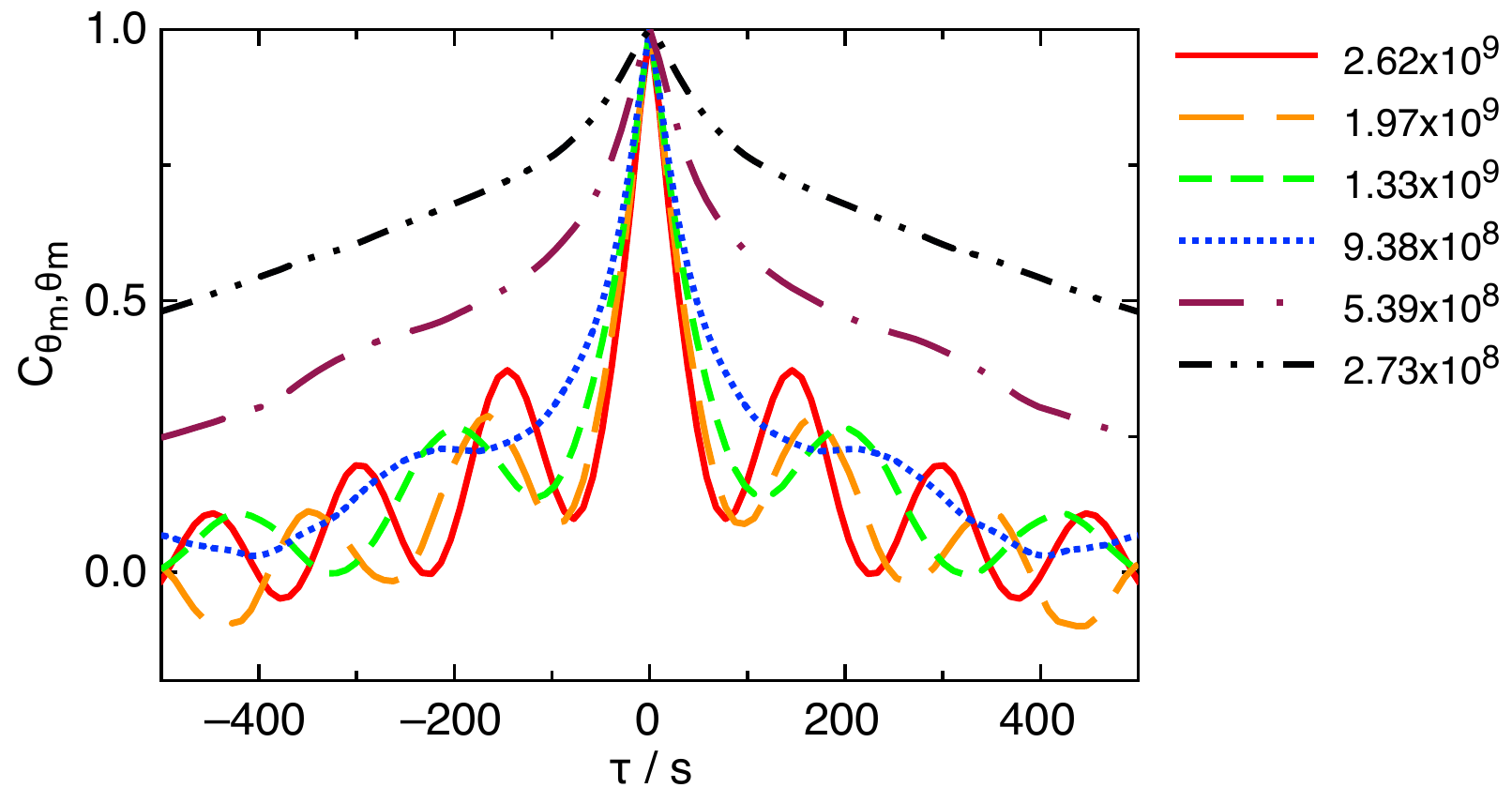} 
\caption{ The autocorrelation $C_{\theta_m,\theta_m}$ for different Ra given in the legend.  We  did not observe any oscillation for $Ra \le 2.73\times10^8$. 
}
\label{corr_example_Ra}
\end{figure}

 To test whether the model can  make quantitative  predictions of the  oscillation period and whether the system is overdamped or underdamped, we compare  predictions and measurements at different $Ra$.   
 
 We show  examples of measured auto-correlations $C_{\theta_m,\theta_m}$ at different Ra in Fig.~\ref{corr_example_Ra}. With decreasing Ra, the oscillation amplitude in the correlation function decreases. For $Ra = 2.73\times10^8$ or lower,  we could not clearly resolve any oscillation. 

To quantitatively calculate the oscillation period $T_{osc}$  from cross-correlation functions, we used the fit function 
\begin{equation}
C_{x,y}(\tau) = b_1\cos\left(\frac{2\pi}{T_{osc}}\tau - \phi_{x,y}\right) + b_2e^{-|\frac{\tau - b_3}{b_4}|^{b_5}} + b_6 \ .
\label{fitting_corr_period}
\end{equation}
The fits of Eq.~\ref{fitting_corr_period} are shown along with the cross-correlations in Figs.~\ref{fig:corrtheta}, \ref{fig:corralphatheta}, and \ref{corr_example_Ra}.  The  parameters $b_2$ through $b_6$  are for  fitting a decaying background which is not analyzed here.   The input error for the fit is  a constant adjusted to get a reduced $\chi^2=1$.    Values of $\phi_{x,y}$ are generally consistent with reported phase shifts in Sec.~\ref{sec:correlations}.  We report the average $T_{osc}$ from the 5 correlation functions reported in Sec.~\ref{sec:correlations} for each Ra.  Error bars represent  the uncertainties on the fits.

\begin{figure}
\includegraphics[width=.475\textwidth]{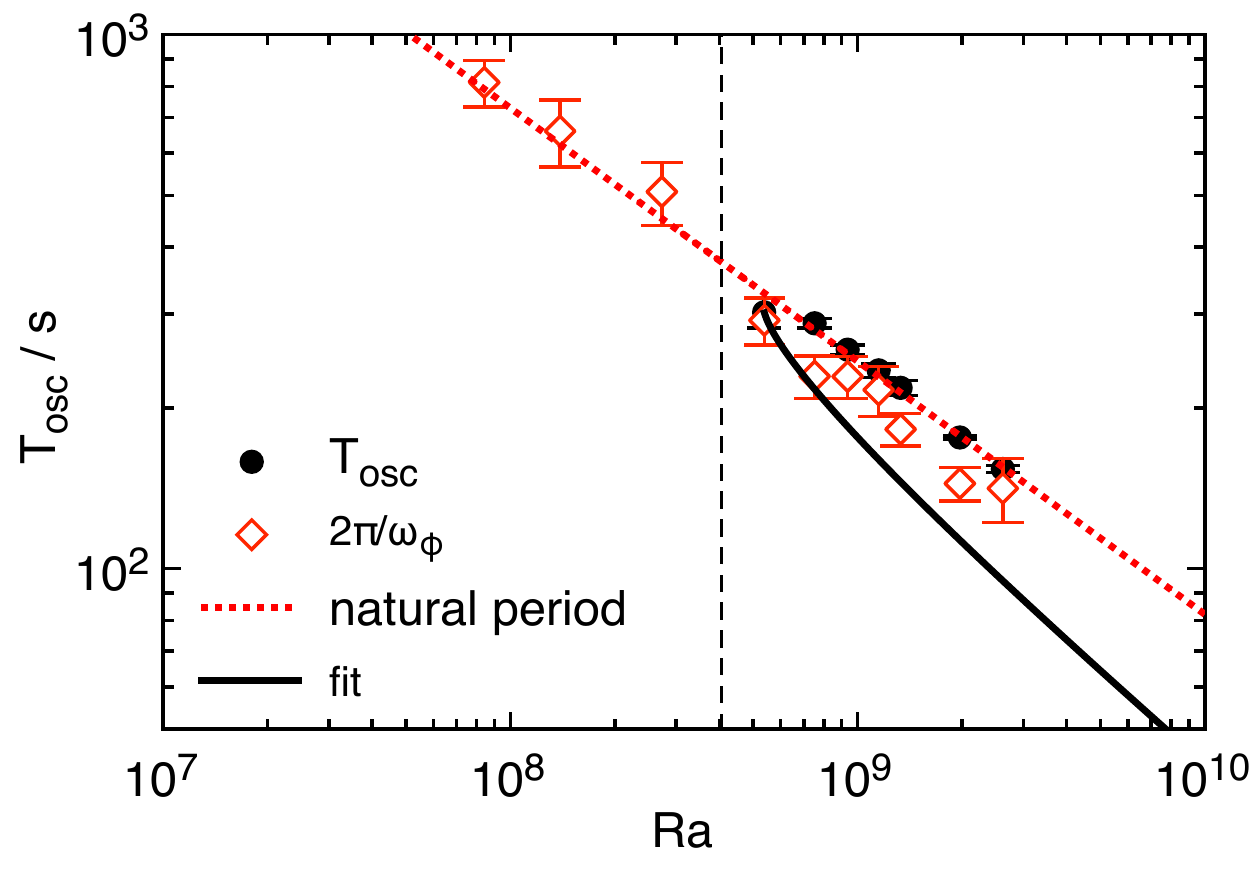} 
\caption{Solid circles: The measured oscillation period $T_{osc}$ as a function of Ra.  Vertical dashed line: Ra  below which oscillations disappeared.  
Solid line: model  prediction using the measured values of $\omega_r$, $\mathcal{T}$, and adjusting $\tau_{\dot\theta}$ by a factor of 2.3 larger than the measured value to fit data. 
Dotted line: the natural period of the potential. Open diamonds: measured turnover time.    The oscillation period is consistent with both the natural period of the potential, and the turnover time.
}
\label{time_scales_Ra_1tilt}
\end{figure}
 

The prediction for the oscillation period $T_{osc}$ is calculated numerically as the frequency of the maximum of the power spectrum  in Eq.~\ref{eqn:power_spec_theory} for $n=1$, since the $n=1$ mode is predicted to be the dominant mode and the oscillation structure observed from the correlation functions in Sec.~\ref{sec:correlations} is consistent with the $n=1$ mode.  We used fits from Sec.~\ref{sec:potential} for the measured Ra-dependence of $\omega_r$, from Appendix 1 for  $\mathcal{T}$ and $\tau_{\dot\theta}$, and adjusted  $\tau_{\dot\theta}$ by a constant factor to best fit the data in Fig.~\ref{time_scales_Ra_1tilt}.  This fit is shown as the solid line in Fig.~\ref{time_scales_Ra_1tilt}.  The average magnitude of the difference between the  fit and data is 28\% of the  measured value, using a fit value of $\tau_{\dot\theta}$ larger than the measured value by a factor of 2.3.  The major constraint in the fit was to obtain the critical $Ra_c$ where oscillations disappear, so the agreement with the measured $T_{osc}$ is an indication that the the natural frequency $\omega_r$ obtained from $p(\theta_0)$ also describes the oscillations, which confirms the self-consistency of the stochastic ODE model.   This same adjustment of $\tau_{\dot\theta}$ also allowed capturing the peak in the power spectrum of $\theta_0$ (red solid curve in Fig.~\ref{power_spec_theta_mid}).  While this suggests  that the model is reasonably consistent with the data within its large error,  this  error is still large enough to span the overdamped-underdamped transition such that the  model cannot correctly predict the observation that the system is the underdamped state rather than the overdamped state. 

A significant consequence of advection in the model is that it eliminates the usual diverging trend of the resonant period near the overdamped-underdamped transition of a damped harmonic oscillator, producing oscillations closer to the natural frequency until the overdamped-underdamped transition is reached.

\subsection{Possible alternate interpretations of the scaling of oscillation period}

We note that in the range of Ra where we observe oscillations, the oscillation period is within 2\% and within error of the natural frequency $2\pi/\omega_r$ (dotted line in Fig.~\ref{time_scales_Ra_1tilt}), corresponding to the limiting solution where $\omega_r$ is dominant in Eq.~\ref{eqn:power_spec_theory}.  The oscillation period is also close the turnover period (open diamonds in Fig.~\ref{time_scales_Ra_1tilt}), larger by an average of $16\%$ in the range of Ra where the oscillation is found, about equal to the 12\% systematic uncertainty on $\mathcal{T}$.  Such a close agreement is expected for the $n=1$ advected mode in the limit where advection is the dominant factor in determining the oscillation frequency.  Either limiting solution may be appropriate for the model depending on parameter values.

The fact that the resonant period $T_{osc}$ is close to $\mathcal{T}$  is also consistent with an alternate model where the  oscillation is driven more directly by the turnover of the LSC -- perhaps by a periodic driving force \cite{Vi95}, rather than  being driven by white noise with frequency determined by the curvature of the potential and advection.  Such a periodic driving force driven at the turnover frequency could lead to the good agreement of the resonant period with the turnover time in the underdamped regime, and still have a transition to overdamping, as observed in Fig.~\ref{time_scales_Ra_1tilt}. It could also lead to the sharper peak in the power spectrum near the turnover period observed in Fig.~\ref{power_spec_theta_mid}.  Since the measured period is close to the turnover time and the prediction of Eq.~\ref{eqn:power_spec_theory}, we cannot distinguish if one interpretation is more correct than the other, or if a combination of both mechanisms exist in this geometry.  Nonetheless, we emphasize a useful feature of the Brown-Ahlers model is that it can correctly  predict different dynamics in different geometries; specifically that the predicted and observed $n=2$ oscillation mode in a circular cross-section cell has twice the frequency of the LSC turnover \cite{BA09}, while the predicted and observed $n=1$ mode in a cubic cell has the same frequency as the LSC turnover.


\section{Summary and conclusions}


  The model of Brown \& Ahlers \cite{BA08b} was able to  correctly predict the 4-well shape of the geometry-dependent potential $V_g(\theta_0)$ for a cubic cell (Fig.~\ref{fig:ptheta_4well}).  This includes the quadratic shape of the potential minima $V_g(\theta_0,\alpha)$ near the corners with equal curvatures in both $\theta_0$ and $\alpha$, indicating they are independent (Figs.~\ref{fig:pdf_potential}, \ref{fig:potential_alpha}, \ref{fig:pdf_joint}, \ref{fig:potential_slices}, and Table \ref{tab:potential_curvatures}).  The natural frequency $\omega_r$ was found to scale with the inverse of the turnover time $\mathcal{T}$ at higher Ra as predicted, although the prediction was larger than measurements by factor of 2.9 (Fig.~\ref{curvature_Ra_mid}), which is a typical error of this model \cite{BA08a}.  The magnitudes of the curvature of the potential near its peak, as well as the potential barrier height, which are relevant to barrier crossing events, were both predicted accurately within a factor of 2 (Fig.~\ref{fig:ptheta_4well_fit}).  Such errors are typical of this modeling approach \cite{BA08a, BA08b, BA09}, as it makes significant approximations about the shape of the LSC, scale separation between the LSC and small-scale turbulent fluctuations, and the distribution of turbulent fluctuations.
  
Oscillation modes centered around corners of the cubic cell were observed above a critical Ra $=4\times10^8$, which appears in the model as a  crossing of an underdamped-overdamped transition.   Above this critical Ra,  the oscillation period is consistent with the natural period of the potential and the  turnover time (Fig.~\ref{time_scales_Ra_1tilt}).   The value of the critical Ra, as well as frequency and background of the power spectrum of $\theta_0$ are consistent with the measured one if the model parameters $\omega_r$, $\mathcal{T}$, $\tau_{\dot\theta}$, and $D_{\dot\theta}$ were adjustable up to a factor of 2.3 away from independently measured values, or a factor of 3 from predicted values (Figs.~\ref{power_spec_theta_mid},\ref{time_scales_Ra_1tilt}), again typical errors of this model \cite{BA08a}.    However, this uncertainty in the model parameters turns out to be too large to correctly predict the whether the system is in its underdamped or overdamped state, as the dynamics are sensitive to the model parameters near the overdamped-underdamped transition.   

The structure of these oscillations is mainly the predicted $n=1$ advected oscillation mode, consisting of out-of-phase oscillations in the top and bottom rows of thermistors of the slosh angle $\alpha$ (i.e.~a rocking mode) and in-phase oscillations in all 3 rows of the modified LSC orientation $\hat \theta_0$ (Figs.~\ref{fig:corrtheta}, \ref{fig:corralphatheta}).  This mode is distinct from the $n=2$ twisting and sloshing mode predicted and observed in a circular cylindrical cell \cite{BA09}. The $n=1$ advected oscillation mode exists in cubic cells because of a restoring force due to the variation of the diameter $D(\theta_0)$ across the cube around a corner \cite{BA08b}.  A weaker $n=2$ advected oscillation mode is also observed in the cubic cell, as predicted by the model.  Weak oscillations in $\alpha_m$ in-phase with the oscillation in $\theta_m$ correspond to breaking of the Boussinesq symmetry  where the cold  side of the LSC oscillates with a larger amplitude than the  hot side (Fig.~\ref{fig:corralphatheta}).  
There are hints of a more complex oscillation structure involving a higher order Fourier moments $A_n$ of the temperature profile (Fig.~\ref{fig:power_spec_momentsA}) which will be addressed in a follow-up work.

 
We consider it remarkable that a low-dimensional model of diffusive motion in a potential can predict many features of a high-dimensional turbulent flow. Non-trivial predictions that were confirmed include the shape of the potential as a function of cell geometry, and the successful prediction of a how the oscillation structure changes with the cell geometry.  The ability of this model to predict how features of the LSC dynamics change from a circular- to square-cross-section containers suggests that such a model could be applied more generally to predict dynamics for different cross section shapes supporting a single convection roll.  Since the modeling approach assumes a robust large-scale structure with a scale-separation from small scale turbulent fluctuations, it is not limited to Rayleigh-B{\'e}nard convection, there is great promise for general models of the dynamics of large-scale coherent structures in turbulent flows.  Extending predictions to multiple convection roll systems and more complex convection roll shapes remain open problems.  While the model has been able to predict features within about a factor of 2,  the sensitivity of features near the  overdamped-underdamped transition to model parameters leads to an inability to predict correctly whether the system will oscillate or not (even though the observations are consistent with the model within the generous errors).  For this modeling approach to be useful in predicting such sensitive features will require more refinement of the  quantitative accuracy of the predictions by allowing more complex functions in the model, as was done by \cite{AAG11}.

\section{Acknowledgments}

We thank the University of California, Santa Barbara machine shop and K. Faysal for helping with construction of the experimental apparatus. This work was supported by Grant CBET-1255541 of the U.S. National Science Foundation.

 \section*{Appendix 1: Measurements of model parameter values}
\label{sec:parameters}

In this section,  we report independent measurements of the parameters that are input into Eqs.~\ref{eqn:delta_model}, \ref{eqn:theta_model}, and \ref{eqn:potential} \, using mainly the methods of Ref.~\cite{BA08a}, so that these parameters can be used  to test model predictions in Secs.~\ref{sec:potential}, \ref{sec:powerspec}, and \ref{sec:period}.  Measurements reported in this section were done with a timestep of 2.16 s to capture faster fluctuations of the LSC.

\subsection{Turnover time}
\label{sec:turnovertime}

\begin{figure}
\includegraphics[width=.475\textwidth]{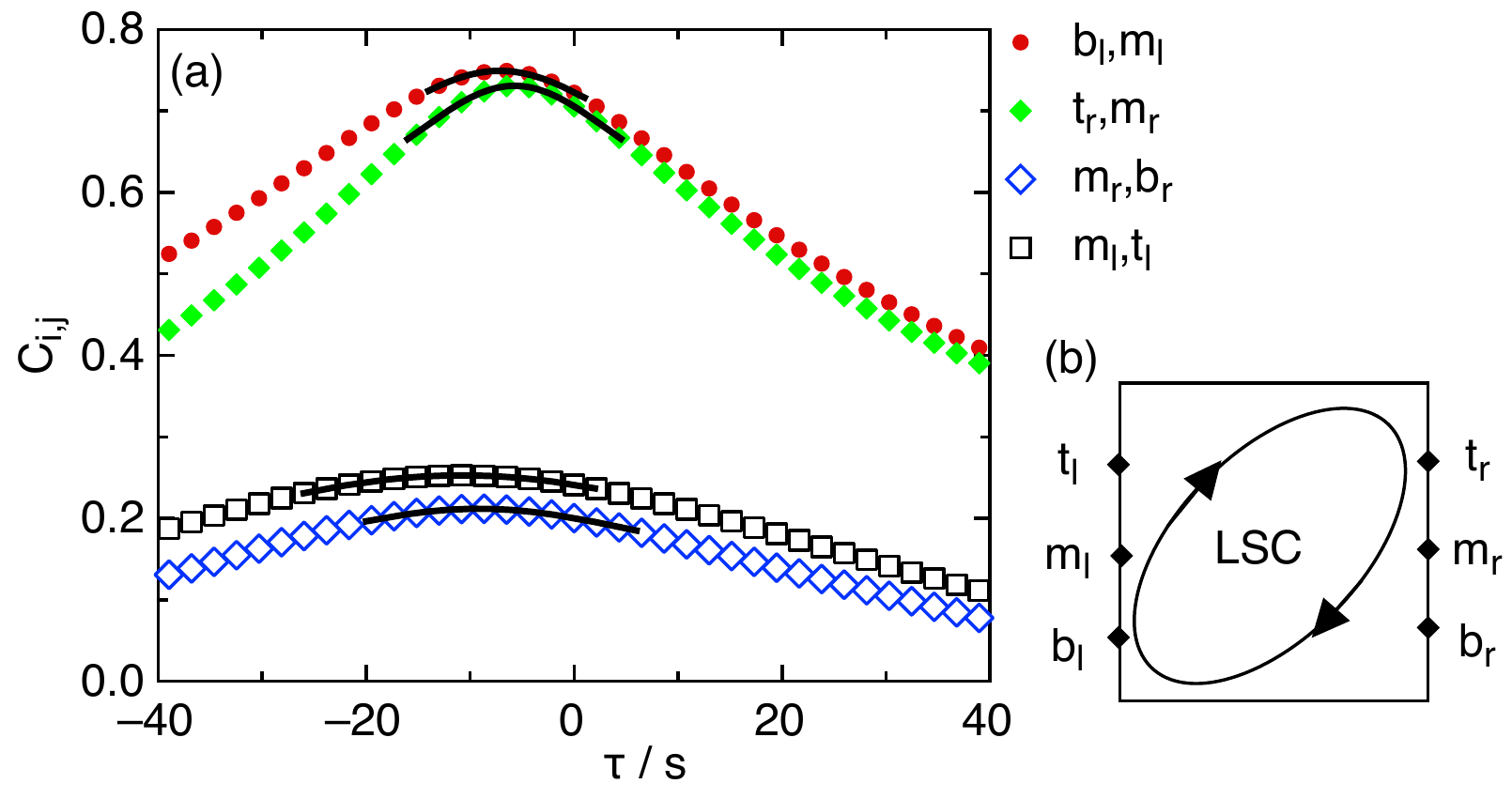} 
\caption{(a) The correlation function between thermistor temperatures vertically separated by $H/4$ along the path of the LSC. The  thermistors correlated in each case are indicated in the legend. Lines:  Gaussian fits to the peak of each data set.  (b)  Illustration of the  path of the LSC and thermistor labels.
 }
\label{turnover_time_calculation}
\end{figure} 

 In Ref.~\cite{BFA07}, the LSC turnover time was calculated from the peak of the cross-correlation between two thermistors mounted on the opposite sides of the side wall. However, in our case, the same calculation yields suspiciously low turnover times 
 and a correlation peak with the opposite sign as in a circular cylinder \cite{BFA07}. This suggests this correlation time may be affected by the different oscillation modes of the LSC structure which change with the cell geometry. Therefore, we took a different approach in this paper to obtain the turnover time using information from more thermistors. 
 
 \begin{figure}
\includegraphics[width=.475\textwidth]{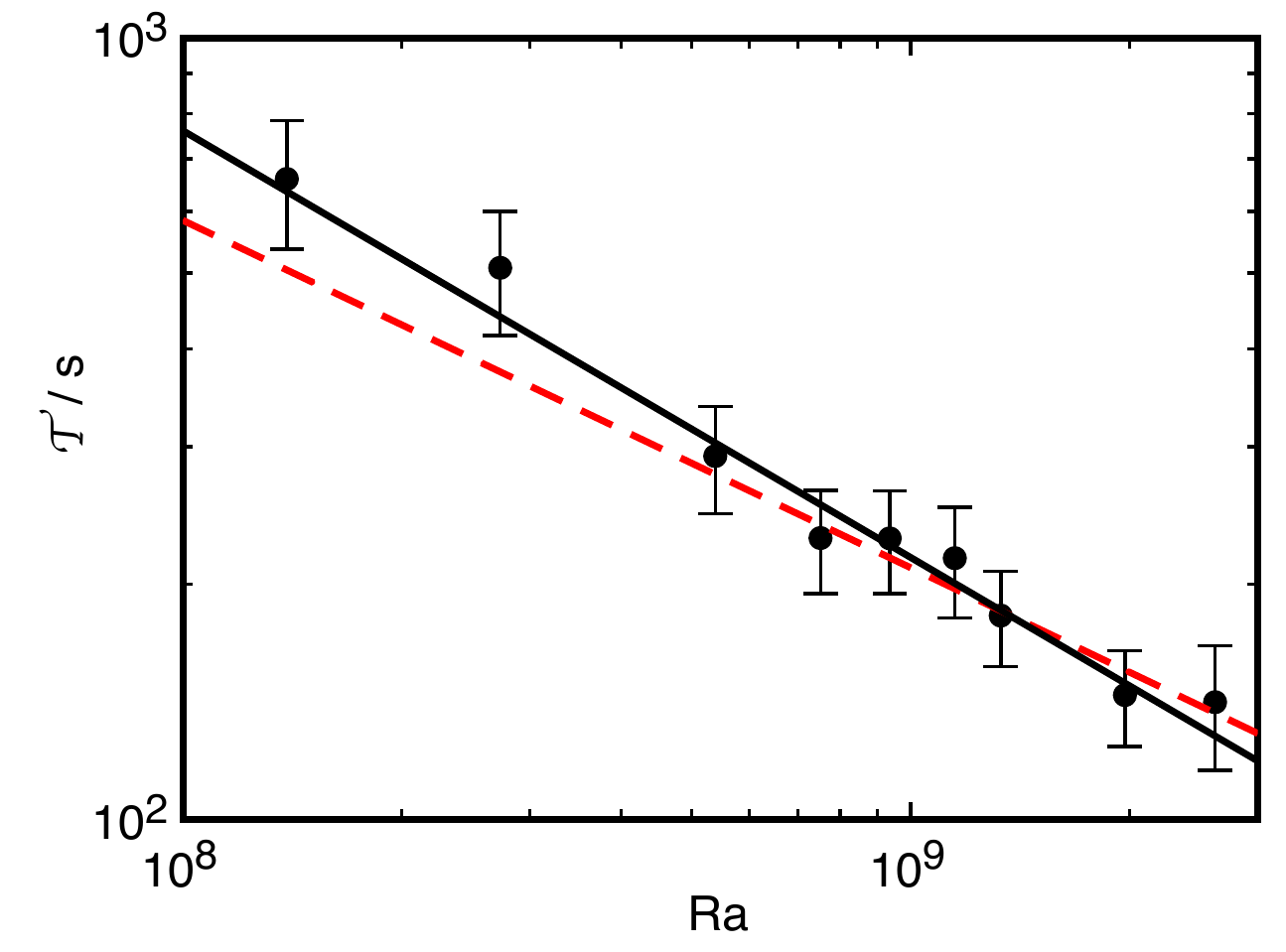} 
\caption{The turnover time $\mathcal{T}$ as a function of Ra. Solid line:  a power law fit. Dashed line: prediction of the Grossmann-Lohse model, which is consistent with the data.
}
\label{turnoverTimeVSRa}
\end{figure} 
 
We measured the correlation times of thermistor pairs vertically separated by $H/4$ and along the path of the LSC. As is illustrated in Fig.~\ref{turnover_time_calculation}b, there are 4 such pairs in the two columns of thermistors most closely aligned with the mean path of the LSC.   The correlation between each pair is shown in Fig.~\ref{turnover_time_calculation}a.  We fitted a Gaussian function to data near each peak, as shown in Fig.~\ref{turnover_time_calculation}a.  We took the average of those four peak locations as the time  the LSC needed to travel the distance $H/4$, with a standard deviation of the mean of 10\%.  While  we do not know the specific path length of the LSC, it can be reasonably bounded between an oval with pathlength $\pi(1+\sqrt{2})/2H$ and a rectangle with pathlength $2(1+\sqrt{2})H$, and  so we take the mean of those two  paths of $\lambda=4.3H$ as our best estimate of the pathlength of the LSC, with a 12\% uncertainty spanning to the two extremes. The turnover time $\mathcal{T}$ was then calculated as the correlation time between the two vertically separated thermistors scaled up by the pathlength $\lambda$ divided by $H/4$.    The resulting turnover time $\mathcal{T}$ for different Ra is shown in Fig.~\ref{turnoverTimeVSRa}.  The error on $\mathcal{T}$ was obtained as the standard deviation of the mean of the fit propagated in quadrature with the error from the uncertainty on the pathlength.   A power law fit to the data yields $\mathcal{T}  = 1.8\times10^7Ra^{-0.55\pm0.05}$.
 
 The Grossmann-Lohse model gives the scaling between Re and Ra  as $Re=0.31 Ra^{4/9}Pr^{-2/3}$  when fit to similar data in the same Ra- and Pr-range in circular cylindrical containers \cite{BFA07}.  For the cubic cell,  we calculate the Reynolds number as $Re = \lambda H/\mathcal{T} \nu$.  The resulting prediction for the turnover time $\mathcal{T} = 4.3H^2/(\nu Re) = 2.1\times 10^6 Ra^{-4/9}$ is shown as the dashed line in Fig.~\ref{turnoverTimeVSRa}.  The prediction is consistent with the error of the data indicating that the Re-Ra relations for cubic and circular cylindrical cells are consistent with each other within the 16\% error of the data.

  \subsection{Stable fixed point temperature amplitude $\delta_0$}
 \label{sec:delta0}

  \begin{figure}
\includegraphics[width=.475\textwidth]{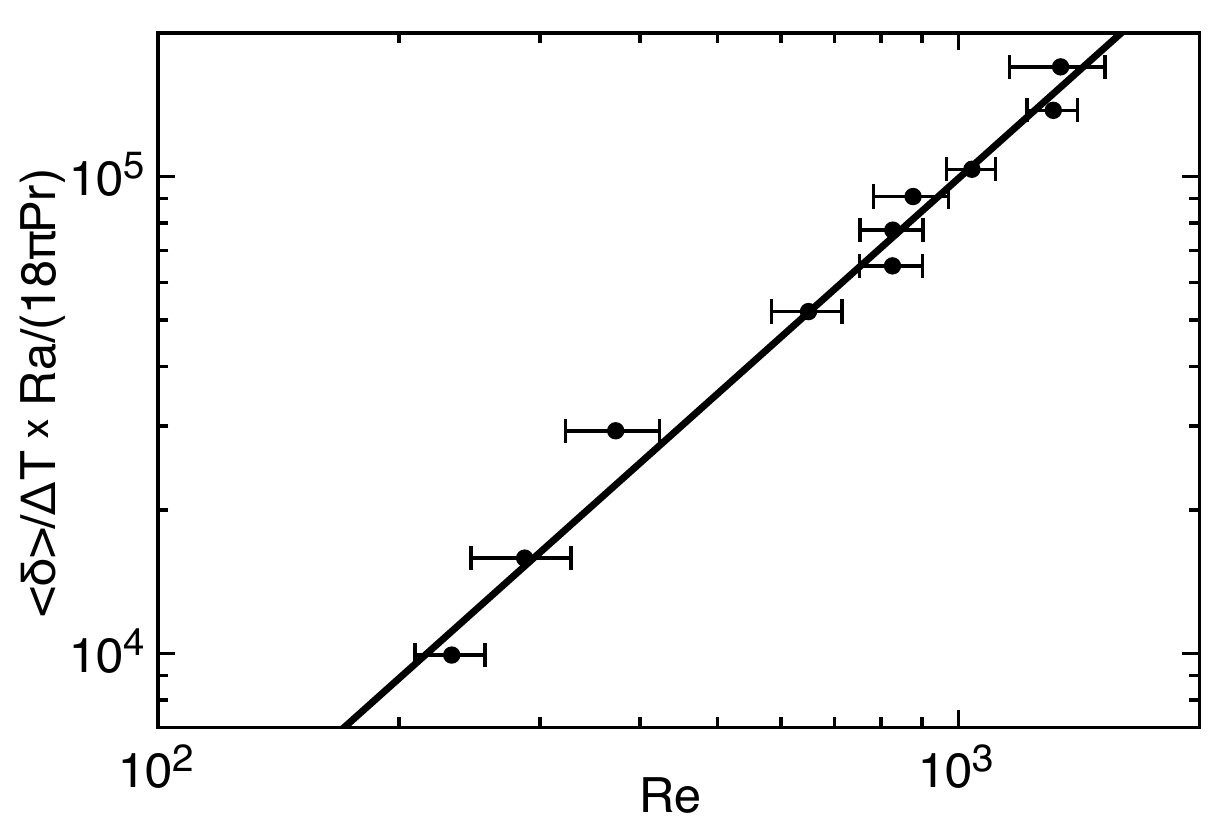} 
\caption{ The  mean temperature amplitude $\langle \delta \rangle$ scaled as in Eq.~\ref{eqn:delta_re} as a function of Re. Solid line: power law fit to $c Re^{3/2}$.  The scaling is consistent with the prediction of Eq.~\ref{eqn:delta_re}, and both the scaling and value of fit coefficient $c$ are consistent with those found in circular cylindrical cells.
}
\label{deltaVSReynolds}
\end{figure}

To present measurements of the stable fixed point temperature amplitude $\delta_0$ in a general form, we make use of the model that was used to derive Eq.~\ref{eqn:delta_model} \cite{BA08a}, which predicted a relationship between $\delta_0$ and the Reynolds number to be 
 
 \begin{equation}
 \frac{\delta_0 Ra}{18\pi \Delta T Pr} = cRe^{3/2} \ , 
 \label{eqn:delta_re}
 \end{equation}
 
\noindent  where $c$ is a  dimensionless fit coefficient of order 1. We approximate $\delta_0 \approx \langle \delta \rangle$, since $p(\delta)$  is nearly symmetric around its stable fixed point $\delta_0$  \cite{BA08a}.   We plot $\langle\delta\rangle$  scaled according to the left side of Eq.~\ref{eqn:delta_re} as a function of Re in Fig.~\ref{deltaVSReynolds}.   We calculate Re $= UH/\nu$ where $U$ is the ratio of the distance $H/4$ between vertically separated thermistors and  the correlation time calculated  in Appendix 1.A, with an error propagated from the standard deviation of the mean of the correlation time.  We fit the data in Fig.~\ref{deltaVSReynolds} to Eq.~\ref{eqn:delta_re} with $c$ as the only free parameter, which yields $c = 3.13\pm0.14$ with a reduced $\chi^2=0.5$.  This prefactor is consistent within a couple of standard deviations of  $c=2.8\pm0.1$ obtained in a circular cylinder \cite{BA08a},  indicating that the same relationship holds between $\delta_0$ and Re in both geometries.

 \subsection{Diffusivity and damping time of the temperature amplitude $\delta$} 
 
 \begin{figure}
\includegraphics[width=.475\textwidth]{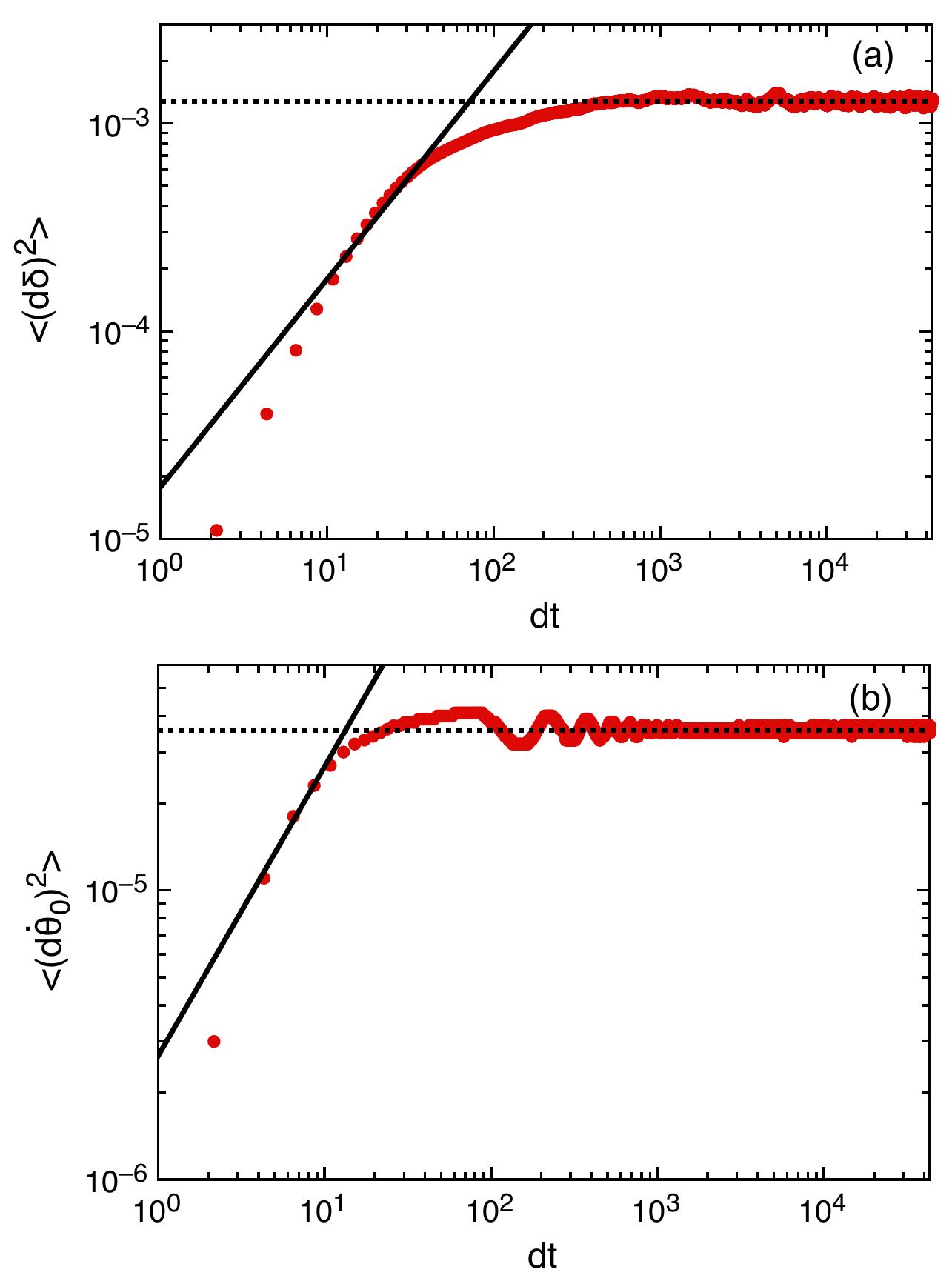} 
\caption{ The mean square change (a)  $\langle (d\delta)^2 \rangle$ and (b) $\langle (d\dot\theta_0)^2 \rangle$ as a function of the time interval $dt$. Solid lines: linear fits to the data for small $dt$ yield the diffusivities $D_\delta$ and $D_{\dot\theta}$, respectively. Dotted line: constant fits  to data for large $dt$ yield $\tau_{\delta}$ and $\tau_{\dot\theta}$, respectively. 
}
\label{diffusivities_fit_example}
\end{figure}

 The diffusivity $D_\delta$ and damping time $\tau_{\delta}$ were obtained by measuring the mean-square change of the temperature amplitude $\langle(d\delta)^2 \rangle$ over a time period $dt$.  The  diffusive behavior of  the noise in Eq.~\ref{eqn:delta_model} leads to the  prediction $\langle(d\delta)^2 \rangle=D_\delta dt$ for small $dt$ \cite{BA08a}. We fit  $\langle(d\delta)^2 \rangle=D_\delta dt$ to data within the range of 0.25$\tau_\delta \leq dt \leq 0.6\tau_\delta$ to obtain $D_\delta$ \cite{BA08a}. An example is shown in Fig.~\ref{diffusivities_fit_example}a.  The data do not follow the diffusive trend very well, indicating that Eq.~\ref{eqn:delta_model} does not capture the short time fluctuations of $\delta$ very well.  Nonetheless, Eq.~\ref{eqn:delta_model}  has been found to capture the qualitative dynamics of $\delta$ in circular cylindrical cells \cite{BA08a}, as the dynamics of stochastic ordinary differential equations are often not very sensitive to the details of the fluctuation distributions, and we still use the fit to obtain a value for $D_{\delta}$.  Different fit ranges could result in different values of $D_{\delta}$, and in the worst case, our fit range overestimates $D_{\delta}$ by as much as a factor of 2.  Fits are of  similar quality at different Ra.

Equation ~\ref{eqn:delta_model} leads to the  prediction $\langle(d\delta)^2 \rangle=2D_\delta\tau_\delta$  in the limit of large time $dt$, assuming small variations in $\delta$ such that the net forcing in Eq.~\ref{eqn:delta_model} is approximately linear in $\delta$ near the stable fixed point $\delta_0$ \cite{BA08a}.  The damping time $\tau_\delta$ was obtained from fitting the plateau value of $\langle(d\delta)^2 \rangle = 2D_\delta \tau_\delta$  in the limit of large time $dt$ after  the value of $D_{\delta}$  was determined \cite{BA08a}.  An example is shown in Fig.~\ref{diffusivities_fit_example}a.   While $\tau_{\delta}$ could be underestimated by as much as a factor of 2 due to the poor fit of the diffusive scaling at short times, the plateau value  $2D_{\delta}\tau_{\delta}$ and thus the variance of $\delta$ are still well-defined by this fit.

\begin{figure}
\includegraphics[width=.475\textwidth]{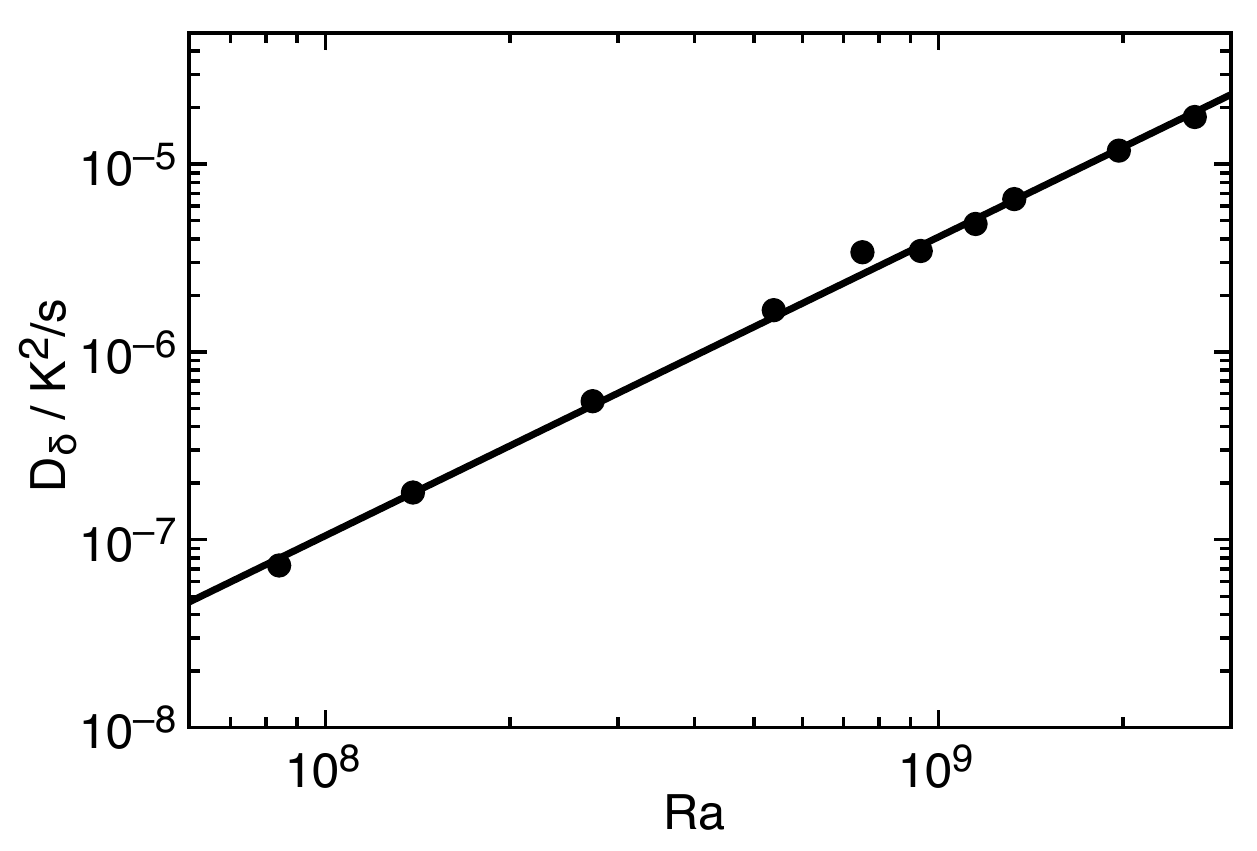} 
\caption{ Measurement of the diffusivity $D_{\delta}$  as a function of Ra. Solid line: power law fit, which yields $D_\delta \propto Ra^{1.59\pm0.03}$ K$^2$/s,  different  from the  scaling found in a circular cylindrical cell. }
\label{Ddelta}
\end{figure}

The Ra-dependence of $D_\delta$ is shown in Fig.~\ref{Ddelta}.  A power law fit yields $D_\delta = 2.0\times10^{-20}Ra^{1.59\pm0.03}$ K$^2$/s with a standard deviation between the data and fit of 9.3\%. 
 This scaling differs from that found in a circular cylindrical cell $D_{\delta} \propto Ra^{1.96}$ at higher Ra \cite{BA08a}. 
 
\begin{figure}
\includegraphics[width=.475\textwidth]{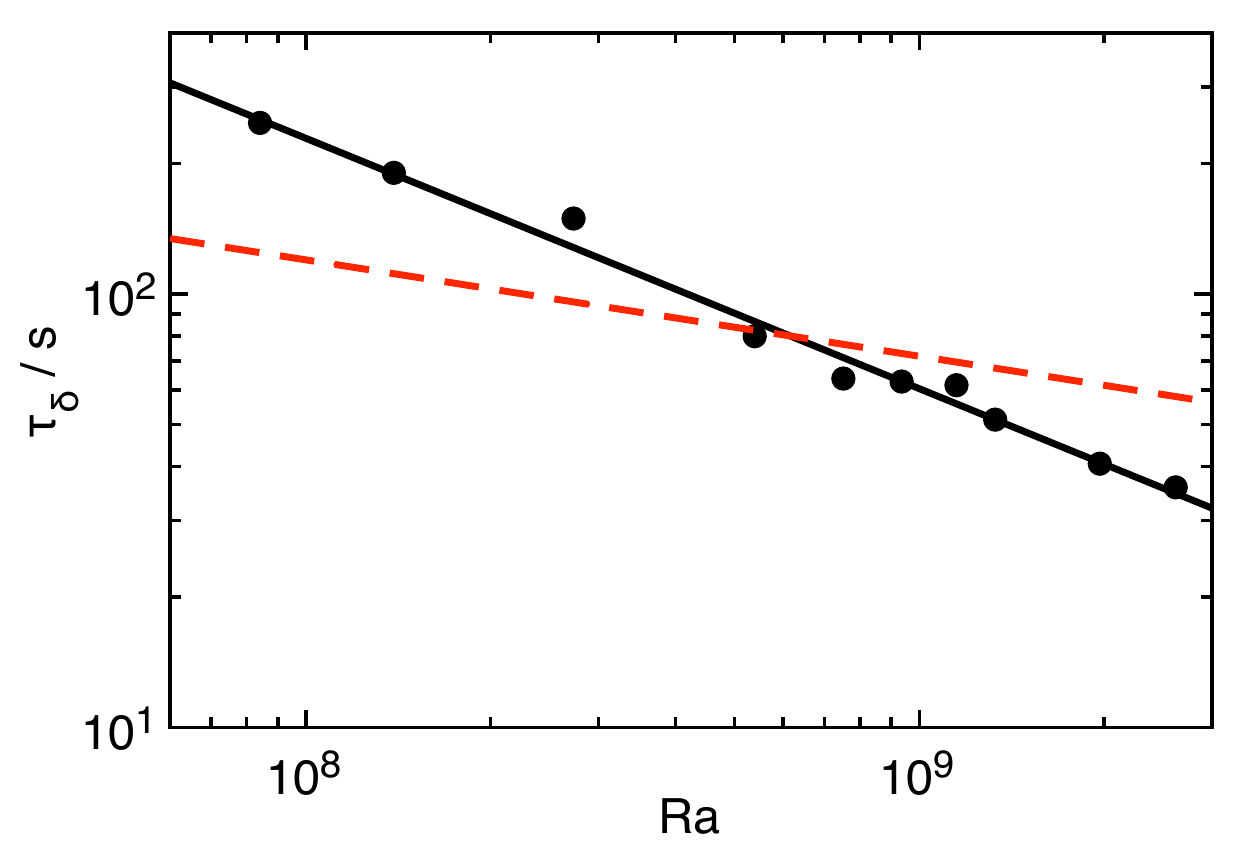} 
\caption{Measurement of the damping time $\tau_{\delta}$ as a function of Ra. Solid line: power law fit, which yields $\tau_\delta \propto Ra^{-0.58\pm0.02}$ s, different  from the  scaling found in a circular cylindrical cell. Dashed line:  prediction of Ref.~\cite{BA08a}}
\label{taudelta}
\end{figure}

The Ra-dependence of $\tau_\delta$ is shown in Fig.~\ref{taudelta}. A power law fit yields $\tau_\delta = 9.5\times10^6Ra^{-0.58\pm0.02}$ s with  a standard deviation between the data and fit of 7.1\%.  Reference  \cite{BA08a} predicted that $\tau_\delta = H^2/18\nu Re^{1/2}$ \cite{BA08a}.  Using the GL model for Re, the prediction is $\tau_\delta =  7.2\times10^{3} Ra^{-2/9}$ s in our range of Ra.  This prediction is shown in Fig.\ref{taudelta}.  While the prediction has a different scaling exponent than the data, the magnitude of the prediction is within a factor of 2 over the measured range.   The scaling also differs from that found in a circular cylindrical cell $\tau_\delta \propto Ra^{-0.43}$,  in which case the model also only predicted the correct order-of-magnitude \cite{BA08a}. As in a circular cylindrical cell, we also find the scaling of the measured  $\tau_{\delta}\propto$ Ra$^{-0.58\pm0.02}$ \cite{BA08a} is consistent with the scaling of the turnover time $\mathcal{T}\propto$ Ra$^{-0.55\pm0.05}$.

\subsection{Diffusivity  and damping time of the angular rotation rate $\dot\theta_0$}

\begin{figure}
\includegraphics[width=.475\textwidth]{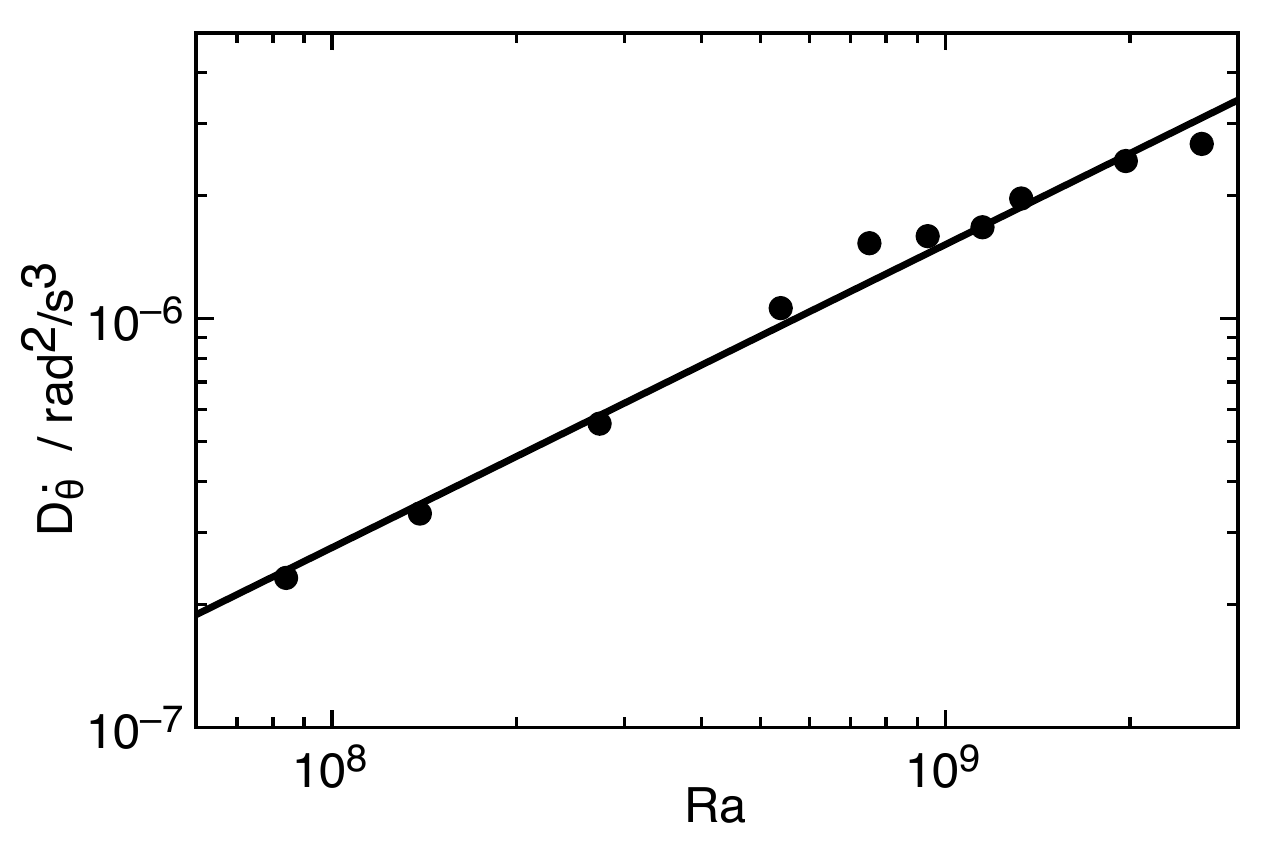} 
\caption{Measurement of the diffusivity $D_{\dot{\theta}}$  as a function of Ra. Solid line: power law fit, which yields $D_{\dot{\theta}} \propto Ra^{0.74\pm0.03}$ rad$^2$/s$^3$, different  from the  scaling found in a circular cylindrical cell.
}
\label{fig:Dthetadot}
\end{figure}

The diffusivity $D_{\dot{\theta}}$ and damping time scale $\tau_{\dot{\theta}}$ were calculated from the mean square change of $\dot\theta_0$ similar to $\delta$, where  $\langle (d\dot\theta_0)^2\rangle = D_{\dot\theta}dt$ was fit for small $dt$ and $\langle (d\dot\theta_0)^2\rangle = D_{\dot\theta}\tau_{\dot\theta}$ was fit for large $dt$. An example is shown in Fig.~\ref{diffusivities_fit_example}b.  This assumes small variations in $\dot\theta_0$ such that the forcing on $\dot\theta_0$ in Eq.~\ref{eqn:theta_model}  is approximately linear in $\dot\theta_0$ -- more specifically, the incremental change in rotation rate from the potential term $\nabla V_g(\theta_0) dt$ is small compared to the contribution from diffusion  $\sqrt{D_{\dot\theta} dt}$ and damping $\dot\theta_0\delta dt/\tau_{\dot\theta}\delta_0$ for small $dt$ \cite{BA08a}.  The validity of this approximation is confirmed by the parameter values reported in this section.

The Ra-dependence of $D_{\dot{\theta}}$ is shown in Fig.~\ref{fig:Dthetadot}, the power law fit yields $D_{\dot{\theta}} = 3.2\times10^{-13} Ra^{0.74\pm0.03}$ rad$^2$/s$^3$ with a standard deviation between the data and fit of  9.6\%.  The scaling is consistent with that found in a circular cylinder  $D_{\dot{\theta}} \propto Ra^{0.76}$ rad$^2$/s$^3$. 

\begin{figure}
\includegraphics[width=.475\textwidth]{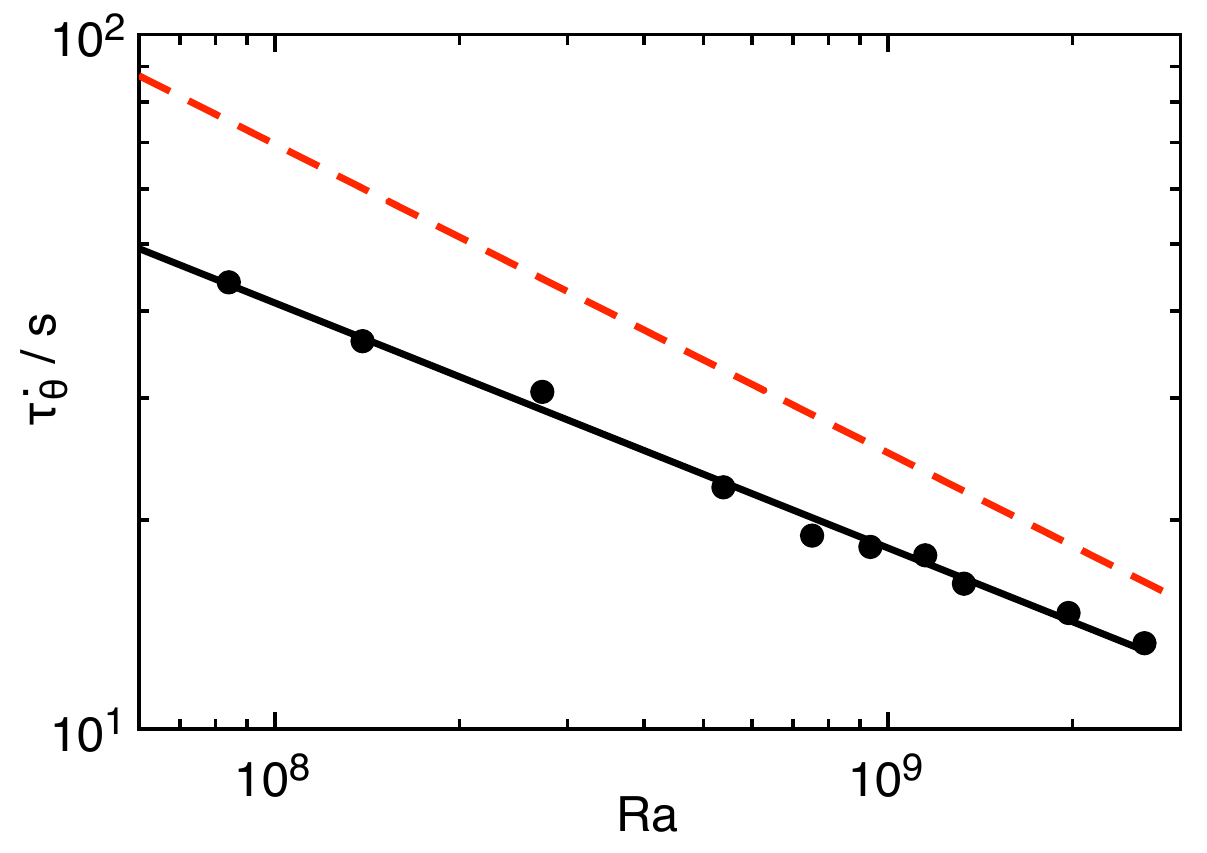} 
\caption{Measurement of the damping time $\tau_{\dot{\theta}}$ as  a function of Ra. Solid line: power law fit, which yields $\tau_{\dot{\theta}} \propto Ra^{-0.35\pm0.01}$ s, different  from the  scaling found in a circular cylindrical cell. Dashed line: prediction of Ref.~\cite{BA08a}.}
\label{tauthetadot}
\end{figure}

The Ra-dependence of $\tau_{\dot{\theta}}$ is shown in Fig.~\ref{tauthetadot}. A power law fit yields $\tau_{\dot{\theta}} = 2.7\times10^{4}Ra^{-0.35\pm0.01}$ s with a standard deviation between the data and fit of 3.2\%.  Brown \& Ahlers  predicted that $\tau_{\dot{\theta}} = H^2/2\nu Re$ \cite{BA08a}.  Using the GL model for Re,  the prediction is $\tau_{\dot{\theta}} = 2.5\times10^5 Ra^{-4/9}$ s, which has a different power law exponent, but is within a factor of 2 of the data in the range of Ra tested.  The scaling also differs from that found in a circular cylindrical cell $\tau_{\dot\theta}\propto Ra^{-0.20}$,  in which case the model also only predicted the correct order-of-magnitude \cite{BA08a}.

\subsection{Applicability of model parameters to other experiments or simulations}

While the parameter values presented here in Appendix 1  are useful for analyzing this experiment, care should be taken when comparing to other experiments or simulations under different conditions, as not all appropriate scalings are known.  

Tilt of the cell relative to gravity could change parameter values.  There is a significant effect of tilt on $D_{\dot\theta}$.  When increasing from $\beta=0$ to $\beta=2^{\circ}$,  $D_{\dot\theta}$ was found to decrease by 18\% per degree, so our reported data at $\beta=1^{\circ}$ have a smaller $D_{\dot\theta}$ than at $\beta=0$ by 18\%. The values of $\tau_{\dot\theta}$, $\tau_{\delta}$, and $D_{\delta}$ were within measurement errors at different tilt angles up to $\beta=2^{\circ}$.

While the spatial variation of plate temperature affected the preferred orientation of the LSC, the variation of plate temperature over time could in principle lead to apparent increases in the diffusivities.  
 Correlations between thermistors in the plates and in the middle row of the sidewall are less than 7\% at the highest $\Delta T=18.35^{\circ}$ C (where the plate temperature fluctuations are strongest) for time delays less than $\tau_{\dot\theta}$ where fluctuations are dominant in the dynamics, suggesting that effects of the plate temperature fluctuations on $D_{\dot\theta}$ are less than 7\%.

  \section*{Appendix 2: calculation of $D(\theta_0,\alpha)$}
  
 To calculate $D(\theta_0,\alpha)$ using the geometry shown in Fig.~\ref{fig:angle_definitions}b, we label the center of the cube as $O$, the location on the wall at the hot side of the LSC as $A$, and the location on the wall at the cold side of the LSC as $B$.  The length of line OA is
 \begin{equation}
\overline{OA} = \frac{\frac{\sqrt{2}}{2}{H}}{|\cos(\theta_0 + \alpha)| + |\sin(\theta_0 + \alpha)|} \ .
\label{eqn:OA}
\end{equation}
Similarly, the length of line OB is
  \begin{equation}
\overline{OB} = \frac{\frac{\sqrt{2}}{2}{H}}{|\cos(\theta_0 - \alpha)| + |\sin(\theta_0 - \alpha)|} \ .
\label{eqn:OB}
\end{equation}
The length of the third side of the triangle of AOB -- which corresponds to $D(\theta_0,\alpha)$ -- is calculated using the law of cosines to be
\begin{equation}
 D(\theta_0,\alpha)^2 = \overline{OA}^2 + \overline{OB}^2 - 2\overline{OA} \ \overline{OB}\cos(\pi - 2\alpha)
 \end{equation}
 Substituting Eqs.~\ref{eqn:OA} and Eq.~\ref{eqn:OB} into this expression for $D(\theta_0,\alpha)$ yields
 \begin{equation}
\begin{multlined}
D(\theta_0,\alpha)^2 = \frac{H^2}{2} \bigg[ (|\cos(\theta_0 + \alpha)| + |\sin(\theta_0 + \alpha)|) ^{-2}  \\ +( |\cos(\theta_0 - \alpha)| + |\sin(\theta_0 - \alpha)|)^{-2} \\  + 2\cos2\alpha (|\cos(\theta_0 + \alpha)| +  |\sin(\theta_0 + \alpha)|)^{-1}  \\ \times\left(|\cos(\theta_0 - \alpha)| + |\sin(\theta_0 - \alpha)|\right)^{-1}\bigg] \ .
\label{eqn:D_theta_alpha}
\end{multlined}
\end{equation}
 
For analytical calculations near corners, this expression simplifies in the small angle limit for both $\theta_0$ and $\alpha$ to
  \begin{equation}
D(\theta_0,\alpha)^2 \approx 2H^2(1 - |\theta_0+\alpha| - |\theta_0-\alpha|) \ .
\end{equation}

\section*{Appendix 3: prediction of oscillation modes and phase shifts $\phi_{x,y}$ for the advected oscillation model}
\label{sec:phaseshift_model}

 In this appendix, we give detailed predictions for the phase shifts $\phi_{x,y}$ in correlation functions. These are predicted from the model  by evaluating the correlation functions of Eq.~\ref{eqn:correlation} in terms of the 
  standing-wave partial solutions $\alpha_{n}(\omega,t)$ and $\theta_{0,n}(\omega,t)$ from Eqs.~\ref{eqn:alpha_travelingwave} and \ref{eqn:theta_travelingwave}.  Since these are  functions of frequency $\omega$, the correlation function is integrated over $d\omega$ to cover all frequencies.  Averaging over time first and using the identity $P_n(\omega) = |a_n(\omega)|^2$ results in simplified expressions which identify specific phase shifts $\phi_{x,y}$ for each cross-correlation.  We address cases for different angle pairs $x,y$ and odd and even $n$ in the following paragraphs.  

For modes with odd-$n$ order and correlations  between different rows of $\theta_0$,  the predicted correlation function is $C_{\theta_i,\theta_j}(\tau) \propto \cos(nk_0 z_i) \cos(nk_0 z_j) \int P_n(\omega) \cos(\omega \tau)d\omega$ where $i$ and $j$ correspond to the top, middle, or bottom rows of thermistors.  As long as the power spectrum $P_n(\omega)$ has a resonance peak (as seen in Fig.~\ref{power_spec_theta_mid}), the integral produces an oscillatory function in $\tau$ with a period the same as the resonant period of $P_n(\omega)$.  Since $\cos(\omega \tau)$ has a peak at $\tau=0$, the integral also has a peak at a time delay $\tau=0$, corresponding to a phase shift $\phi_{\theta_i,\theta_j} = 0$.   When $\cos(nk_0z_i) > 0$,  the overall correlation is positive.  This occurs for example at the middle row where $z_i=0$, and for $n=1$ at the top and bottom rows for the value of $k_0=\pi/2H$ in an aspect ratio 1 circular cylinder \cite{BA09} or  $k_0=\pi/[(1+\sqrt{2})H]$ in a cube.   Thus, the  predicted structure for the dominant $n=1$ mode is an oscillation where $\theta_0$ is in-phase at all  rows, similar to that found in a tilted circular cylinder \cite{BA08b}.

For modes with odd-$n$ order and correlations  between different rows of $\alpha$, the predicted correlation function is $C_{\alpha_i,\alpha_j}(\tau) \propto \sin(nk_0z_i) \sin(nk_0z_j)  \int P_n(\omega) \cos(\omega \tau)d\omega$. The integral again is an oscillatory function in $\tau$ as long as $P_n(\omega)$ has a resonance peak, and has a peak at a time delay $\tau=0$, corresponding to a  phase shift $\phi_{\alpha_i,\alpha_j} = 0$.  However, the sign of the correlation now varies with the row chosen.  For example, when calculating $C_{\alpha_b,\alpha_t}(\tau)$, then $z_i= - z_j$, so the product $\sin(nk_0z_i) \sin(nk_0z_j)$ is negative for all odd $n$, resulting in a negative correlation.  For correlations with the middle row at $z=0$,  the corresponding sinusoid  in front of the integral is 0, resulting in no oscillation.  These correspond to a predicted oscillation in $\alpha$ where the top and bottom rows are out-of-phase, corresponding to an LSC rocking back and forth around the horizontal axis in the LSC plane, similar to the rocking mode found in a horizontal cylinder \cite{SBHT14}.

For cross  correlations between $\theta_0$ and $\alpha$, the predicted correlation function is $C_{\alpha_i,\theta_j}(\tau) \propto - \cos(nk_0z_j) \sin(nk_0z_i) \int P_n(\omega) \sin(\omega \tau)d\omega$.  The integral again is an oscillatory function in $\tau$ as long as $P_n(\omega)$ has a resonance peak, but now with a peak near $\omega\tau = \pi/2$ for a phase shift of $\pi/2$.  If $\cos(nk_0z_i) > 0$ (e.g.~at the middle row, or for modes $n=1$ and $n=3$ at the top and bottom rows), then the cosine term is positive and the sign of the correlation is determined by the row of $\alpha$.  The correlation between $\theta_m$ with $\alpha_t$ is negative since $\sin(nk_0z_i) < 0$, while the correlation between $\theta_m$ with $\alpha_b$ is positive since $\sin(nk_0z_i) > 0$. For correlations with the middle row of $\alpha$ at $z=0$,  the corresponding sinusoid  in front of the integral would again be 0, resulting in no oscillation.  

For modes with even-$n$ order, the places of $\theta_0$ and $\alpha$ are switched from the odd $n$ calculation in this subsection.  For the $n=2$ mode, this reproduces the result found in circular cylinders, with an out-of-phase (twisting) oscillation in $\theta_0$, and and in-phase (sloshing) oscillation in $\alpha$ \cite{BA09}. 

\section*{Appendix 4: Non-Boussinesq effects resulting in asymmetry of the oscillation}

The oscillation in $\alpha_m$ that is $\pi$ rad out of phase with $\theta_m$ shown in Fig.~\ref{fig:corralphatheta} corresponds to an asymmetry in which $\alpha_c$ osicllates with a larger amplitude than $\alpha_h$.  This appears to be a non-Boussinesq effect in which a difference in material parameters at different temperatures on the hot and cold sides of the LSC  leads to different local effective values of model parameters that affect the motion of $\alpha_h$ and $\alpha_c$ (for example, the parameters  $\omega_r$, $D_{\dot\theta}$ and $\tau_{\dot\theta}$ in Eq.~\ref{eqn:model_alphah}).  

To test whether this asymmetry is a non-Boussinesq effect rather than due to some other asymmetry of the system, we compare the distributions of $\alpha_h$ and $\alpha_c$ when the flow direction has aligned with a different corner of the cell, since different results from corner to corner would come from asymmetries of the setup. Using the dataset that sampled all 4 potential wells  from Fig.~\ref{fig:ptheta_4well}, we calculate root-mean-square values of $\alpha_h$ and $\alpha_c$ for each well.  For each corner, we find $\alpha_{c,rms} > \alpha_{h,rms}$.  The means $\pm$ standard deviations of the four distributions are $\langle\alpha_{h,rms}\rangle = 0.32 \pm 0.07$ rad and $\langle\alpha_{c,rms}\rangle=0.44\pm0.06$ rad. This corresponds to a systematically larger variation of $\alpha_c$ by $36\%$.  This systematic difference is apparently a non-Boussinesq effect due to the temperature difference of the hot and cold sides.  The variation in $\alpha_{h,rms}$ and $\alpha_{c,rms}$ of 0.06 or 0.07 rad are much larger than the expected random variation for thousands of measurement points in each well, which could correspond to an asymmetry of the experimental setup.  Alternatively, this asymmetry could be due to a lack of ergodicity in which the system does not sample different asymmetric solutions long enough to cancel out those asymmetries in the average, although our primary dataset where this asymmetry was observed was 10 days long.

While the appearance of this non-Boussinesq oscillation mode may be of interest in its own right, since it does not have the symmetry between the hot and cold sides of the LSC of the idealized model we are testing in this paper, it will not be discussed further here.


\begin{thebibliography}{59}%
\makeatletter
\providecommand \@ifxundefined [1]{%
 \@ifx{#1\undefined}
}%
\providecommand \@ifnum [1]{%
 \ifnum #1\expandafter \@firstoftwo
 \else \expandafter \@secondoftwo
 \fi
}%
\providecommand \@ifx [1]{%
 \ifx #1\expandafter \@firstoftwo
 \else \expandafter \@secondoftwo
 \fi
}%
\providecommand \natexlab [1]{#1}%
\providecommand \enquote  [1]{``#1''}%
\providecommand \bibnamefont  [1]{#1}%
\providecommand \bibfnamefont [1]{#1}%
\providecommand \citenamefont [1]{#1}%
\providecommand \href@noop [0]{\@secondoftwo}%
\providecommand \href [0]{\begingroup \@sanitize@url \@href}%
\providecommand \@href[1]{\@@startlink{#1}\@@href}%
\providecommand \@@href[1]{\endgroup#1\@@endlink}%
\providecommand \@sanitize@url [0]{\catcode `\\12\catcode `\$12\catcode
  `\&12\catcode `\#12\catcode `\^12\catcode `\_12\catcode `\%12\relax}%
\providecommand \@@startlink[1]{}%
\providecommand \@@endlink[0]{}%
\providecommand \url  [0]{\begingroup\@sanitize@url \@url }%
\providecommand \@url [1]{\endgroup\@href {#1}{\urlprefix }}%
\providecommand \urlprefix  [0]{URL }%
\providecommand \Eprint [0]{\href }%
\providecommand \doibase [0]{http://dx.doi.org/}%
\providecommand \selectlanguage [0]{\@gobble}%
\providecommand \bibinfo  [0]{\@secondoftwo}%
\providecommand \bibfield  [0]{\@secondoftwo}%
\providecommand \translation [1]{[#1]}%
\providecommand \BibitemOpen [0]{}%
\providecommand \bibitemStop [0]{}%
\providecommand \bibitemNoStop [0]{.\EOS\space}%
\providecommand \EOS [0]{\spacefactor3000\relax}%
\providecommand \BibitemShut  [1]{\csname bibitem#1\endcsname}%
\let\auto@bib@innerbib\@empty
\bibitem [{\citenamefont {Lorenz}(1963)}]{Lo63}%
  \BibitemOpen
  \bibfield  {author} {\bibinfo {author} {\bibfnamefont {E.~N.}\ \bibnamefont
  {Lorenz}},\ }\bibfield  {title} {\enquote {\bibinfo {title} {Deterministic
  nonperiodic flow},}\ }\href@noop {} {\bibfield  {journal} {\bibinfo
  {journal} {J. Atmos. Sci}\ }\textbf {\bibinfo {volume} {20}},\ \bibinfo
  {pages} {130--141} (\bibinfo {year} {1963})}\BibitemShut {NoStop}%
\bibitem [{\citenamefont {Brown}\ and\ \citenamefont {Ahlers}(2007)}]{BA07a}%
  \BibitemOpen
  \bibfield  {author} {\bibinfo {author} {\bibfnamefont {E.}~\bibnamefont
  {Brown}}\ and\ \bibinfo {author} {\bibfnamefont {G.}~\bibnamefont {Ahlers}},\
  }\bibfield  {title} {\enquote {\bibinfo {title} {Large-scale circulation
  model of turbulent {{Rayleigh-B\'enard}} convection},}\ }\href@noop {}
  {\bibfield  {journal} {\bibinfo  {journal} {Phys. Rev. Lett.}\ }\textbf
  {\bibinfo {volume} {98}},\ \bibinfo {pages} {134501--1--4} (\bibinfo {year}
  {2007})}\BibitemShut {NoStop}%
\bibitem [{\citenamefont {de~la Torre}\ and\ \citenamefont
  {Burguete}(2007)}]{TB07}%
  \BibitemOpen
  \bibfield  {author} {\bibinfo {author} {\bibfnamefont {A.}~\bibnamefont
  {de~la Torre}}\ and\ \bibinfo {author} {\bibfnamefont {J.}~\bibnamefont
  {Burguete}},\ }\bibfield  {title} {\enquote {\bibinfo {title} {Slow
  {Dynamics} in a {Turbulent} von {K}{\textbackslash}'arm{\textbackslash}'an
  {Swirling} {Flow}},}\ }\href@noop {} {\bibfield  {journal} {\bibinfo
  {journal} {Phy. Rev. Lett.}\ }\textbf {\bibinfo {volume} {99}},\ \bibinfo
  {pages} {054101} (\bibinfo {year} {2007})}\BibitemShut {NoStop}%
\bibitem [{\citenamefont {Thual}\ \emph {et~al.}(2014)\citenamefont {Thual},
  \citenamefont {Majda},\ and\ \citenamefont {Stechmann}}]{TMS14}%
  \BibitemOpen
  \bibfield  {author} {\bibinfo {author} {\bibfnamefont {S.}~\bibnamefont
  {Thual}}, \bibinfo {author} {\bibfnamefont {A.J.}\ \bibnamefont {Majda}}, \
  and\ \bibinfo {author} {\bibfnamefont {S.N.}\ \bibnamefont {Stechmann}},\
  }\bibfield  {title} {\enquote {\bibinfo {title} {A stochastic skeleton model
  for the mjo},}\ }\href@noop {} {\bibfield  {journal} {\bibinfo  {journal} {J.
  of the Atmospheric Sciences}\ }\textbf {\bibinfo {volume} {71}},\ \bibinfo
  {pages} {697} (\bibinfo {year} {2014})}\BibitemShut {NoStop}%
\bibitem [{\citenamefont {Rigas}\ \emph {et~al.}(2015)\citenamefont {Rigas},
  \citenamefont {Morgans}, , \citenamefont {Brackston},\ and\ \citenamefont
  {Morrison}}]{RMBM15}%
  \BibitemOpen
  \bibfield  {author} {\bibinfo {author} {\bibfnamefont {Georgios}\
  \bibnamefont {Rigas}}, \bibinfo {author} {\bibfnamefont {Aimee~S.}\
  \bibnamefont {Morgans}}, , \bibinfo {author} {\bibfnamefont {R.D.}\
  \bibnamefont {Brackston}}, \ and\ \bibinfo {author} {\bibfnamefont
  {Jonathan~F.}\ \bibnamefont {Morrison}},\ }\bibfield  {title} {\enquote
  {\bibinfo {title} {Diffusive dynamics and stochastic models of turbulent
  axisymmetric wakes},}\ }\href@noop {} {\bibfield  {journal} {\bibinfo
  {journal} {J. Fluid Mech. (in review)}\ } (\bibinfo {year}
  {2015})}\BibitemShut {NoStop}%
\bibitem [{\citenamefont {Holmes}\ \emph {et~al.}(2012)\citenamefont {Holmes},
  \citenamefont {Lumley}, \citenamefont {Berkooz},\ and\ \citenamefont
  {Rowley}}]{HL96}%
  \BibitemOpen
  \bibfield  {author} {\bibinfo {author} {\bibfnamefont {Philip}\ \bibnamefont
  {Holmes}}, \bibinfo {author} {\bibfnamefont {John~L.}\ \bibnamefont
  {Lumley}}, \bibinfo {author} {\bibfnamefont {Gahl}\ \bibnamefont {Berkooz}},
  \ and\ \bibinfo {author} {\bibfnamefont {Clancy.~W.}\ \bibnamefont
  {Rowley}},\ }\href@noop {} {\emph {\bibinfo {title} {Turbulence, Coherent
  Structures, Dynamical Systems, and Symmetry}}}\ (\bibinfo  {publisher}
  {Cambridge University Press},\ \bibinfo {year} {2012})\BibitemShut {NoStop}%
\bibitem [{\citenamefont {Ahlers}\ \emph {et~al.}(2009)\citenamefont {Ahlers},
  \citenamefont {Grossmann},\ and\ \citenamefont {Lohse}}]{AGL09}%
  \BibitemOpen
  \bibfield  {author} {\bibinfo {author} {\bibfnamefont {G.}~\bibnamefont
  {Ahlers}}, \bibinfo {author} {\bibfnamefont {S.}~\bibnamefont {Grossmann}}, \
  and\ \bibinfo {author} {\bibfnamefont {D.}~\bibnamefont {Lohse}},\ }\bibfield
   {title} {\enquote {\bibinfo {title} {Heat transfer and large-scale dynamics
  in turbulent {{Rayleigh-B\'enard}} convection},}\ }\href@noop {} {\bibfield
  {journal} {\bibinfo  {journal} {Reviews of Modern Physics}\ }\textbf
  {\bibinfo {volume} {81}},\ \bibinfo {pages} {503--537} (\bibinfo {year}
  {2009})}\BibitemShut {NoStop}%
\bibitem [{\citenamefont {Lohse}\ and\ \citenamefont {Xia}(2010)}]{LX10}%
  \BibitemOpen
  \bibfield  {author} {\bibinfo {author} {\bibfnamefont {D.}~\bibnamefont
  {Lohse}}\ and\ \bibinfo {author} {\bibfnamefont {K.-Q.}\ \bibnamefont
  {Xia}},\ }\bibfield  {title} {\enquote {\bibinfo {title} {Small-scale
  properties of turbulent {{Rayleigh-B\'enard}} convection},}\ }\href@noop {}
  {\bibfield  {journal} {\bibinfo  {journal} {Annual Reviews of Fluid
  Mechanics}\ }\textbf {\bibinfo {volume} {42}},\ \bibinfo {pages} {335--364}
  (\bibinfo {year} {2010})}\BibitemShut {NoStop}%
\bibitem [{\citenamefont {Krishnamurti}\ and\ \citenamefont
  {Howard}(1981)}]{KH81}%
  \BibitemOpen
  \bibfield  {author} {\bibinfo {author} {\bibfnamefont {R.}~\bibnamefont
  {Krishnamurti}}\ and\ \bibinfo {author} {\bibfnamefont {L.~N.}\ \bibnamefont
  {Howard}},\ }\bibfield  {title} {\enquote {\bibinfo {title} {Large scale flow
  generation in turbulent convection},}\ }\href@noop {} {\bibfield  {journal}
  {\bibinfo  {journal} {Proc. Natl. Acad. Sci.}\ }\textbf {\bibinfo {volume}
  {78}},\ \bibinfo {pages} {1981--1985} (\bibinfo {year} {1981})}\BibitemShut
  {NoStop}%
\bibitem [{\citenamefont {Brown}\ and\ \citenamefont
  {Ahlers}(2006{\natexlab{a}})}]{BA06a}%
  \BibitemOpen
  \bibfield  {author} {\bibinfo {author} {\bibfnamefont {E.}~\bibnamefont
  {Brown}}\ and\ \bibinfo {author} {\bibfnamefont {G.}~\bibnamefont {Ahlers}},\
  }\bibfield  {title} {\enquote {\bibinfo {title} {Rotations and cessations of
  the large-scale circulation in turbulent {{Rayleigh-B\'enard}} convection},}\
  }\href@noop {} {\bibfield  {journal} {\bibinfo  {journal} {J. Fluid Mech.}\
  }\textbf {\bibinfo {volume} {568}},\ \bibinfo {pages} {351} (\bibinfo {year}
  {2006}{\natexlab{a}})}\BibitemShut {NoStop}%
\bibitem [{\citenamefont {Xi}\ and\ \citenamefont {Xia}(2007)}]{XX07}%
  \BibitemOpen
  \bibfield  {author} {\bibinfo {author} {\bibfnamefont {H.-D.}\ \bibnamefont
  {Xi}}\ and\ \bibinfo {author} {\bibfnamefont {K.-Q.}\ \bibnamefont {Xia}},\
  }\bibfield  {title} {\enquote {\bibinfo {title} {Cessations and reversals of
  the large-scale circulation in turbulent thermal convection},}\ }\href@noop
  {} {\bibfield  {journal} {\bibinfo  {journal} {Phys. Rev. E}\ }\textbf
  {\bibinfo {volume} {75}},\ \bibinfo {pages} {066307--1--5} (\bibinfo {year}
  {2007})}\BibitemShut {NoStop}%
\bibitem [{\citenamefont {Heslot}\ \emph {et~al.}(1987)\citenamefont {Heslot},
  \citenamefont {Castaing},\ and\ \citenamefont {Libchaber}}]{HCL87}%
  \BibitemOpen
  \bibfield  {author} {\bibinfo {author} {\bibfnamefont {F.}~\bibnamefont
  {Heslot}}, \bibinfo {author} {\bibfnamefont {B.}~\bibnamefont {Castaing}}, \
  and\ \bibinfo {author} {\bibfnamefont {A.}~\bibnamefont {Libchaber}},\
  }\bibfield  {title} {\enquote {\bibinfo {title} {Transition to turbulence in
  helium gas},}\ }\href@noop {} {\bibfield  {journal} {\bibinfo  {journal}
  {Phys. Rev. A}\ }\textbf {\bibinfo {volume} {36}},\ \bibinfo {pages}
  {5870--5873} (\bibinfo {year} {1987})}\BibitemShut {NoStop}%
\bibitem [{\citenamefont {Sano}\ \emph {et~al.}(1989)\citenamefont {Sano},
  \citenamefont {Wu},\ and\ \citenamefont {Libchaber}}]{SWL89}%
  \BibitemOpen
  \bibfield  {author} {\bibinfo {author} {\bibfnamefont {M.}~\bibnamefont
  {Sano}}, \bibinfo {author} {\bibfnamefont {X.~Z.}\ \bibnamefont {Wu}}, \ and\
  \bibinfo {author} {\bibfnamefont {A.}~\bibnamefont {Libchaber}},\ }\href@noop
  {} {\bibfield  {journal} {\bibinfo  {journal} {Phys. Rev. A}\ }\textbf
  {\bibinfo {volume} {40}},\ \bibinfo {pages} {6421} (\bibinfo {year}
  {1989})}\BibitemShut {NoStop}%
\bibitem [{\citenamefont {Castaing}\ \emph {et~al.}(1989)\citenamefont
  {Castaing}, \citenamefont {Gunaratne}, \citenamefont {Heslot}, \citenamefont
  {Kadanoff}, \citenamefont {Libchaber}, \citenamefont {Thomae}, \citenamefont
  {Wu}, \citenamefont {Zaleski},\ and\ \citenamefont {Zanetti}}]{CGHKLTWZZ89}%
  \BibitemOpen
  \bibfield  {author} {\bibinfo {author} {\bibfnamefont {B.}~\bibnamefont
  {Castaing}}, \bibinfo {author} {\bibfnamefont {G.}~\bibnamefont {Gunaratne}},
  \bibinfo {author} {\bibfnamefont {F.}~\bibnamefont {Heslot}}, \bibinfo
  {author} {\bibfnamefont {L.}~\bibnamefont {Kadanoff}}, \bibinfo {author}
  {\bibfnamefont {A.}~\bibnamefont {Libchaber}}, \bibinfo {author}
  {\bibfnamefont {S.}~\bibnamefont {Thomae}}, \bibinfo {author} {\bibfnamefont
  {X.~Z.}\ \bibnamefont {Wu}}, \bibinfo {author} {\bibfnamefont
  {S.}~\bibnamefont {Zaleski}}, \ and\ \bibinfo {author} {\bibfnamefont
  {G.}~\bibnamefont {Zanetti}},\ }\bibfield  {title} {\enquote {\bibinfo
  {title} {Scaling of hard thermal turbulence in {{Rayleigh-B\'enard}}
  convection},}\ }\href@noop {} {\bibfield  {journal} {\bibinfo  {journal} {J.
  Fluid Mech.}\ }\textbf {\bibinfo {volume} {204}},\ \bibinfo {pages} {1--30}
  (\bibinfo {year} {1989})}\BibitemShut {NoStop}%
\bibitem [{\citenamefont {Ciliberto}\ \emph {et~al.}(1996)\citenamefont
  {Ciliberto}, \citenamefont {Cioni},\ and\ \citenamefont {Laroche}}]{CCL96}%
  \BibitemOpen
  \bibfield  {author} {\bibinfo {author} {\bibfnamefont {S.}~\bibnamefont
  {Ciliberto}}, \bibinfo {author} {\bibfnamefont {S.}~\bibnamefont {Cioni}}, \
  and\ \bibinfo {author} {\bibfnamefont {C.}~\bibnamefont {Laroche}},\
  }\bibfield  {title} {\enquote {\bibinfo {title} {Large-scale flow properties
  of turbulent thermal convection},}\ }\href@noop {} {\bibfield  {journal}
  {\bibinfo  {journal} {Phys. Rev. E}\ }\textbf {\bibinfo {volume} {54}},\
  \bibinfo {pages} {R5901--R5904} (\bibinfo {year} {1996})}\BibitemShut
  {NoStop}%
\bibitem [{\citenamefont {Takeshita}\ \emph {et~al.}(1996)\citenamefont
  {Takeshita}, \citenamefont {Segawa}, \citenamefont {Glazier},\ and\
  \citenamefont {Sano}}]{TSGS96}%
  \BibitemOpen
  \bibfield  {author} {\bibinfo {author} {\bibfnamefont {T.}~\bibnamefont
  {Takeshita}}, \bibinfo {author} {\bibfnamefont {T.}~\bibnamefont {Segawa}},
  \bibinfo {author} {\bibfnamefont {J.~A.}\ \bibnamefont {Glazier}}, \ and\
  \bibinfo {author} {\bibfnamefont {M.}~\bibnamefont {Sano}},\ }\bibfield
  {title} {\enquote {\bibinfo {title} {Thermal turbulence in mercury},}\
  }\href@noop {} {\bibfield  {journal} {\bibinfo  {journal} {Phys. Rev. Lett.}\
  }\textbf {\bibinfo {volume} {76}},\ \bibinfo {pages} {1465--1468} (\bibinfo
  {year} {1996})}\BibitemShut {NoStop}%
\bibitem [{\citenamefont {Cioni}\ \emph {et~al.}(1997)\citenamefont {Cioni},
  \citenamefont {Ciliberto},\ and\ \citenamefont {Sommeria}}]{CCS97}%
  \BibitemOpen
  \bibfield  {author} {\bibinfo {author} {\bibfnamefont {S.}~\bibnamefont
  {Cioni}}, \bibinfo {author} {\bibfnamefont {S.}~\bibnamefont {Ciliberto}}, \
  and\ \bibinfo {author} {\bibfnamefont {J.}~\bibnamefont {Sommeria}},\
  }\bibfield  {title} {\enquote {\bibinfo {title} {Strongly turbulent
  {{Rayleigh-B\'enard}} convection in mercury: comparison with results at
  moderate {{Prandtl}} number},}\ }\href@noop {} {\bibfield  {journal}
  {\bibinfo  {journal} {J. Fluid Mech.}\ }\textbf {\bibinfo {volume} {335}},\
  \bibinfo {pages} {111--140} (\bibinfo {year} {1997})}\BibitemShut {NoStop}%
\bibitem [{\citenamefont {Qiu}\ \emph {et~al.}(2000)\citenamefont {Qiu},
  \citenamefont {Yao},\ and\ \citenamefont {Tong}}]{QYT00}%
  \BibitemOpen
  \bibfield  {author} {\bibinfo {author} {\bibfnamefont {X.~L.}\ \bibnamefont
  {Qiu}}, \bibinfo {author} {\bibfnamefont {S.~H.}\ \bibnamefont {Yao}}, \ and\
  \bibinfo {author} {\bibfnamefont {P.}~\bibnamefont {Tong}},\ }\bibfield
  {title} {\enquote {\bibinfo {title} {Large-scale coherent rotation and
  oscillation in turbulent thermal convection},}\ }\href@noop {} {\bibfield
  {journal} {\bibinfo  {journal} {Phys. Rev. E}\ }\textbf {\bibinfo {volume}
  {61}},\ \bibinfo {pages} {R6075} (\bibinfo {year} {2000})}\BibitemShut
  {NoStop}%
\bibitem [{\citenamefont {Qiu}\ and\ \citenamefont {Tong}(2001)}]{QT01b}%
  \BibitemOpen
  \bibfield  {author} {\bibinfo {author} {\bibfnamefont {X.~L.}\ \bibnamefont
  {Qiu}}\ and\ \bibinfo {author} {\bibfnamefont {P.}~\bibnamefont {Tong}},\
  }\bibfield  {title} {\enquote {\bibinfo {title} {Onset of coherent
  oscillations in turbulent {{Rayleigh-B\'enard}} convection},}\ }\href@noop {}
  {\bibfield  {journal} {\bibinfo  {journal} {Phys. Rev. Lett}\ }\textbf
  {\bibinfo {volume} {87}},\ \bibinfo {pages} {094501} (\bibinfo {year}
  {2001})}\BibitemShut {NoStop}%
\bibitem [{\citenamefont {Niemela}\ \emph {et~al.}(2001)\citenamefont
  {Niemela}, \citenamefont {Skrbek}, \citenamefont {Sreenivasan},\ and\
  \citenamefont {Donnelly}}]{NSSD01}%
  \BibitemOpen
  \bibfield  {author} {\bibinfo {author} {\bibfnamefont {J.}~\bibnamefont
  {Niemela}}, \bibinfo {author} {\bibfnamefont {L.}~\bibnamefont {Skrbek}},
  \bibinfo {author} {\bibfnamefont {K.~R.}\ \bibnamefont {Sreenivasan}}, \ and\
  \bibinfo {author} {\bibfnamefont {R.~J.}\ \bibnamefont {Donnelly}},\
  }\bibfield  {title} {\enquote {\bibinfo {title} {The wind in confined thermal
  turbulence},}\ }\href@noop {} {\bibfield  {journal} {\bibinfo  {journal} {J.
  Fluid Mech.}\ }\textbf {\bibinfo {volume} {449}},\ \bibinfo {pages}
  {169--178} (\bibinfo {year} {2001})}\BibitemShut {NoStop}%
\bibitem [{\citenamefont {Qiu}\ and\ \citenamefont {Tong}(2002)}]{QT02}%
  \BibitemOpen
  \bibfield  {author} {\bibinfo {author} {\bibfnamefont {X.~L.}\ \bibnamefont
  {Qiu}}\ and\ \bibinfo {author} {\bibfnamefont {P.}~\bibnamefont {Tong}},\
  }\bibfield  {title} {\enquote {\bibinfo {title} {Temperature oscillations in
  turbulent rayleigh-benard convection},}\ }\href@noop {} {\bibfield  {journal}
  {\bibinfo  {journal} {Phys. Rev. E}\ }\textbf {\bibinfo {volume} {66}},\
  \bibinfo {pages} {026308} (\bibinfo {year} {2002})}\BibitemShut {NoStop}%
\bibitem [{\citenamefont {Qiu}\ \emph {et~al.}(2004)\citenamefont {Qiu},
  \citenamefont {Shang}, \citenamefont {Tong},\ and\ \citenamefont
  {Xia}}]{QSTX04}%
  \BibitemOpen
  \bibfield  {author} {\bibinfo {author} {\bibfnamefont {X.~L.}\ \bibnamefont
  {Qiu}}, \bibinfo {author} {\bibfnamefont {X.~D.}\ \bibnamefont {Shang}},
  \bibinfo {author} {\bibfnamefont {P.}~\bibnamefont {Tong}}, \ and\ \bibinfo
  {author} {\bibfnamefont {K.-Q.}\ \bibnamefont {Xia}},\ }\bibfield  {title}
  {\enquote {\bibinfo {title} {Velocity oscillations in turbulent
  {{Rayleigh-B\'enard}} convection},}\ }\href@noop {} {\bibfield  {journal}
  {\bibinfo  {journal} {Phys. Fluids.}\ }\textbf {\bibinfo {volume} {16}},\
  \bibinfo {pages} {412--423} (\bibinfo {year} {2004})}\BibitemShut {NoStop}%
\bibitem [{\citenamefont {Sun}\ \emph {et~al.}(2005)\citenamefont {Sun},
  \citenamefont {Xia},\ and\ \citenamefont {Tong}}]{SXT05}%
  \BibitemOpen
  \bibfield  {author} {\bibinfo {author} {\bibfnamefont {C.}~\bibnamefont
  {Sun}}, \bibinfo {author} {\bibfnamefont {K.~Q.}\ \bibnamefont {Xia}}, \ and\
  \bibinfo {author} {\bibfnamefont {P.}~\bibnamefont {Tong}},\ }\bibfield
  {title} {\enquote {\bibinfo {title} {Three-dimensional flow structures and
  dynamics of turbulent thermal convection in a cylindrical cell},}\
  }\href@noop {} {\bibfield  {journal} {\bibinfo  {journal} {Phys. Rev. E}\
  }\textbf {\bibinfo {volume} {72}},\ \bibinfo {pages} {026302} (\bibinfo
  {year} {2005})}\BibitemShut {NoStop}%
\bibitem [{\citenamefont {Tsuji}\ \emph {et~al.}(2005)\citenamefont {Tsuji},
  \citenamefont {Mizuno}, \citenamefont {Mashiko},\ and\ \citenamefont
  {Sano}}]{TMMS05}%
  \BibitemOpen
  \bibfield  {author} {\bibinfo {author} {\bibfnamefont {Y.}~\bibnamefont
  {Tsuji}}, \bibinfo {author} {\bibfnamefont {T.}~\bibnamefont {Mizuno}},
  \bibinfo {author} {\bibfnamefont {T.}~\bibnamefont {Mashiko}}, \ and\
  \bibinfo {author} {\bibfnamefont {M.}~\bibnamefont {Sano}},\ }\bibfield
  {title} {\enquote {\bibinfo {title} {Mean wind in convective turbulence of
  mercury},}\ }\href@noop {} {\bibfield  {journal} {\bibinfo  {journal} {Phys.
  Rev. Lett.}\ }\textbf {\bibinfo {volume} {94}},\ \bibinfo {pages} {034501}
  (\bibinfo {year} {2005})}\BibitemShut {NoStop}%
\bibitem [{\citenamefont {Funfschilling}\ and\ \citenamefont
  {Ahlers}(2004)}]{FA04}%
  \BibitemOpen
  \bibfield  {author} {\bibinfo {author} {\bibfnamefont {D.}~\bibnamefont
  {Funfschilling}}\ and\ \bibinfo {author} {\bibfnamefont {G.}~\bibnamefont
  {Ahlers}},\ }\bibfield  {title} {\enquote {\bibinfo {title} {Plume motion and
  large scale circulation in a cylindrical {{Rayleigh-B\'enard}} cell},}\
  }\href@noop {} {\bibfield  {journal} {\bibinfo  {journal} {Phys. Rev. Lett.}\
  }\textbf {\bibinfo {volume} {92}},\ \bibinfo {pages} {194502} (\bibinfo
  {year} {2004})}\BibitemShut {NoStop}%
\bibitem [{\citenamefont {Xi}\ \emph {et~al.}(2009)\citenamefont {Xi},
  \citenamefont {Zhou}, \citenamefont {Zhou}, \citenamefont {Chan},\ and\
  \citenamefont {Xia}}]{XZZCX09}%
  \BibitemOpen
  \bibfield  {author} {\bibinfo {author} {\bibfnamefont {H.-D.}\ \bibnamefont
  {Xi}}, \bibinfo {author} {\bibfnamefont {S.-Q.}\ \bibnamefont {Zhou}},
  \bibinfo {author} {\bibfnamefont {Q.}~\bibnamefont {Zhou}}, \bibinfo {author}
  {\bibfnamefont {T.-S.}\ \bibnamefont {Chan}}, \ and\ \bibinfo {author}
  {\bibfnamefont {K.-Q.}\ \bibnamefont {Xia}},\ }\bibfield  {title} {\enquote
  {\bibinfo {title} {Origin of the temperature oscillation in turbulent thermal
  convection},}\ }\href@noop {} {\bibfield  {journal} {\bibinfo  {journal}
  {Phys. Rev. Lett.}\ }\textbf {\bibinfo {volume} {102}},\ \bibinfo {pages}
  {044503--1--4} (\bibinfo {year} {2009})}\BibitemShut {NoStop}%
\bibitem [{\citenamefont {Zhou}\ \emph {et~al.}(2009)\citenamefont {Zhou},
  \citenamefont {Xi}, \citenamefont {Zhou}, \citenamefont {Sun},\ and\
  \citenamefont {Xia}}]{ZXZSX09}%
  \BibitemOpen
  \bibfield  {author} {\bibinfo {author} {\bibfnamefont {Q.}~\bibnamefont
  {Zhou}}, \bibinfo {author} {\bibfnamefont {H.-D.}\ \bibnamefont {Xi}},
  \bibinfo {author} {\bibfnamefont {S.-Q.}\ \bibnamefont {Zhou}}, \bibinfo
  {author} {\bibfnamefont {C.}~\bibnamefont {Sun}}, \ and\ \bibinfo {author}
  {\bibfnamefont {K.-Q.}\ \bibnamefont {Xia}},\ }\bibfield  {title} {\enquote
  {\bibinfo {title} {Oscillations of the large-scale circulation in turbulent
  {{Rayleigh-B\'enard}} convection: the sloshing mode and its relationship with
  the torsional mode},}\ }\href@noop {} {\bibfield  {journal} {\bibinfo
  {journal} {J. Fluid Mech.}\ }\textbf {\bibinfo {volume} {630}},\ \bibinfo
  {pages} {367--390} (\bibinfo {year} {2009})}\BibitemShut {NoStop}%
\bibitem [{\citenamefont {Brown}\ and\ \citenamefont {Ahlers}(2009)}]{BA09}%
  \BibitemOpen
  \bibfield  {author} {\bibinfo {author} {\bibfnamefont {E.}~\bibnamefont
  {Brown}}\ and\ \bibinfo {author} {\bibfnamefont {G.}~\bibnamefont {Ahlers}},\
  }\bibfield  {title} {\enquote {\bibinfo {title} {The origin of oscillations
  of the large-scale circulation of turbulent {{Rayleigh-B\'enard}}
  convection},}\ }\href@noop {} {\bibfield  {journal} {\bibinfo  {journal} {J.
  Fluid Mech.}\ }\textbf {\bibinfo {volume} {638}},\ \bibinfo {pages}
  {383--400} (\bibinfo {year} {2009})}\BibitemShut {NoStop}%
\bibitem [{\citenamefont {Vogt}\ \emph {et~al.}(2018)\citenamefont {Vogt},
  \citenamefont {Horn}, \citenamefont {Grannan},\ and\ \citenamefont
  {Aurnou}}]{VHGA18}%
  \BibitemOpen
  \bibfield  {author} {\bibinfo {author} {\bibfnamefont {T.}~\bibnamefont
  {Vogt}}, \bibinfo {author} {\bibfnamefont {S.}~\bibnamefont {Horn}}, \bibinfo
  {author} {\bibfnamefont {A.~M.}\ \bibnamefont {Grannan}}, \ and\ \bibinfo
  {author} {\bibfnamefont {J.~M.}\ \bibnamefont {Aurnou}},\ }\bibfield  {title}
  {\enquote {\bibinfo {title} {Jump rope vortex in liquid metal convection},}\
  }\href@noop {} {\bibfield  {journal} {\bibinfo  {journal} {Proc. Nat. Acad.
  Sciences}\ }\textbf {\bibinfo {volume} {115}},\ \bibinfo {pages}
  {12674--12679} (\bibinfo {year} {2018})}\BibitemShut {NoStop}%
\bibitem [{\citenamefont {Brown}\ and\ \citenamefont
  {Ahlers}(2006{\natexlab{b}})}]{BA06b}%
  \BibitemOpen
  \bibfield  {author} {\bibinfo {author} {\bibfnamefont {E.}~\bibnamefont
  {Brown}}\ and\ \bibinfo {author} {\bibfnamefont {G.}~\bibnamefont {Ahlers}},\
  }\bibfield  {title} {\enquote {\bibinfo {title} {Effect of the earth's
  coriolis force on turbulent {{Rayleigh-B\'enard}} convection in the
  laboratory},}\ }\href@noop {} {\bibfield  {journal} {\bibinfo  {journal}
  {Phys. Fluids}\ }\textbf {\bibinfo {volume} {18}},\ \bibinfo {pages}
  {125108--1--15} (\bibinfo {year} {2006}{\natexlab{b}})}\BibitemShut {NoStop}%
\bibitem [{\citenamefont {Zhong}\ \emph {et~al.}(2017)\citenamefont {Zhong},
  \citenamefont {Li},\ and\ \citenamefont {Wang}}]{ZLW17}%
  \BibitemOpen
  \bibfield  {author} {\bibinfo {author} {\bibfnamefont {J.-Q.}\ \bibnamefont
  {Zhong}}, \bibinfo {author} {\bibfnamefont {H.-M.}\ \bibnamefont {Li}}, \
  and\ \bibinfo {author} {\bibfnamefont {X.-Y.}\ \bibnamefont {Wang}},\
  }\bibfield  {title} {\enquote {\bibinfo {title} {Enhanced azimuthal rotation
  of the large-scale flow through stochastic cessations in turbulent rotating
  convection with large rossby numbers},}\ }\href@noop {} {\bibfield  {journal}
  {\bibinfo  {journal} {Phys. Rev. Fluids.}\ }\textbf {\bibinfo {volume} {2}},\
  \bibinfo {pages} {044602} (\bibinfo {year} {2017})}\BibitemShut {NoStop}%
\bibitem [{\citenamefont {Sterl}\ \emph {et~al.}(2016)\citenamefont {Sterl},
  \citenamefont {Li},\ and\ \citenamefont {Zhong}}]{SLZ16}%
  \BibitemOpen
  \bibfield  {author} {\bibinfo {author} {\bibfnamefont {S.}~\bibnamefont
  {Sterl}}, \bibinfo {author} {\bibfnamefont {H.-M.}\ \bibnamefont {Li}}, \
  and\ \bibinfo {author} {\bibfnamefont {J.-Q.}\ \bibnamefont {Zhong}},\
  }\bibfield  {title} {\enquote {\bibinfo {title} {Dynamical and statistical
  phenomena of circulation and heat transfer in periodically forced rotating
  turbulent rayleigh-b{'e}nard convection},}\ }\href@noop {} {\bibfield
  {journal} {\bibinfo  {journal} {Phys. Rev. Fluids.}\ }\textbf {\bibinfo
  {volume} {1}},\ \bibinfo {pages} {084401} (\bibinfo {year}
  {2016})}\BibitemShut {NoStop}%
\bibitem [{\citenamefont {Sreenivasan}\ \emph {et~al.}(2002)\citenamefont
  {Sreenivasan}, \citenamefont {Bershadski},\ and\ \citenamefont
  {Niemela}}]{SBN02}%
  \BibitemOpen
  \bibfield  {author} {\bibinfo {author} {\bibfnamefont {K.~R.}\ \bibnamefont
  {Sreenivasan}}, \bibinfo {author} {\bibfnamefont {A.}~\bibnamefont
  {Bershadski}}, \ and\ \bibinfo {author} {\bibfnamefont {J.J.}\ \bibnamefont
  {Niemela}},\ }\bibfield  {title} {\enquote {\bibinfo {title} {Mean wind and
  its reversals in thermal convection},}\ }\href@noop {} {\bibfield  {journal}
  {\bibinfo  {journal} {Phys. Rev. E}\ }\textbf {\bibinfo {volume} {65}},\
  \bibinfo {pages} {056306} (\bibinfo {year} {2002})}\BibitemShut {NoStop}%
\bibitem [{\citenamefont {Benzi}(2005)}]{Be05}%
  \BibitemOpen
  \bibfield  {author} {\bibinfo {author} {\bibfnamefont {R.}~\bibnamefont
  {Benzi}},\ }\bibfield  {title} {\enquote {\bibinfo {title} {Flow reversal in
  a simple dynamical model of turbulence},}\ }\href@noop {} {\bibfield
  {journal} {\bibinfo  {journal} {Phys. Rev. Lett.}\ }\textbf {\bibinfo
  {volume} {95}},\ \bibinfo {pages} {024502--1--4} (\bibinfo {year}
  {2005})}\BibitemShut {NoStop}%
\bibitem [{\citenamefont {{{Fontenele Araujo}}}\ \emph
  {et~al.}(2005)\citenamefont {{{Fontenele Araujo}}}, \citenamefont
  {Grossmann},\ and\ \citenamefont {Lohse}}]{FGL05}%
  \BibitemOpen
  \bibfield  {author} {\bibinfo {author} {\bibfnamefont {F.}~\bibnamefont
  {{{Fontenele Araujo}}}}, \bibinfo {author} {\bibfnamefont {S.}~\bibnamefont
  {Grossmann}}, \ and\ \bibinfo {author} {\bibfnamefont {D.}~\bibnamefont
  {Lohse}},\ }\bibfield  {title} {\enquote {\bibinfo {title} {Wind reversals in
  turbulent {{Rayleigh-B\'enard}} convection},}\ }\href@noop {} {\bibfield
  {journal} {\bibinfo  {journal} {Phys. Rev. Lett.}\ }\textbf {\bibinfo
  {volume} {95}},\ \bibinfo {pages} {084502} (\bibinfo {year}
  {2005})}\BibitemShut {NoStop}%
\bibitem [{\citenamefont {Resagk}\ \emph {et~al.}(2006)\citenamefont {Resagk},
  \citenamefont {du~Puits}, , \citenamefont {Thess}, \citenamefont
  {Dolzhansky}, \citenamefont {Grossmann}, \citenamefont {{Fontenele Araujo}},\
  and\ \citenamefont {Lohse}}]{RPTDGFL06}%
  \BibitemOpen
  \bibfield  {author} {\bibinfo {author} {\bibnamefont {Resagk}}, \bibinfo
  {author} {\bibfnamefont {R.}~\bibnamefont {du~Puits}}, , \bibinfo {author}
  {\bibfnamefont {A.}~\bibnamefont {Thess}}, \bibinfo {author} {\bibfnamefont
  {F.V.}\ \bibnamefont {Dolzhansky}}, \bibinfo {author} {\bibfnamefont
  {S.}~\bibnamefont {Grossmann}}, \bibinfo {author} {\bibfnamefont
  {F.}~\bibnamefont {{Fontenele Araujo}}}, \ and\ \bibinfo {author}
  {\bibfnamefont {D.}~\bibnamefont {Lohse}},\ }\bibfield  {title} {\enquote
  {\bibinfo {title} {Oscillations of the large scale wind in turbulent thermal
  convection},}\ }\href@noop {} {\bibfield  {journal} {\bibinfo  {journal}
  {Phys. Fluids}\ }\textbf {\bibinfo {volume} {18}},\ \bibinfo {pages}
  {095105--1 -- 095105--15} (\bibinfo {year} {2006})}\BibitemShut {NoStop}%
\bibitem [{\citenamefont {Brown}\ and\ \citenamefont
  {Ahlers}(2008{\natexlab{a}})}]{BA08a}%
  \BibitemOpen
  \bibfield  {author} {\bibinfo {author} {\bibfnamefont {E.}~\bibnamefont
  {Brown}}\ and\ \bibinfo {author} {\bibfnamefont {G.}~\bibnamefont {Ahlers}},\
  }\bibfield  {title} {\enquote {\bibinfo {title} {A model of diffusion in a
  potential well for the dynamics of the large-scale circulation in turbulent
  {{Rayleigh-B\'enard}} convection},}\ }\href@noop {} {\bibfield  {journal}
  {\bibinfo  {journal} {Phys. Fluids}\ }\textbf {\bibinfo {volume} {20}},\
  \bibinfo {pages} {075101--1--16} (\bibinfo {year}
  {2008}{\natexlab{a}})}\BibitemShut {NoStop}%
\bibitem [{\citenamefont {Chandra}\ and\ \citenamefont {Verma}(2011)}]{CV11}%
  \BibitemOpen
  \bibfield  {author} {\bibinfo {author} {\bibfnamefont {M.}~\bibnamefont
  {Chandra}}\ and\ \bibinfo {author} {\bibfnamefont {M.~K.}\ \bibnamefont
  {Verma}},\ }\bibfield  {title} {\enquote {\bibinfo {title} {Dynamics and
  symmetries of flow reversals in turbulent convection},}\ }\href@noop {}
  {\bibfield  {journal} {\bibinfo  {journal} {Physical Review E}\ }\textbf
  {\bibinfo {volume} {83}},\ \bibinfo {pages} {067303} (\bibinfo {year}
  {2011})}\BibitemShut {NoStop}%
\bibitem [{\citenamefont {Podvin}\ and\ \citenamefont {Sergent}(2015)}]{PS15}%
  \BibitemOpen
  \bibfield  {author} {\bibinfo {author} {\bibfnamefont {B.}~\bibnamefont
  {Podvin}}\ and\ \bibinfo {author} {\bibfnamefont {A.}~\bibnamefont
  {Sergent}},\ }\bibfield  {title} {\enquote {\bibinfo {title} {A large-scale
  investigation of wind reversal in a square rayleighÐ b{\'e}nard},}\
  }\href@noop {} {\bibfield  {journal} {\bibinfo  {journal} {J. Fluid Mech.}\
  }\textbf {\bibinfo {volume} {766}},\ \bibinfo {pages} {172--201} (\bibinfo
  {year} {2015})}\BibitemShut {NoStop}%
\bibitem [{\citenamefont {Giannakis}\ \emph {et~al.}(2018)\citenamefont
  {Giannakis}, \citenamefont {Kolchinskaya}, \citenamefont {Krasnov},\ and\
  \citenamefont {Schumacher}}]{GKKS18}%
  \BibitemOpen
  \bibfield  {author} {\bibinfo {author} {\bibfnamefont {D.}~\bibnamefont
  {Giannakis}}, \bibinfo {author} {\bibfnamefont {A.}~\bibnamefont
  {Kolchinskaya}}, \bibinfo {author} {\bibfnamefont {D.}~\bibnamefont
  {Krasnov}}, \ and\ \bibinfo {author} {\bibfnamefont {J.}~\bibnamefont
  {Schumacher}},\ }\bibfield  {title} {\enquote {\bibinfo {title} {Koopman
  analysis of the long-term evolution in a turbulent convection cell},}\
  }\href@noop {} {\bibfield  {journal} {\bibinfo  {journal} {J. Fluid Mech.}\
  }\textbf {\bibinfo {volume} {847}},\ \bibinfo {pages} {735--767} (\bibinfo
  {year} {2018})}\BibitemShut {NoStop}%
\bibitem [{\citenamefont {Vasiliev}\ \emph {et~al.}(2019)\citenamefont
  {Vasiliev}, \citenamefont {Frick}, \citenamefont {Kumar}, \citenamefont
  {Stepanov}, \citenamefont {Sukhanovskii},\ and\ \citenamefont
  {Verma}}]{VFKSSV19}%
  \BibitemOpen
  \bibfield  {author} {\bibinfo {author} {\bibfnamefont {A.}~\bibnamefont
  {Vasiliev}}, \bibinfo {author} {\bibfnamefont {P.}~\bibnamefont {Frick}},
  \bibinfo {author} {\bibfnamefont {A.}~\bibnamefont {Kumar}}, \bibinfo
  {author} {\bibfnamefont {R.}~\bibnamefont {Stepanov}}, \bibinfo {author}
  {\bibfnamefont {A.}~\bibnamefont {Sukhanovskii}}, \ and\ \bibinfo {author}
  {\bibfnamefont {M.~K.}\ \bibnamefont {Verma}},\ }\bibfield  {title} {\enquote
  {\bibinfo {title} {Transient flows and reorientations of large-scale
  convection in a cubic cell},}\ }\href@noop {} {\bibfield  {journal} {\bibinfo
   {journal} {International Comunications in Heat and Mass Transfer}\ }\textbf
  {\bibinfo {volume} {108}},\ \bibinfo {pages} {104319} (\bibinfo {year}
  {2019})}\BibitemShut {NoStop}%
\bibitem [{\citenamefont {Brown}\ and\ \citenamefont
  {Ahlers}(2008{\natexlab{b}})}]{BA08b}%
  \BibitemOpen
  \bibfield  {author} {\bibinfo {author} {\bibfnamefont {E.}~\bibnamefont
  {Brown}}\ and\ \bibinfo {author} {\bibfnamefont {G.}~\bibnamefont {Ahlers}},\
  }\bibfield  {title} {\enquote {\bibinfo {title} {Azimuthal asymmetries of the
  large-scale circulation in turbulent {{Rayleigh-B\'enard}} convection},}\
  }\href@noop {} {\bibfield  {journal} {\bibinfo  {journal} {Phys. Fluids}\
  }\textbf {\bibinfo {volume} {20}},\ \bibinfo {pages} {105105--1--15}
  (\bibinfo {year} {2008}{\natexlab{b}})}\BibitemShut {NoStop}%
\bibitem [{\citenamefont {Assaf}\ \emph {et~al.}(2011)\citenamefont {Assaf},
  \citenamefont {Angheluta},\ and\ \citenamefont {N.Goldenfeld}}]{AAG11}%
  \BibitemOpen
  \bibfield  {author} {\bibinfo {author} {\bibfnamefont {M.}~\bibnamefont
  {Assaf}}, \bibinfo {author} {\bibfnamefont {L.}~\bibnamefont {Angheluta}}, \
  and\ \bibinfo {author} {\bibnamefont {N.Goldenfeld}},\ }\bibfield  {title}
  {\enquote {\bibinfo {title} {Rare fluctuations and large-scale circulation
  cessations in turbulent convection},}\ }\href@noop {} {\bibfield  {journal}
  {\bibinfo  {journal} {Phys. Rev. Lett.}\ }\textbf {\bibinfo {volume} {107}},\
  \bibinfo {pages} {044502} (\bibinfo {year} {2011})}\BibitemShut {NoStop}%
\bibitem [{\citenamefont {Song}\ \emph {et~al.}(2014)\citenamefont {Song},
  \citenamefont {Brown}, \citenamefont {Hawkins},\ and\ \citenamefont
  {Tong}}]{SBHT14}%
  \BibitemOpen
  \bibfield  {author} {\bibinfo {author} {\bibfnamefont {H.}~\bibnamefont
  {Song}}, \bibinfo {author} {\bibfnamefont {E.}~\bibnamefont {Brown}},
  \bibinfo {author} {\bibfnamefont {R.}~\bibnamefont {Hawkins}}, \ and\
  \bibinfo {author} {\bibfnamefont {P}~\bibnamefont {Tong}},\ }\bibfield
  {title} {\enquote {\bibinfo {title} {Dynamics of large-scale circulation of
  turbulent thermal convection in a horizontal cylinder},}\ }\href@noop {}
  {\bibfield  {journal} {\bibinfo  {journal} {J. Fluid Mech}\ }\textbf
  {\bibinfo {volume} {740}},\ \bibinfo {pages} {136--167} (\bibinfo {year}
  {2014})}\BibitemShut {NoStop}%
\bibitem [{\citenamefont {Liu}\ and\ \citenamefont {Ecke}(2009)}]{LE09}%
  \BibitemOpen
  \bibfield  {author} {\bibinfo {author} {\bibfnamefont {Y.}~\bibnamefont
  {Liu}}\ and\ \bibinfo {author} {\bibfnamefont {R.~E.}\ \bibnamefont {Ecke}},\
  }\bibfield  {title} {\enquote {\bibinfo {title} {Heat transport measurements
  in turbulent rotating {{Rayleigh-B\'enard}} convection},}\ }\href {\doibase
  10.1103/PhysRevE.80.036314} {\bibfield  {journal} {\bibinfo  {journal} {Phys.
  Rev. E}\ }\textbf {\bibinfo {volume} {80}},\ \bibinfo {pages} {036314}
  (\bibinfo {year} {2009})}\BibitemShut {NoStop}%
\bibitem [{\citenamefont {Bai}\ \emph {et~al.}(2016)\citenamefont {Bai},
  \citenamefont {Ji},\ and\ \citenamefont {Brown}}]{BJB16}%
  \BibitemOpen
  \bibfield  {author} {\bibinfo {author} {\bibfnamefont {K.}~\bibnamefont
  {Bai}}, \bibinfo {author} {\bibfnamefont {D.}~\bibnamefont {Ji}}, \ and\
  \bibinfo {author} {\bibfnamefont {E.}~\bibnamefont {Brown}},\ }\bibfield
  {title} {\enquote {\bibinfo {title} {Ability of a low-dimensional model to
  predict geometry-dependent dynamics of large-scale coherent structures in
  turbulence},}\ }\href@noop {} {\bibfield  {journal} {\bibinfo  {journal}
  {Phys. Rev. E)}\ }\textbf {\bibinfo {volume} {93}},\ \bibinfo {pages}
  {023117--1--5} (\bibinfo {year} {2016})}\BibitemShut {NoStop}%
\bibitem [{\citenamefont {Foroozani}\ \emph {et~al.}(2017)\citenamefont
  {Foroozani}, \citenamefont {Niemela}, \citenamefont {Armenio},\ and\
  \citenamefont {Sreenivasan}}]{FNAS17}%
  \BibitemOpen
  \bibfield  {author} {\bibinfo {author} {\bibfnamefont {N.}~\bibnamefont
  {Foroozani}}, \bibinfo {author} {\bibfnamefont {J.J.}\ \bibnamefont
  {Niemela}}, \bibinfo {author} {\bibfnamefont {V.}~\bibnamefont {Armenio}}, \
  and\ \bibinfo {author} {\bibfnamefont {K.R.}\ \bibnamefont {Sreenivasan}},\
  }\bibfield  {title} {\enquote {\bibinfo {title} {Reorientations of the
  large-scale flow in turbulent convection in a cube},}\ }\href@noop {}
  {\bibfield  {journal} {\bibinfo  {journal} {Phys. Rev. E}\ }\textbf {\bibinfo
  {volume} {95}},\ \bibinfo {pages} {033107} (\bibinfo {year}
  {2017})}\BibitemShut {NoStop}%
\bibitem [{\citenamefont {Vasiliev}\ \emph {et~al.}(2016)\citenamefont
  {Vasiliev}, \citenamefont {Sukhanovskii}, \citenamefont {Frick},
  \citenamefont {Budnikov}, \citenamefont {Fomichev}, \citenamefont
  {Bolshukhin},\ and\ \citenamefont {Romanov}}]{VSFBFBR16}%
  \BibitemOpen
  \bibfield  {author} {\bibinfo {author} {\bibfnamefont {A.}~\bibnamefont
  {Vasiliev}}, \bibinfo {author} {\bibfnamefont {A.}~\bibnamefont
  {Sukhanovskii}}, \bibinfo {author} {\bibfnamefont {P.}~\bibnamefont {Frick}},
  \bibinfo {author} {\bibfnamefont {A.}~\bibnamefont {Budnikov}}, \bibinfo
  {author} {\bibfnamefont {V.}~\bibnamefont {Fomichev}}, \bibinfo {author}
  {\bibfnamefont {M.}~\bibnamefont {Bolshukhin}}, \ and\ \bibinfo {author}
  {\bibfnamefont {R.}~\bibnamefont {Romanov}},\ }\bibfield  {title} {\enquote
  {\bibinfo {title} {High rayleigh number convection in a cubic cell with
  adiabatic sidewalls},}\ }\href {\doibase
  https://doi.org/10.1016/j.ijheatmasstransfer.2016.06.015} {\bibfield
  {journal} {\bibinfo  {journal} {International Journal of Heat and Mass
  Transfer}\ }\textbf {\bibinfo {volume} {102}},\ \bibinfo {pages} {201 -- 212}
  (\bibinfo {year} {2016})}\BibitemShut {NoStop}%
\bibitem [{\citenamefont {Zimin}\ and\ \citenamefont
  {Ketov}(1978)}]{Zimin1978}%
  \BibitemOpen
  \bibfield  {author} {\bibinfo {author} {\bibfnamefont {V.~D.}\ \bibnamefont
  {Zimin}}\ and\ \bibinfo {author} {\bibfnamefont {A.~I.}\ \bibnamefont
  {Ketov}},\ }\bibfield  {title} {\enquote {\bibinfo {title} {Turbulent
  convection in a cubic cavity heated from below},}\ }\href {\doibase
  10.1007/BF01055112} {\bibfield  {journal} {\bibinfo  {journal} {Fluid
  Dynamics}\ }\textbf {\bibinfo {volume} {13}},\ \bibinfo {pages} {594--599}
  (\bibinfo {year} {1978})}\BibitemShut {NoStop}%
\bibitem [{\citenamefont {Zocchi}\ \emph {et~al.}(1990)\citenamefont {Zocchi},
  \citenamefont {Moses},\ and\ \citenamefont {Libchaber}}]{ZML90}%
  \BibitemOpen
  \bibfield  {author} {\bibinfo {author} {\bibfnamefont {G.}~\bibnamefont
  {Zocchi}}, \bibinfo {author} {\bibfnamefont {E.}~\bibnamefont {Moses}}, \
  and\ \bibinfo {author} {\bibfnamefont {A.}~\bibnamefont {Libchaber}},\
  }\bibfield  {title} {\enquote {\bibinfo {title} {Coherent structures in
  turbulent convection: an experimental study},}\ }\href@noop {} {\bibfield
  {journal} {\bibinfo  {journal} {Physica A}\ }\textbf {\bibinfo {volume}
  {166}},\ \bibinfo {pages} {387--407} (\bibinfo {year} {1990})}\BibitemShut
  {NoStop}%
\bibitem [{\citenamefont {Qiu}\ and\ \citenamefont {Xia}(1998)}]{QX98}%
  \BibitemOpen
  \bibfield  {author} {\bibinfo {author} {\bibfnamefont {X.~L.}\ \bibnamefont
  {Qiu}}\ and\ \bibinfo {author} {\bibfnamefont {K.-Q.}\ \bibnamefont {Xia}},\
  }\bibfield  {title} {\enquote {\bibinfo {title} {Viscous boundary layers at
  the sidewall of a convection cell},}\ }\href@noop {} {\bibfield  {journal}
  {\bibinfo  {journal} {Phys. Rev. E}\ }\textbf {\bibinfo {volume} {58}},\
  \bibinfo {pages} {486--491} (\bibinfo {year} {1998})}\BibitemShut {NoStop}%
\bibitem [{\citenamefont {Valencia}\ \emph {et~al.}(2007)\citenamefont
  {Valencia}, \citenamefont {Pallares}, \citenamefont {Cuesta},\ and\
  \citenamefont {Grau}}]{VPCG07}%
  \BibitemOpen
  \bibfield  {author} {\bibinfo {author} {\bibfnamefont {Leonardo}\
  \bibnamefont {Valencia}}, \bibinfo {author} {\bibfnamefont {Jordi}\
  \bibnamefont {Pallares}}, \bibinfo {author} {\bibfnamefont {Ildefonso}\
  \bibnamefont {Cuesta}}, \ and\ \bibinfo {author} {\bibfnamefont
  {Francesc~Xavier}\ \bibnamefont {Grau}},\ }\bibfield  {title} {\enquote
  {\bibinfo {title} {Turbulent {{Rayleigh-B\'enard}} convection of water in
  cubical cavities: A numerical and experimental study},}\ }\href {\doibase
  https://doi.org/10.1016/j.ijheatmasstransfer.2007.01.013} {\bibfield
  {journal} {\bibinfo  {journal} {International Journal of Heat and Mass
  Transfer}\ }\textbf {\bibinfo {volume} {50}},\ \bibinfo {pages} {3203 --
  3215} (\bibinfo {year} {2007})}\BibitemShut {NoStop}%
\bibitem [{\citenamefont {Funfschilling}\ \emph {et~al.}(2008)\citenamefont
  {Funfschilling}, \citenamefont {Brown},\ and\ \citenamefont
  {Ahlers}}]{FBA08}%
  \BibitemOpen
  \bibfield  {author} {\bibinfo {author} {\bibfnamefont {D.}~\bibnamefont
  {Funfschilling}}, \bibinfo {author} {\bibfnamefont {E.}~\bibnamefont
  {Brown}}, \ and\ \bibinfo {author} {\bibfnamefont {G.}~\bibnamefont
  {Ahlers}},\ }\bibfield  {title} {\enquote {\bibinfo {title} {Torsional
  oscillations of the large-scale circulation in turbulent
  {{Rayleigh-B\'enard}} convection},}\ }\href@noop {} {\bibfield  {journal}
  {\bibinfo  {journal} {J. Fluid. Mech.}\ }\textbf {\bibinfo {volume} {607}},\
  \bibinfo {pages} {119--139} (\bibinfo {year} {2008})}\BibitemShut {NoStop}%
\bibitem [{\citenamefont {Brown}\ \emph
  {et~al.}(2005{\natexlab{a}})\citenamefont {Brown}, \citenamefont
  {Nikolaenko},\ and\ \citenamefont {Ahlers}}]{BNA05}%
  \BibitemOpen
  \bibfield  {author} {\bibinfo {author} {\bibfnamefont {E.}~\bibnamefont
  {Brown}}, \bibinfo {author} {\bibfnamefont {A.}~\bibnamefont {Nikolaenko}}, \
  and\ \bibinfo {author} {\bibfnamefont {G.}~\bibnamefont {Ahlers}},\
  }\bibfield  {title} {\enquote {\bibinfo {title} {Reorientation of the
  large-scale circulation in turbulent {{Rayleigh-B\'enard}} convection},}\
  }\href@noop {} {\bibfield  {journal} {\bibinfo  {journal} {Phys. Rev. Lett}\
  }\textbf {\bibinfo {volume} {95}},\ \bibinfo {pages} {084503} (\bibinfo
  {year} {2005}{\natexlab{a}})}\BibitemShut {NoStop}%
\bibitem [{\citenamefont {Brown}\ \emph
  {et~al.}(2005{\natexlab{b}})\citenamefont {Brown}, \citenamefont
  {Funfschilling}, \citenamefont {Nikolaenko},\ and\ \citenamefont
  {Ahlers}}]{BFNA05}%
  \BibitemOpen
  \bibfield  {author} {\bibinfo {author} {\bibfnamefont {E.}~\bibnamefont
  {Brown}}, \bibinfo {author} {\bibfnamefont {D.}~\bibnamefont
  {Funfschilling}}, \bibinfo {author} {\bibfnamefont {A.}~\bibnamefont
  {Nikolaenko}}, \ and\ \bibinfo {author} {\bibfnamefont {G.}~\bibnamefont
  {Ahlers}},\ }\bibfield  {title} {\enquote {\bibinfo {title} {Heat transport
  by turbulent {{Rayleigh-B\'enard}} convection: Effect of finite top- and
  bottom conductivity},}\ }\href@noop {} {\bibfield  {journal} {\bibinfo
  {journal} {Phys. Fluids}\ }\textbf {\bibinfo {volume} {17}},\ \bibinfo
  {pages} {075108} (\bibinfo {year} {2005}{\natexlab{b}})}\BibitemShut
  {NoStop}%
\bibitem [{\citenamefont {Ji}\ and\ \citenamefont {Brown}(2020)}]{JB20b}%
  \BibitemOpen
  \bibfield  {author} {\bibinfo {author} {\bibfnamefont {D.}~\bibnamefont
  {Ji}}\ and\ \bibinfo {author} {\bibfnamefont {E.}~\bibnamefont {Brown}},\
  }\bibfield  {title} {\enquote {\bibinfo {title} {Oscillation in the
  temperature profile of the large-scale circulation of turbulent convection
  induced by a cubic container},}\ }\href@noop {} {\bibfield  {journal}
  {\bibinfo  {journal} {arXiv:2003.00067}\ } (\bibinfo {year}
  {2020})}\BibitemShut {NoStop}%
\bibitem [{\citenamefont {Kramers}(1940)}]{Kr40}%
  \BibitemOpen
  \bibfield  {author} {\bibinfo {author} {\bibfnamefont {H.A.}\ \bibnamefont
  {Kramers}},\ }\bibfield  {title} {\enquote {\bibinfo {title} {Brownian motion
  in a filed of force and the diffusion model of chemical reactions},}\
  }\href@noop {} {\bibfield  {journal} {\bibinfo  {journal} {Physica}\ }\textbf
  {\bibinfo {volume} {7}},\ \bibinfo {pages} {284--304} (\bibinfo {year}
  {1940})}\BibitemShut {NoStop}%
\bibitem [{\citenamefont {Villermaux}(1995)}]{Vi95}%
  \BibitemOpen
  \bibfield  {author} {\bibinfo {author} {\bibfnamefont {E.}~\bibnamefont
  {Villermaux}},\ }\bibfield  {title} {\enquote {\bibinfo {title}
  {Memory-induced low frequency oscillations in closed convection boxes},}\
  }\href@noop {} {\bibfield  {journal} {\bibinfo  {journal} {Phys. Rev. Lett.}\
  }\textbf {\bibinfo {volume} {75}},\ \bibinfo {pages} {4618--4621} (\bibinfo
  {year} {1995})}\BibitemShut {NoStop}%
\bibitem [{\citenamefont {Brown}\ \emph {et~al.}(2007)\citenamefont {Brown},
  \citenamefont {Funfschilling},\ and\ \citenamefont {Ahlers}}]{BFA07}%
  \BibitemOpen
  \bibfield  {author} {\bibinfo {author} {\bibfnamefont {E.}~\bibnamefont
  {Brown}}, \bibinfo {author} {\bibfnamefont {D.}~\bibnamefont
  {Funfschilling}}, \ and\ \bibinfo {author} {\bibfnamefont {G.}~\bibnamefont
  {Ahlers}},\ }\bibfield  {title} {\enquote {\bibinfo {title} {Anomalous
  reynolds-number scaling in turbulent {{Rayleigh-B\'enard}} convection},}\
  }\href@noop {} {\bibfield  {journal} {\bibinfo  {journal} {J. Stat. Mech.}\
  ,\ \bibinfo {pages} {P10005}} (\bibinfo {year} {2007})}\BibitemShut {NoStop}%
\end{thebibliography}

%

\end{document}